\documentclass[12pt,letterpaper]{article}
\usepackage{jheppub}
\pdfoutput=1

\usepackage{amsmath, amsthm, amsfonts, amssymb}
\usepackage{graphicx}
\usepackage{multirow}
\usepackage{float}
\usepackage{afterpage}

\newcommand\Lame {Lam\'e\ }
\newcommand\Backlund {B\"{a}cklund }
\newcommand\Schrodinger {Schr\"{o}dinger }
\newcommand\Poschl  {P\"{o}schl}
\newcommand\diag {\mathrm{diag}}

\title{On Elliptic String Solutions in AdS$_3$ and dS$_3$}
\author[]{Ioannis Bakas}
\author[]{and Georgios Pastras}
\affiliation[]{Department of Physics, School of Applied Mathematics and Physical Sciences\\ National Technical University, Athens 15780, Greece}
\emailAdd{bakas@mail.ntua.gr}
\emailAdd{pastras@mail.ntua.gr}

\abstract{Classical string actions in AdS$_3$ and dS$_3$ can be connected to the sinh-Gordon and cosh-Gordon equations through Pohlmeyer reduction. We show that the problem of constructing a classical string solution with a given static or translationally invariant Pohlmeyer counterpart is equivalent to solving four pairs of effective \Schrodinger problems. Each pair consists of a flat potential and an $n = 1$ \Lame potential whose eigenvalues are connected, and, additionally, the four solutions satisfy a set of constraints. An approach for solving this system is developed by employing an interesting connection between the specific class of classical string solutions and the band structure of the \Lame potential. This method is used for the construction of several families of classical string solutions, one of which turns out to be the spiky strings in AdS$_3$. New solutions include circular rotating strings in AdS$_3$ with singular time evolution of their radius and angular velocity as well as classical string solutions in dS$_3$.}

\keywords{Classical string solutions, Pohlmeyer reduction, AdS/CFT, integrable systems}
\arxivnumber{}

\begin{document}

\maketitle


\section{Introduction}
\label{sec:Introduction}

It has been known for a long time that two-dimensional non-linear sigma models (NLSM) defined on a symmetric target space can be reduced to integrable systems of the type of sine-Gordon equation. The first such example is the correspondence of the O$(3)$ NLSM to the sine-Gordon equation \cite{Pohlmeyer:1975nb,Zakharov:1973pp}, known as Pohlmeyer reduction, which was subsequently generalized to O$(N)$ sigma models \cite{Eichenherr:1979yw,Pohlmeyer:1979ch} and CP$(N)$ models \cite{Eichenherr:1979uk}. Since then, the understanding of Pohlmeyer reduction has significantly advanced. The connection between the target space being a symmetric space and the integrability of the sigma model has been understood \cite{Eichenherr:1979ci,Eichenherr:1979hz} and interpreted geometrically as Gauss-Codazzi equations for the embedding of the submanifold of the NLSM solution within the target space \cite{Lund:1976xd}. Although Pohlmeyer reduction leads to an integrable Hamiltonian system, it is a non-trivial question whether the dynamics of this system can be derived from a local Lagrangian. This question was answered through the relation of the integrable system with gauged WZW models with an additional integrable potential \cite{Bakas:1993xh,Bakas:1995bm,FernandezPousa:1996hi,Miramontes:2008wt}, which allows to develop a systematic approach for a local action of the reduced system.

Pohlmeyer reduction can be applied to the NLSM describing strings propagating in constant curvature spaces \cite{Barbashov:1980kz,DeVega:1992xc,Larsen:1996gn}. In the context of holographic dualities \cite{Maldacena:1997re,Gubser:1998bc,Witten:1998qj}, it is particularly interesting to study the reduction of string actions in AdS$_5 \times$S$^5$ \cite{Grigoriev:2007bu,Mikhailov:2007xr,Grigoriev:2008jq}, where kink solutions of the reduced integrable systems were understood as magnon solutions \cite{Hofman:2006xt,Chen:2006gea} and likewise in AdS$_4 \times$CP$^3$ \cite{Rashkov:2008rm} in the framework of the ABJM conjecture \cite{Aharony:2008ug}.

Classical string solutions in AdS spaces have shed light to several aspects of the AdS/CFT correspondence \cite{Gubser:2002tv,Tseytlin:2004xa,Plefka:2005bk}. Interest in those solutions was further inspired following the work by Alday and Maldacena that provided a prescription for computing gluon scattering amplitudes in terms of classical string solutions with boundary conditions related to the gluon momenta \cite{Alday:2007hr,Alday:2009yn}. Based on this prescription, the classical string solution found in \cite{Kruczenski:2002fb} was used to calculate the four gluon scattering amplitude \cite{Alday:2007hr} and to demonstrate that it matches the conjectured iteration relations for perturbative calculations of scattering amplitudes at several loops \cite{Bern:2005iz}.

The reduced models for strings propagating in constant curvature spaces are typically some sort of multi-component generalization of the sine-Gordon or sinh-Gordon equation, which are characterized by a set of first order relations connecting pairs of solutions, known as \Backlund transformations. These transformations can be used inductively to construct families of kink solutions starting from a given solution, such as the vacuum solution. Special emphasis has been placed on the construction of the NLSM counterparts of kink solutions in AdS backgrounds \cite{Jevicki:2007aa,Klose:2008rx,Jevicki:2008mm,Hollowood:2009tw,Dorn:2009kq}.

The simplest examples for string propagation in constant curvature spaces are related to AdS$_3$ and dS$_3$, as their Pohlmeyer reduction leads to a single component integrable system. In AdS$_3$, a family of interesting solutions, possessing a finite number of singular points, were discovered some years ago \cite{Kruczenski:2004wg,Mosaffa:2007ty,Kruczenski:2008bs,Jevicki:2009uz}. These solutions, which are called spiky strings, are interesting in the framework of AdS/CFT, as they can be related to single trace operators of the dual CFT \cite{Kruczenski:2004wg}. Spiky strings correspond to elliptic solutions of the sinh-Gordon equation. Their study in the Pohlmeyer reduced theory poses an interesting question and, as it turns out, it can also lead to new families of solutions.

The relation between the degrees of freedom of the NLSM and those of the reduced theory is highly non-local, resulting in great difficulty to invert the procedure. In particular, it is not clear how exactly a solution of the Pohlmeyer reduced theory corresponds to one or more physically distinct solutions of the original NLSM. It is our prime directive to provide a mapping of known classical string solutions to solutions of the Pohlmeyer reduced theory, develop generalizations, and discuss the inversion problem in more general terms than they have been discussed in the literature so far. In this paper, we develop a systematic method to build classical string solutions in AdS$_3$ or dS$_3$ that correspond to a special class of solutions of the reduced two-dimensional, integrable system, which are either static or translationally invariant and they can be expressed in terms of elliptic functions. Our method is based on an interesting connection between a certain class of string solutions and the band structure of the $n = 1$ \Lame potential.

In section \ref{sec:Pohlmeyer_AdS3}, we review the Pohlmeyer reduction of string actions in AdS$_3$ and dS$_3$ that results to the sinh-Gordon or cosh-Gordon equation, depending on circumstances. In section \ref{sec:Elliptic_Solutions}, we study solutions of the sinh-Gordon and cosh-Gordon equations that depend solely on one of the two coordinates and develop a uniform description for all in terms of Weierstrass elliptic functions. In section \ref{sec:Spiky_Strings}, we revisit the construction of spiky string solutions in AdS$_3$ and find their counterpart in the Pohlmeyer reduced theory. Next, in section \ref{sec:SUSYQM}, we show that the construction of a classical string solution with a given static or translationally invariant Pohlmeyer reduced counterpart can be reformulated in terms of four pairs of effective \Schrodinger problems. Each pair consists of a flat potential together with an $n = 1$ \Lame potential, whose eigenvalues are interrelated, and, additionally, the four solutions satisfy a set of constraints. We study the particular form of eigenstates of those problems, and, in section \ref{sec:construction}, we use them as building blocks to construct classical string solutions. Finally, in section \ref{sec:discussion}, we summarize our results and discuss possible extensions and applications to other directions. There is also an appendix containing the basic mathematical results about the Weierstrass functions used in the main text.

\section{Pohlmeyer Reduction of String Actions in AdS$_3$ and dS$_3$}
\label{sec:Pohlmeyer_AdS3}

Pohlmeyer reduction relates string actions in symmetric spaces like AdS and dS to integrable systems such as the sinh-Gordon or cosh-Gordon equation and multi-component generalizations thereof. In this section, following closely the literature \cite{Rashkov:2008rm,Jevicki:2007aa,Dorn:2009kq}, we review the Pohlmeyer reduction of string actions in AdS$_3$ and dS$_3$ spaces that result in the sinh-Gordon or cosh-Gordon equation, depending on circumstances, which are specified below.

The starting point of Pohlmeyer reduction is the embedding of the two-dimensional (world-sheet) manifold  into the symmetric target space of the NLSM, which is in turn embedded in a higher-dimensional flat space. Since we are considering strings propagating in AdS$_3$ or dS$_3$, this higher dimensional flat space is four-dimensional. We will denote the coordinates in the enhanced space as $Y^{-1}$, $Y^0$, $Y^1$ and $Y^2$.

In the case of AdS an extra time-like dimension is required, while in the case of dS an extra space-like dimension is required. Then, the enhanced four-dimensional flat space is $\mathbb{R}^{(2,2)}$ or $\mathbb{R}^{(1,3)}$, respectively, with metric $\eta_{\mu \nu} = \diag\{-1,s,1,1\}$, setting $s = -1$ for AdS and $s = + 1$ for dS. Then, AdS$_3$ or dS$_3$ is the submanifold
\begin{equation}
Y \cdot Y = s \Lambda^2,
\label{eq:Pohlmeyer_constraint}
\end{equation}
using the common notation
\begin{equation}
A \cdot B \equiv \eta_{\mu \nu} {A^\mu }{B^\nu } .
\end{equation}

\vspace{3pt}
\subsection{Action, Equations of Motion and Virasoro Conditions}
\label{subsec:Pohlmeyer_setup}
A two-dimensional time-like surface in AdS$_3$ or dS$_3$ parametrized with a time-like coordinate ${\xi_0}$ and a space-like coordinate ${\xi_1}$ has area given by the Polyakov action
\begin{equation}
S = \frac{1}{2}\int {d{\xi_0} d{\xi_1} \sqrt { - \gamma } \left( {\gamma ^{ab}}{\partial _a}Y \cdot {\partial _b}Y + \lambda \left( {Y \cdot Y - s {\Lambda ^2}} \right) \right) } .
\end{equation}
One can select the conformal gauge ${\gamma _{ab}} = {e^\omega }{\eta _{ab}}$, where $\eta_{ab} = \diag\{-1,1\}$. In this gauge ${\gamma ^{ab}} = {e^{ - \omega }}{\eta ^{ab}}$ and $ - \gamma  = {e^{2\omega }}$, thus the action can be written as
\begin{equation}
S = \frac{1}{2}\int {d{\xi_0} d{\xi_1} \left( {\eta ^{ab}}{\partial _a}Y \cdot {\partial _b}Y + \lambda \left( {Y \cdot Y - s {\Lambda ^2}} \right) \right) } .
\end{equation}
Finally, using the left- and right-moving coordinates ${\xi_ \pm } = \left( {{\xi_1}  \pm {\xi_0} } \right) / 2$, the action takes the form
\begin{equation}
S = \int {d{\xi_ + }d{\xi_ - } \left( {\partial _ + }Y \cdot {\partial _ - }Y + \lambda \left( {Y \cdot Y - s {\Lambda ^2}} \right) \right) } .
\end{equation}

Having set up the notation, the equations of motion for the field $Y$ are
\begin{equation}
{\partial _ + }{\partial _ - }Y = \lambda Y,
\end{equation}
whereas the equation of motion for the Lagrange multiplier $\lambda$ yields the constraint \eqref{eq:Pohlmeyer_constraint},
\begin{equation}
Y \cdot Y = s \Lambda^2 .
\end{equation}
In order to eliminate the Lagrange multiplier $\lambda$ from the equations of motion of the field $Y$ we note, as direct consequence of the constraint equation, that
\begin{equation}
{\partial _ \pm }Y \cdot Y = 0,
\label{eq:Pohlmeyer_v1_orth_v23}
\end{equation}
which upon differentiation yields
\begin{equation}
{\partial _ + }Y \cdot {\partial _ - }Y + {\partial _ + }{\partial _ - }Y \cdot Y = 0 .
\end{equation}
The equations of motion for the field $Y$ suggest that
\begin{equation}
{\partial _ + }{\partial _ - }Y \cdot Y = \lambda Y \cdot Y = s \lambda \Lambda^2,
\end{equation}
resulting to
\begin{equation}
\lambda  = - s \frac{1}{{{\Lambda ^2}}} {\partial _ + }Y \cdot {\partial _ - }Y.
\end{equation}
In turn, it allows to rewrite the equations of motion of $Y$ as
\begin{equation}
{\partial _ + }{\partial _ - }Y = - s \frac{1}{{{\Lambda ^2}}} \left( {{\partial _ + }Y \cdot {\partial _ - }Y} \right)Y .
\label{eq:eom}
\end{equation}

Varying the action with respect to the metric, one calculates the stress-energy tensor whose elements are
\begin{align}
{T_{ \pm  \pm }} &= {\partial _ \pm }Y \cdot {\partial _ \pm }Y ,\\
{T_{ +  - }} &= 0 .
\end{align}
${T_{ +  - }}$ vanishes identically as direct consequence of Weyl invariance, whereas the form of ${T_{ \pm  \pm }}$ implies that the Virasoro constraints take the form
\begin{equation}
{\partial _ \pm }Y \cdot {\partial _ \pm }Y = 0 ,
\label{eq:Pohlmeyer_Virasoro}
\end{equation}
so that both vectors ${\partial _ \pm }Y$ are null.

\subsection{The Pohlmeyer Reduction}
\label{subsec:Pohlmeyer_reduction}
It is convenient to use the notation $v_i$, $i=1,2,3,4$, to combine the vectors $Y$, ${\partial _ + }Y$ and ${\partial _ - }Y$ with one more vector $v_4$ as
\begin{equation}
{v_i} = \left\{ {Y,\,{\partial _ + }Y,\,{\partial _ - }Y,\,{v_4}} \right\} .
\end{equation}
The vectors ${\partial _ + }Y$ and ${\partial _ - }Y$ span the tangent space of the embedded two-dimensional surface in the enhanced target space. Since we study a string world-sheet, the tangent space contains one time-like and one space-like direction. Clearly, the basis of vectors $v_i$ should contain two time-like and two space-like vectors in the case of AdS$_3$ and one time-like and three space-like vectors in the case of dS$_3$. The vector $v_1$ is time-like in AdS$_3$ and space-like in dS$_3$, following equation \eqref{eq:Pohlmeyer_constraint} and so in both cases $v_4$ is space-like. Demanding that it is a unit vector orthogonal to $v_1$, $v_2$ and $v_3$, we have
\begin{align}
{v_4} \cdot Y = {v_4} \cdot {\partial _ \pm }Y &= 0 ,\\
{v_4} \cdot {v_4} &= 1 .
\end{align}

Since we restrict our attention to time-like surfaces, the inner product ${\partial _ + }Y \cdot {\partial _ - }Y$ is positive and we may define
\begin{equation}
{e^a} : = {\partial _ + }Y \cdot {\partial _ - }Y .
\label{eq:Pohlmeyer_a_definition}
\end{equation}
Thus, any vector $X$ in the enhanced target space admits the following decomposition,
\begin{equation}
X = s \frac{1}{{{\Lambda ^2}}}\left( {X \cdot {v_1}} \right){v_1} + {e^{ - a}}\left( {X \cdot {v_3}} \right){v_2} + {e^{ - a}}\left( {X \cdot {v_2}} \right){v_3} + \left( {X \cdot {v_4}} \right){v_4} .
\end{equation}

The derivatives of the vectors $v_i$ can be expressed as linear combination of the vectors $v_i$ themselves as follows, using the $4\times 4$ matrices $A^\pm$,
\begin{align}
{\partial _ + }{v_i} = A_{ij}^ + {v_j} ,\\
{\partial _ - }{v_i} = A_{ij}^ - {v_j} .
\end{align}
The derivatives of $v_1$ are the simplest and follow from the definition,
\begin{align}
{\partial _ + }{v_1} &= {\partial _ + }Y = {v_2} ,\\
{\partial _ - }{v_1} &= {\partial _ - }Y = {v_3} .
\end{align}
For the remaining components we have the following general relations
\begin{align}
{\partial _ + }{v_2} &= \partial _ + ^2Y = {a_0^{\left( + \right)}}Y + {a_ + ^{\left( + \right)}}{\partial _ + }Y + {a_ - ^{\left( + \right)}}{\partial _ - }Y + {a_4^{\left( + \right)}}{v_4} ,\\
{\partial _ - }{v_3} &= \partial _ - ^2Y = {a_0^{\left( - \right)}}Y + {a_ + ^{\left( - \right)}}{\partial _ + }Y + {a_ - ^{\left( - \right)}}{\partial _ - }Y + {a_4^{\left( - \right)}}{v_4} ,\\
{\partial _ + }{v_3} = {\partial _ - }{v_2} &= {\partial _ + }{\partial _ - }Y = {c_0}Y + {c_ + }{\partial _ + }Y + {c_ - }{\partial _ - }Y + {c_4}{v_4} .
\end{align}
Using the equations of motion for the field $Y$
\begin{equation}
{\partial _ + }{\partial _ - }Y = - s \frac{1}{{{\Lambda ^2}}}\left( {{\partial _ + }Y \cdot {\partial _ - }Y} \right)Y = - s \frac{1}{{{\Lambda ^2}}}{e^a}Y ,
\label{eq:Pohlmeyer_equation_fields_a}
\end{equation}
we obtain
\begin{equation}
{c_0} = - s \frac{1}{{{\Lambda ^2}}}{e^a},\quad {c_ + } = {c_ - } = {c_4} = 0 .
\end{equation}
Differentiating equation \eqref{eq:Pohlmeyer_v1_orth_v23}, ${\partial _ \pm }Y \cdot Y = 0$ and using the fact that $\partial _ \pm Y$ are null vectors, we get
\begin{equation}
\partial _ \pm ^2Y \cdot Y = 0 ,
\end{equation}
which implies that
\begin{equation}
{a_0^{\left( \pm \right)}} = 0 .
\end{equation}
Also, differentiating the Virasoro constraints \eqref{eq:Pohlmeyer_Virasoro}, we get
\begin{equation}
\partial _ \pm ^2 Y \cdot {\partial _ \pm } Y = 0 ,
\end{equation}
which implies that
\begin{equation}
{a_ \mp ^{\left( \pm \right)}} = 0 .
\end{equation}
Finally, differentiating the defining relation \eqref{eq:Pohlmeyer_a_definition} for the field $a$ and making use of the field equations and the orthogonality between $Y$ and $\partial _ \pm Y$, we get
\begin{equation}
\partial _ \pm ^2 Y \cdot {\partial _ \mp }Y = {\partial _ \pm }a{e^a} ,
\end{equation}
which implies that
\begin{equation}
{a_ \pm ^{\left( \pm \right)}} = {\partial _ \pm }a .
\end{equation}
Summarizing, we have the following decomposition for the derivatives of the vectors $v_2$ and $v_3$
\begin{align}
{\partial _ + }{v_2} &= \left({\partial _ + }a\right){v_2} + {a_4^{\left( + \right)}}{v_4} ,\\
{\partial _ - }{v_3} &= \left({\partial _ - }a\right){v_3} + {a_4^{\left( - \right)}}{v_4} ,\\
{\partial _ + }{v_3} = {\partial _ - }{v_2} &= - s {\frac{1}{{{\Lambda ^2}}}{e^a}}{v_1} .
\end{align}

As for the derivatives of the vector $v_4$, we make use of the orthogonality conditions to obtain
\begin{align}
{v_4} \cdot {v_4} &= 1 \Rightarrow {\partial _ \pm }{v_4} \cdot {v_4} = 0 , \\
{v_4} \cdot Y &= 0 \Rightarrow {\partial _ \pm }{v_4} \cdot Y =  - {v_4} \cdot {\partial _ \pm }Y = 0 , \\
{v_4} \cdot {\partial _ \pm }Y &= 0 \Rightarrow {\partial _ \pm }{v_4} \cdot {\partial _ \pm }Y =  - {v_4} \cdot \partial _ \pm ^2Y =  - {a_4^{\left( \pm \right)}} , \\
{v_4} \cdot {\partial _ \pm }Y &= 0 \Rightarrow {\partial _ \mp }{v_4} \cdot {\partial _ \pm }Y =  - {v_4} \cdot {\partial _ + }{\partial _ - }Y = 0 .
\end{align}
Thus, the derivatives of $v_4$ are equal to
\begin{align}
{\partial _ + }{v_4} &=  - {a_4^{\left( + \right)}}{e^{ - a}}{v_3} ,\\
{\partial _ - }{v_4} &=  - {a_4^{\left( - \right)}}{e^{ - a}}{v_2} .
\end{align}

Putting everything together, the matrices $A^\pm$ take the following form,
\begin{equation}
{A^ + } = \left( {\begin{array}{*{20}{c}}
0&1&0&0\\
0&{{\partial _ + }a}&0&{{a_4^{\left( + \right)}}}\\
{ - s \frac{1}{{{\Lambda ^2}}}{e^a}}&0&0&0\\
0&0&{ - {a_4^{\left( + \right)}}{e^{ - a}}}&0
\end{array}} \right) ,\quad
{A^ - } = \left( {\begin{array}{*{20}{c}}
0&0&1&0\\
{ - s \frac{1}{{{\Lambda ^2}}}{e^a}}&0&0&0\\
0&0&{{\partial _ - }a}&{{a_4^{\left( - \right)}}}\\
0&{ - {a_4^{\left( - \right)}}{e^{ - a}}}&0&0
\end{array}} \right) .
\end{equation}
They obey the compatibility condition
\begin{equation}
{\partial _ - }\left( {A_{ij}^ + {v_j}} \right) = {\partial _ + }\left( {A_{ij}^ - {v_j}} \right) \Rightarrow \left( {{\partial _ - }A_{ij}^ + } \right){v_j} + A_{ik}^ + A_{kj}^ - {v_j} = \left( {{\partial _ + }A_{ij}^ - } \right){v_j} + A_{ik}^ - A_{kj}^ + {v_j} ,
\end{equation}
which is equivalently written in matrix form as the zero-curvature condition
\begin{equation}
{\partial _ - }{A^ + } - {\partial _ + }{A^ - } + \left[ {{A^ + },{A^ - }} \right] = 0 .
\end{equation}

The zero-curvature condition imposes the following equations for the field variables $a$ and ${a_4^{\left( \pm \right)}}$,
\begin{align}
{\partial _ + }{\partial _ - }a &= - s \frac{1}{{{\Lambda ^2}}}{e^a} + {{a_4^{\left( + \right)}}{a_4^{\left( - \right)}}} {e^{ - a}}  , \label{eq:Pohlmeyer_zcc_1}\\
{\partial _ \mp }{a_4^{\left( \pm \right)}} &= 0 . \label{eq:Pohlmeyer_zcc_2}
\end{align}
Equations \eqref{eq:Pohlmeyer_zcc_2} are immediately solved as
\begin{equation}
{a_4^{\left( \pm \right)}} = f^{\left( \pm \right)}\left( {{\xi_ \pm }} \right) 
\end{equation}
and equation \eqref{eq:Pohlmeyer_zcc_1} takes the final form
\begin{equation}
{\partial _ + }{\partial _ - }a = - s \frac{1}{{{\Lambda ^2}}}{e^a} + f^{\left( + \right)}\left( {{\xi_ + }} \right)f^{\left( - \right)}\left( {{\xi_ - }} \right){e^{ - a}}  .
\label{eq:Pohlmeyer_equation_1}
\end{equation}

At this point, we distinguish three possible cases, depending on the sign of the product $s f^{\left( + \right)}\left( {{\xi_ + }} \right)f^{\left( - \right)}\left( {{\xi_ - }} \right) $.
If this combination does not vanish, we define
\begin{equation}
\varphi  := a - \frac{1}{2}\ln \left( {{\Lambda ^2} \left| f^{\left( + \right)}\left( {{\xi_ + }} \right)f^{\left( - \right)}\left( {{\xi_ - }} \right) \right|} \right) .
\label{eq:Pohlmeyer_field_redef}
\end{equation}
It is also convenient to transform the coordinates ${\xi_ \pm }' = {\xi_ \pm }'\left( {{\xi_ \pm }} \right)$ as
\begin{equation}
\frac{{d{\xi_ \pm }'}}{{d{\xi_ \pm }}} = \sqrt {\Lambda \left| {f^{\left( \pm \right)}\left( {{\xi_ \pm }} \right)} \right|} 
\label{eq:coordinate_rescaling}
\end{equation}
and then, drop the primes in the following for notational convenience. If the product $s f^{\left( + \right)}\left( {{\xi_ + }} \right)f^{\left( - \right)}\left( {{\xi_ - }} \right)$ is negative, equation \eqref{eq:Pohlmeyer_equation_1} will take the form
\begin{equation}
{\partial _ + }{\partial _ - }\varphi = - s \frac{2}{{{\Lambda ^2}}} { \cosh \varphi } ,
\label{eq:Pohlmeyer_cosh}
\end{equation}
which is the cosh-Gordon equation, whereas in the case the product $s f^{\left( + \right)}\left( {{\xi_ + }} \right)f^{\left( - \right)}\left( {{\xi_ - }} \right)$ is positive, equation \eqref{eq:Pohlmeyer_equation_1} is written as
\begin{equation}
{\partial _ + }{\partial _ - }\varphi = - s \frac{2}{{{\Lambda ^2}}} {\sinh \varphi } ,
\label{eq:Pohlmeyer_sinh}
\end{equation}
which is the sinh-Gordon equation. Finally, if the product $f^{\left( + \right)}\left( {{\xi_ + }} \right)f^{\left( - \right)}\left( {{\xi_ - }} \right)$ vanishes, one simply defines $\varphi  := a$, in which case equation \eqref{eq:Pohlmeyer_equation_1} takes the form
\begin{equation}
{\partial _ + }{\partial _ - }\varphi = - s \frac{1}{{{\Lambda ^2}}}{e^\varphi} ,
\label{eq:Pohlmeyer_Liouville}
\end{equation}
which is the Liouville equation.
Table \ref{tb:pohlmeyer_cases} summarizes all three cases for AdS$_3$ and dS$_3$ target spaces.
\begin{table}[ht]
\vspace{10pt}
\begin{center}
\begin{tabular}{ | r || c | c | c | }
\hline
& $f^{\left( + \right)}f^{\left( - \right)} < 0$ & $f^{\left( + \right)}f^{\left( - \right)} > 0$ & $f^{\left( + \right)}f^{\left( - \right)} = 0$ \\
\hline\hline		
AdS$_3$ & ${\partial _ + }{\partial _ - }\varphi = 2 {\sinh \varphi } / {\Lambda ^2}$ & ${\partial _ + }{\partial _ - }\varphi = 2 {\cosh \varphi } / {\Lambda ^2}$ & ${\partial _ - }{\partial _ + }\varphi = {e^\varphi} / {\Lambda ^2}$ \\
\hline
dS$_3$ & ${\partial _ + }{\partial _ - }\varphi = - 2 {\cosh \varphi } / {\Lambda ^2}$ & ${\partial _ + }{\partial _ - }\varphi = - 2 {\sinh \varphi } / {\Lambda ^2}$ & ${\partial _ - }{\partial _ + }\varphi = - {e^\varphi} / {\Lambda ^2}$ \\
\hline
\end{tabular}
\vspace{3pt}
\caption{The reduced equation for all possible values of $f^{\left( + \right)} f^{\left( - \right)}$}
\label{tb:pohlmeyer_cases}
\end{center}
\end{table}

In all cases, the Pohlmeyer reduced equation can be derived from a local Lagrangian density, which reads
\begin{align}
\mathcal{L} &= \frac{1}{2} {\partial _ + }\varphi {\partial _ - }\varphi - s \frac{2}{{{\Lambda ^2}}} {\sinh \varphi } ,\\
\mathcal{L} &= \frac{1}{2} {\partial _ + }\varphi {\partial _ - }\varphi - s \frac{2}{{{\Lambda ^2}}} {\cosh \varphi } ,\\
\mathcal{L} &= \frac{1}{2} {\partial _ + }\varphi {\partial _ - }\varphi - s \frac{1}{{{\Lambda ^2}}}{e^\varphi } ,
\end{align}
corresponding to equations \eqref{eq:Pohlmeyer_cosh}, \eqref{eq:Pohlmeyer_sinh} and \eqref{eq:Pohlmeyer_Liouville}, respectively.
\pagebreak

\section{Elliptic Solutions of the Sinh- and Cosh-Gordon Equations}
\label{sec:Elliptic_Solutions}

In this section, we find a family of solutions of the reduced equations that can be expressed in terms of elliptic functions and for which we can construct the corresponding classical string solutions, as we will do later in section \ref{sec:construction}.

All cases of equations shown in table \ref{tb:pohlmeyer_cases} can be rewritten in unified form
\begin{equation}
{\partial _ + }{\partial _ - }\varphi  = - s\frac{{{m^2}}}{2}\left( {{e^\varphi } + t{e^{ - \varphi }}} \right) ,
\label{eq:elliptic_cosh-sinh}
\end{equation}
where $t$ takes values $\pm 1$ which is opposite to the sign of $s f^{\left( + \right)} f^{\left( - \right)}$ and $m$ is the mass scale of the sinh or cosh-Gordon equation, which equals $\sqrt{2 / \Lambda}$.

Both sinh-Gordon and cosh-Gordon equations are integrable systems. The usual approach for finding solutions of the sinh-Gordon equation, such as the kinks, is to use the corresponding \Backlund transformation starting from the vacuum as seed solution. This method, however, cannot be applied to the cosh-Gordon equation; although it possesses \Backlund transformations similar to those of the sinh-Gordon equation, it does not admit a vacuum solution.

In this work, we follow an alternative approach, focusing on solutions of the sinh-Gordon or cosh-Gordon equations that depend on only one of the two world-sheet coordinates $\xi_0$ and $\xi_1$. The motivation is provided by the inverse Pohlmeyer reduction, namely by the need to find a classical string configuration whose Pohlmeyer counterpart is a given solution $\varphi \left( \xi_0 , \xi_1 \right)$ of the sinh- or the cosh-Gordon equation. As will be seen later, this requires to solve equations
\begin{equation}
\frac{{{\partial^2 Y^\mu}}}{{\partial{\xi_1 ^2}}} - \frac{{{\partial^2 Y^\mu}}}{{\partial{\xi_0 ^2}}} = - s \frac{1}{{{\Lambda ^2}}}{e^{\varphi\left( \xi_0 , \xi_1 \right)}}{Y^\mu } ,
\end{equation}
plus the Virasoro constraints. The latter will be significantly simplified via separation of variables if $\varphi \left( \xi_0 , \xi_1 \right)$ depends only on $\xi_0$ or $\xi_1$.  

\subsection{The Effective One-dimensional Mechanical Problem}
\label{subsec:elliptic_mechanical}
\begin{figure}[ht]
\vspace{10pt}
\begin{center}
\begin{picture}(85,35)
\put(0,0){\includegraphics[width = 0.5\textwidth]{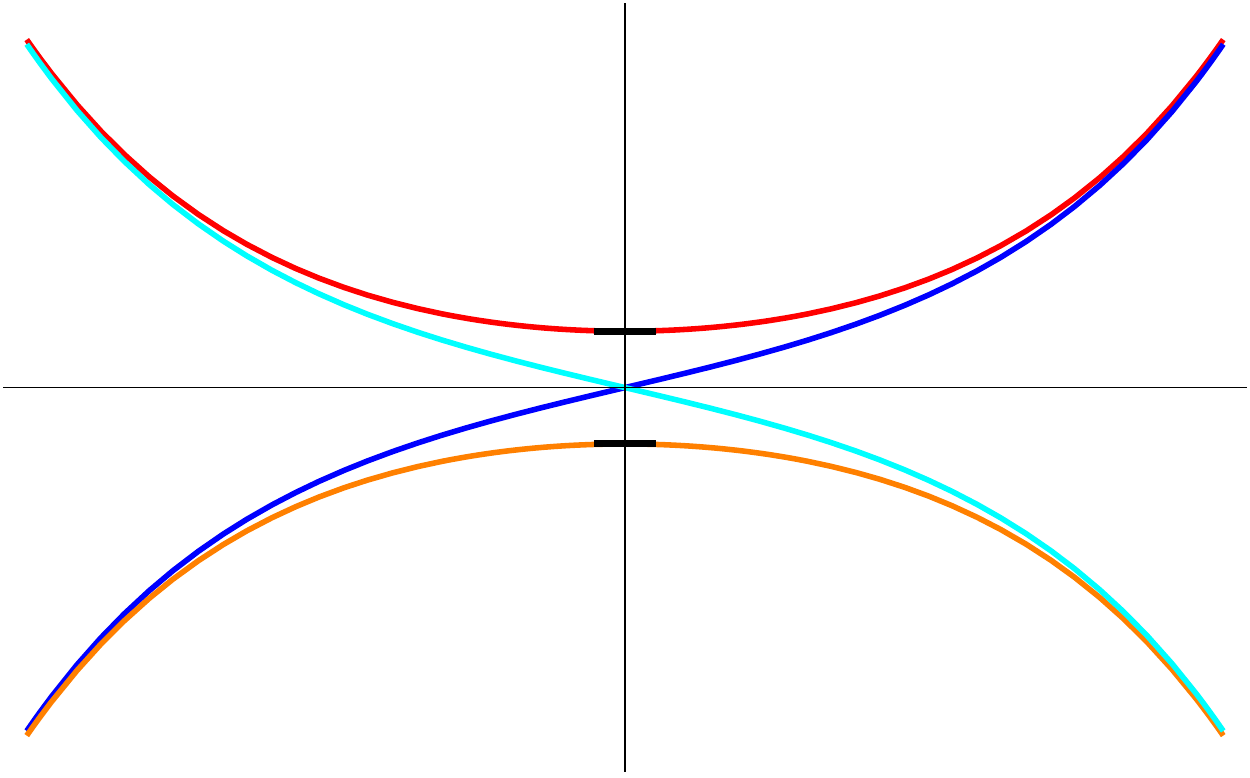}}
\put(54,1){\includegraphics[width = 0.1\textwidth]{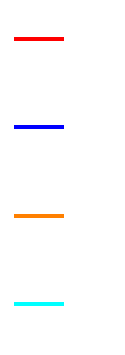}}
\put(19,10){$-m^2$}
\put(26,19){$m^2$}
\put(50.5,15){$\varphi$}
\put(22.5,32.5){$V \left( \varphi \right)$}
\put(60,11.5){$\partial_+ \partial_- \varphi = m^2 \sinh \varphi$}
\put(60,4.5){$\partial_+ \partial_- \varphi = m^2 \cosh \varphi$}
\put(60,25.5){$\partial_+ \partial_- \varphi = - m^2 \sinh \varphi$}
\put(60,18.5){$\partial_+ \partial_- \varphi = - m^2 \cosh \varphi$}
\put(54.25,3){\line(0,1){25.5}}
\put(54.25,3){\line(1,0){31}}
\put(85.25,3){\line(0,1){25.5}}
\put(54.25,28.5){\line(1,0){31}}
\end{picture}
\end{center}
\vspace{-5pt}
\caption{The potential of the one-dimensional mechanical analogue}
\vspace{5pt}
\label{fig:1dpotential}
\end{figure}
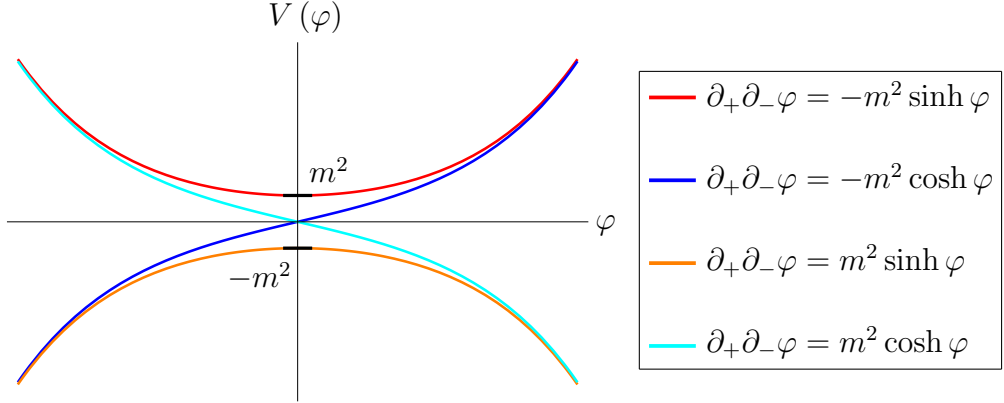

We start searching for solutions of the form $\varphi \left( \xi_0 , \xi_1 \right) = \varphi_1 \left( \xi_1 \right)$, namely static solutions. In this case, equation \eqref{eq:elliptic_cosh-sinh} reduces to the ordinary differential equation
\begin{equation}
\frac{{{d^2}\varphi_1 }}{{d{{\xi_1} ^2}}} = - s \frac{{{m^2}}}{2}\left( {{e^{\varphi_1} } + t{e^{ - {\varphi_1} }}} \right) ,
\end{equation}
which can be integrated to
\begin{equation}
\frac{1}{2}{\left( {\frac{{d\varphi_1 }}{{d{\xi_1} }}} \right)^2} + s\frac{{{m^2}}}{2}\left( {{e^{\varphi_1} } - t{e^{ - {\varphi_1} }}} \right) = E .
\label{eq:elliptic_energy_conservation_1}
\end{equation}
The latter can be viewed as the conservation of energy for an effective one-dimensional mechanical problem describing the motion of a particle with potential
\begin{equation}
U_1 \left( \varphi_1 \right) = s\frac{{{m^2}}}{2}\left( {{e^{\varphi_1} } - t{e^{ - {\varphi_1} }}} \right) ,
\label{eq:elliptic_potential_1}
\end{equation}
letting ${\xi_1}$ play the role of time and $\varphi_1$ the role of the particle coordinate. The potential \eqref{eq:elliptic_potential_1} is plotted in figure \ref{fig:1dpotential} for all four cases we are studying.
Considering this effective mechanical problem, we obtain a qualitative picture for the behaviour of the solutions. For the sinh-Gordon equation with an overall minus sign, we expect to find oscillating solutions with energy $E > m^2$ and no solutions for $E < m^2$. For the sinh-Gordon equation with an overall plus sign, we expect to find two different classes of solutions; for $E < - m^2$ we expect to find reflecting scattering solutions since the effective particle does not have enough energy to overcome the potential barrier, whereas for $E > - m^2$, we expect to find transmitting scattering solutions since the particle overcomes the potential barrier. Finally, for the cosh-Gordon equation, we expect to find reflecting scattering solutions for all energies.

Introducing the quantity
\begin{equation}
V_1 :=  - s\frac{{{m^2}}}{2}{e^\varphi } ,
\label{eq:elliptic_V_definition_1}
\end{equation}
allows to rewrite equation \eqref{eq:elliptic_energy_conservation_1} as
\begin{equation}
{\left( {\frac{{dV_1}}{{d{\xi _1}}}} \right)^2} = 2{V_1^3} + 2E{V_1^2} - t\frac{{{m^4}}}{2}V_1 .
\label{eq:elliptic_V_equation_1}
\end{equation}
As such it transforms the one-dimensional problem of motion for a particle with energy $E$ under the influence of a hyperbolic potential to yet another one-dimensional problem, describing the motion of a particle with zero energy under the influence of a cubic potential. This formulation is advantageous since $V_1$ can be easily expressed in terms of the Weierstrass elliptic function $\wp$ (see appendix \ref{sec:Weierstrass_functions}). Actually, using the change of variable
\begin{equation}
V_1 = 2y - \frac{E}{3} ,
\label{eq:elliptic_y_definition}
\end{equation}
equation \eqref{eq:elliptic_V_equation_1} takes the standard Weierstrass form
\begin{equation}
{\left( {\frac{{dy}}{{d{\xi _1}}}} \right)^2} = 4{y^3} - \left( {\frac{1}{3}{E^2} + t\frac{{{m^4}}}{4}} \right)y + \frac{E}{3}\left( {\frac{1}{9}{E^2} + t\frac{{{m^4}}}{8}} \right) .
\label{eq:elliptic_p_equation}
\end{equation}

Translationally invariant solutions of the sinh- or cosh-Gordon equation are similar to the static ones. In that case the solutions are of the form $\varphi \left( \xi_0 , \xi_1 \right) = \varphi_0 \left( \xi_0 \right)$ and equation \eqref{eq:elliptic_cosh-sinh} reduces to the ordinary differential equation
\begin{equation}
\frac{{{d^2}\varphi_0 }}{{d{{\xi_0} ^2}}} = s \frac{{{m^2}}}{2}\left( {{e^{\varphi_0} } + t{e^{ - {\varphi_0} }}} \right) ,
\end{equation}
which can be integrated once to yield
\begin{equation}
\frac{1}{2}{\left( {\frac{{d\varphi_0 }}{{d{\xi_1} }}} \right)^2} - s\frac{{{m^2}}}{2}\left( {{e^{\varphi_0} } - t{e^{ - {\varphi_0} }}} \right) = E .
\label{eq:elliptic_energy_conservation_0}
\end{equation}
As before, this equation can be viewed as energy conservation for a one-dimensional problem describing the motion of a particle with potential
\begin{equation}
U_0 \left( \varphi_0 \right) = - s \frac{{{m^2}}}{2}\left( {{e^{\varphi_0} } - t{e^{ - {\varphi_0} }}} \right) .
\label{eq:elliptic_potential_0}
\end{equation}
It is identical to the problem of static configurations, letting $s \to - s$. Then, if we define
\begin{equation}
V_0 := s\frac{{{m^2}}}{2}{e^{\varphi_0} } ,
\label{eq:elliptic_V_definition_0}
\end{equation}
we arrive at the exact same equation for $V_0$, as the one satisfied by $V_1$, \eqref{eq:elliptic_V_equation_1}. Thus, translationally invariant solutions follow from the static ones and vice versa using the identification
\begin{align}
V_0\left( {{\xi_0} ;E, s, t} \right) &= V_1\left( {{\xi_0} ;E, s, t} \right),\\
\varphi_0 \left( {{\xi_0} ;E, s, t} \right) &= \varphi_1 \left( {{\xi_0} ;E, - s, t} \right).
\end{align}
This implies that the static solutions of the Pohlmeyer reduced system for string propagation in AdS$_3$ are identical to the translationally invariant solutions of the Pohlmeyer reduced system for string propagation in dS$_3$ and vice versa.

Summarizing, the problem of finding static or translationally invariant solutions of the sinh- and cosh-Gordon equations is reduced to the Weierstrass equation \eqref{eq:elliptic_p_equation}. It is interesting to understand how the same equation can be used to describe a variety of solutions that exhibit qualitatively different behaviour, as suggested by the effective one-dimensional mechanical point particle problem.

\subsection{Some Properties of the Weierstrass Function $\wp$}
\label{subsec:elliptic_wp}

We review some properties of the Weierstrass function $\wp$ that will be important in the following sections. The function $\wp \left( {{x} ;{g_2},{g_3}} \right)$ is a doubly periodic complex function defined in terms of one complex variable $z$, satisfying the equation
\begin{equation}
{\left( {\frac{{dy }}{{dz}}} \right)^2} = 4{y^3} - {g_2}y  - {g_3} .
\label{eq:elliptic_wp_equation}
\end{equation}
The periods of $\wp$ are related with the three roots $e_1$, $e_2$ and $e_3$ of the cubic polynomial
\begin{equation}
Q \left( y \right) = 4 y^3 - g_2 y -g_3.
\label{eq:elliptic_wp_cubic_polynomial}
\end{equation}
The absence of a quadratic term implies the relation $e_1 + e_2 + e_3 = 0$. Introducing the discriminant \begin{equation}
\Delta : = g_2^3 - 27g_3^2 ,
\end{equation}
we distinguish two cases:

If $\Delta > 0$, the cubic polynomial \eqref{eq:elliptic_wp_cubic_polynomial} will have three real roots. Ordering the roots as $e_1 > e_2 > e_3$, the function $\wp$ has one real period $2 \omega_1$ and one imaginary period $2 \omega_2$ which are related to the roots as follows,
\begin{equation}
{\omega _1} = \frac{{K\left( k \right)}}{{\sqrt {{e_1} - {e_3}} }},\quad {\omega _2} = \frac{{iK\left( {k'} \right)}}{{\sqrt {{e_1} - {e_3}} }},
\label{eq:elliptic_periods_D_pos}
\end{equation}
where $K\left( k \right)$ is the complete elliptic integral of the first kind and
\begin{equation}
k^2 = \frac{{{e_2} - {e_3}}}{{{e_1} - {e_3}}},\quad {k'}^2 = \frac{{{e_1} - {e_2}}}{{{e_1} - {e_3}}},\quad {k^2} + k{'^2} = 1 .
\label{eq:elliptic_moluli_D_pos}
\end{equation}

If $\Delta < 0$, the cubic polynomial \eqref{eq:elliptic_wp_cubic_polynomial} will have one real root and two complex, which are conjugate to each other. We let $e_2$ to be the real root and $e_{1,3} = a \pm i b$ with $b > 0$. Then, the function $\wp$ has one real period and one complex, which, however, is not purely imaginary. In this case, it is more convenient to consider as fundamental periods the complex one and its conjugate,
\begin{equation}
{\omega _1} = \frac{{K\left( k \right) - iK\left( {k'} \right)}}{{2\sqrt[4]{{9{a^2} + {b^2}}}}},\quad {\omega _2} = \frac{{K\left( k \right) + iK\left( {k'} \right)}}{{2\sqrt[4]{{9{a^2} + {b^2}}}}},
\label{eq:elliptic_periods_D_neg}
\end{equation}
with
\begin{equation}
k^2 = \frac{1}{2} - \frac{{3{e_2}}}{{4\sqrt {9{a^2} + {b^2}} }} ,\quad {k'}^2 = \frac{1}{2} + \frac{{3{e_2}}}{{4\sqrt {9{a^2} + {b^2}} }} ,\quad {k^2} + k{'^2} = 1 .
\label{eq:elliptic_moluli_D_neg}
\end{equation}
The real period is just the sum of the two fundamental periods $2 \omega_1$ and $2 \omega_2$.

In all cases, the Weierstrass function $\wp$ obeys the following half-period relations
\begin{equation}
\wp \left( {{\omega _1}} \right) = {e_1},\quad \wp \left( {{\omega _2}} \right) = {e_3},\quad \wp \left( {\omega _3} \right) = {e_2},
\label{eq:elliptic_half_period}
\end{equation}
where $\omega_3 := \omega_1 + \omega_2$.

If $\Delta = 0$, then at least two of the roots are equal and the Weierstrass function $\wp$ takes a special form, which is not doubly periodic, but trigonometric. More information on those limits is provided in appendix \ref{sec:Weierstrass_functions}.

\begin{figure}[ht]
\vspace{10pt}
\begin{center}
\begin{picture}(100,36)
\put(2.5,5){\includegraphics[width = 0.45\textwidth]{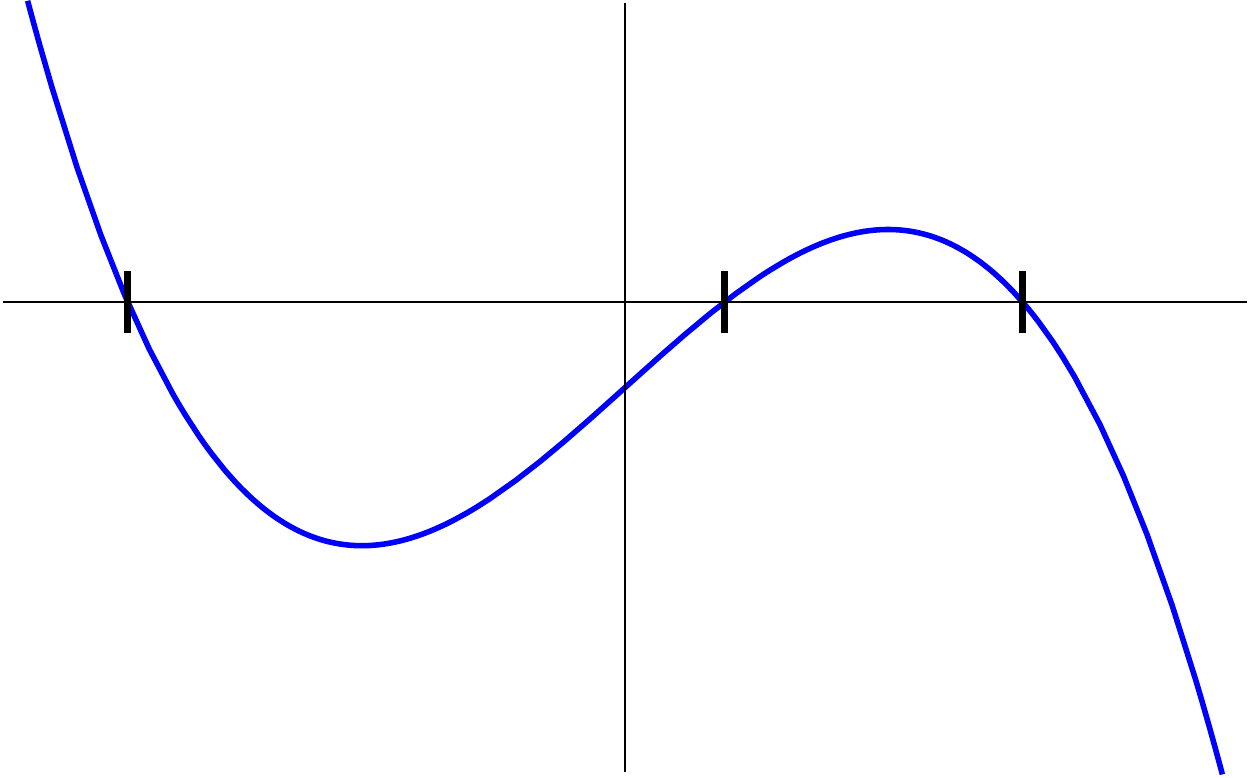}}
\put(52.5,5){\includegraphics[width = 0.45\textwidth]{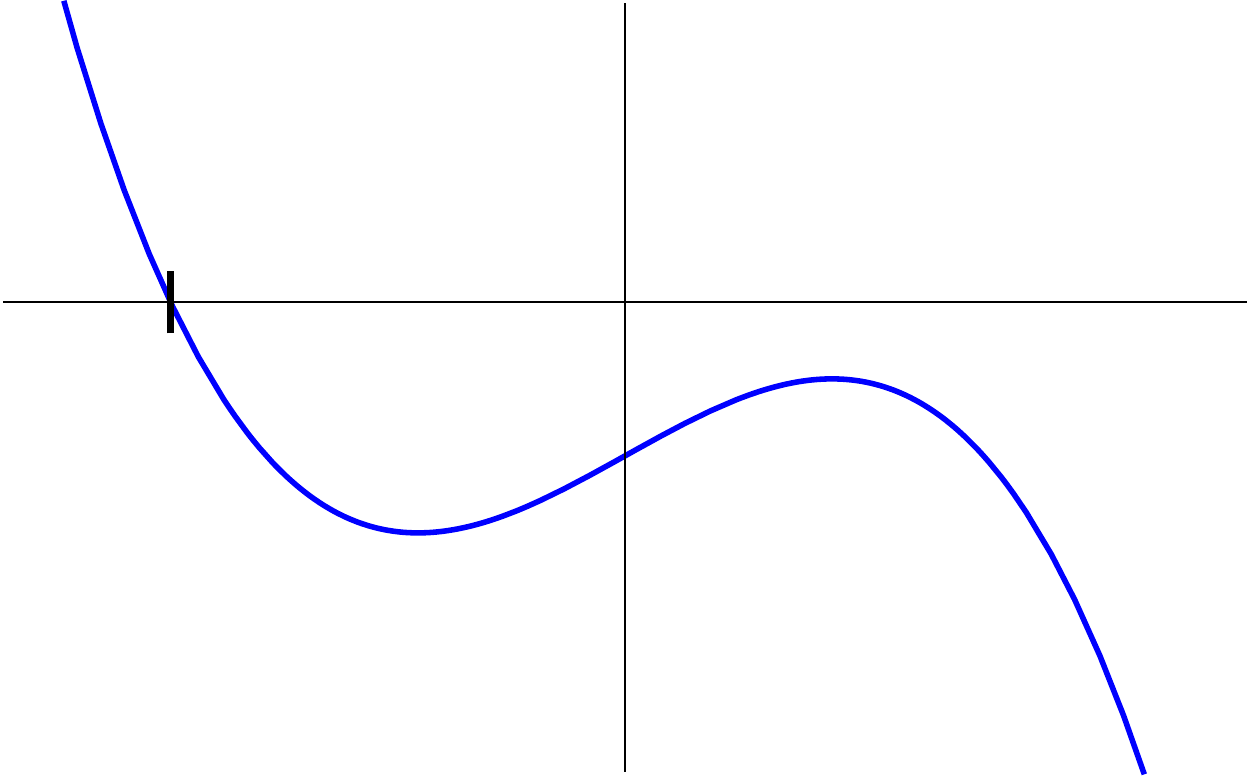}}
\put(4.5,19.5){$e_3$}
\put(29,19.5){$e_2$}
\put(37,19.5){$e_1$}
\put(56,19.5){$e_2$}
\put(23.5,34){$V_\wp$}
\put(73.5,34){$V_\wp$}
\put(21,0){$\Delta > 0$}
\put(71,0){$\Delta < 0$}
\end{picture}
\end{center}
\vspace{-5pt}
\caption{The form of the effective potential $V_\wp$}
\label{fig:vwp}
\end{figure}
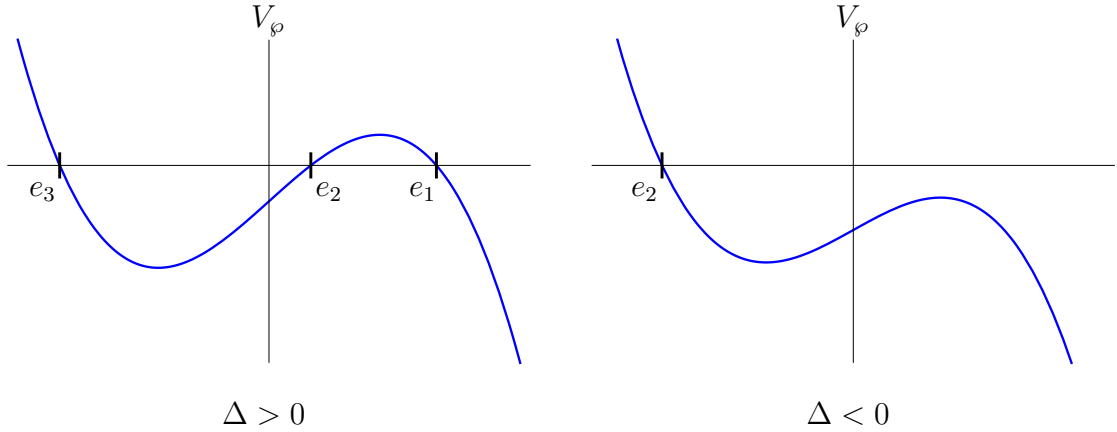

Although $\wp$ has many interesting properties in the complex domain, we should not forget that equation \eqref{eq:elliptic_p_equation} is real, and, thus, we are looking for real solutions of equation \eqref{eq:elliptic_wp_equation} defined in the real domain. Similarly to the analysis of section \ref{subsec:elliptic_mechanical}, the form of the real solutions of equation \eqref{eq:elliptic_wp_equation} can be visualized in terms of an one-dimensional mechanical problem, describing the motion of a point particle with vanishing energy moving under the influence of the cubic potential $V_\wp \left( y \right) = - 4 y^3 +g_2 y + g_3 = - Q \left( y \right)$. As figure \ref{fig:vwp} suggests, when $\Delta > 0$, we expect to have two real solutions, one being unbounded with $y > e_1$ and a bounded one with $e_3 < y < e_2$.
When $\Delta < 0$, we expect to have only one real solution which is unbounded with $y > e_2$. The Weierstrass function $\wp$ has a second order pole at the origin of the complex plane and it is real on the real axis. These properties suffice to determine the unbounded solution by restricting $z$ to the real axis. Indeed, the unbound solutions are given by $y = \wp \left( x \right)$, where $x \in \mathbb{R}$ and the half-period relations \eqref{eq:elliptic_half_period} imply that the ``time of flight'' is equal to the real period of $\wp \left( z \right)$.

Next, we consider the bounded real solution. Note at this point that the argument $z$ of the Weierstrass function can be shifted by an arbitrary constant and still solve equation \eqref{eq:elliptic_wp_equation}. Using the periodic properties of $\wp$ and the fact that $\omega_2$ is purely imaginary in the case $\Delta > 0$, we have the relation
\begin{equation}
\overline {\wp \left( {x + {\omega _2}} \right)}  = \wp \left( {x + \overline{\omega_2}} \right) = \wp \left( {x - {\omega _2}} \right) = \wp \left( {x + {\omega _2}} \right).
\end{equation}
Since the Weierstrass function obeys the half-period identities $\wp \left( {{\omega _1} + {\omega _2}} \right) = {e_2}$ and $\wp \left( {{\omega _2}} \right) = {e_3}$, we conclude that $\wp \left( {x + {\omega _2}} \right)$, where $x \in \mathbb{R}$, is a real periodic solution that oscillates between $e_2$ and $e_3$, with period equal to $2 \omega_1$. As such, it describes the bounded solution corresponding to $\Delta > 0$, as qualitatively understood from figure \ref{fig:vwp}.

In the language of real differential equations, we may say that when $\Delta > 0$ equation \eqref{eq:elliptic_wp_equation} has two real solutions, 
\begin{align}
y_1 \left( x \right) &= \wp \left( x \right) ,\label{eq:elliptic_y_solution_unbound}\\
y_2 \left( x \right) &= \wp \left( x + \omega_2 \right) ,\label{eq:elliptic_y_solution_bound}
\end{align}
corresponding to the unbounded and bounded solutions respectively, whereas for $\Delta < 0$ there is only one real solution given by \eqref{eq:elliptic_y_solution_unbound}.

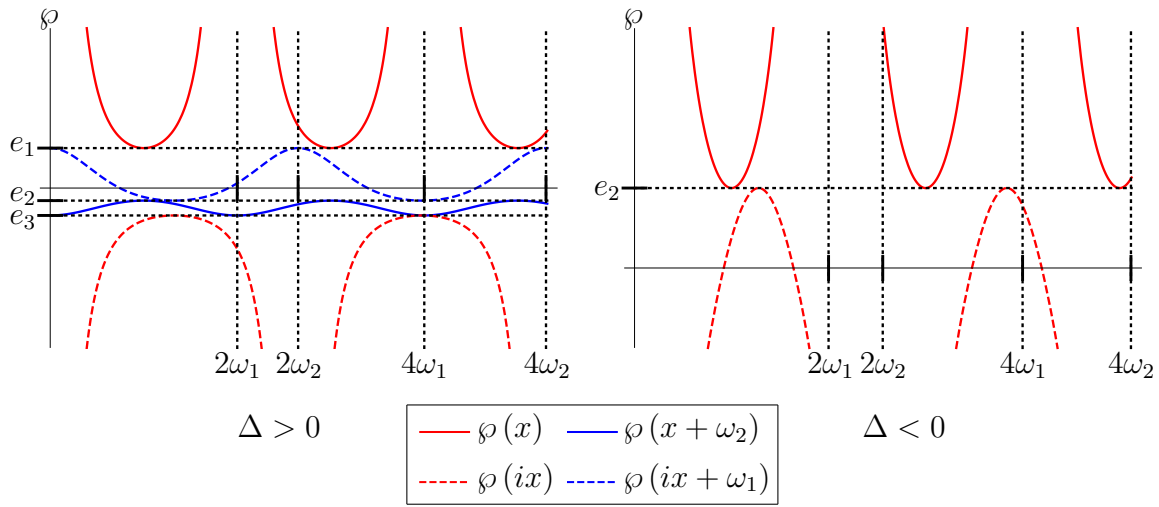
\begin{figure}[ht]
\begin{center}
\begin{picture}(100,43)(0,-6)
\put(2.5,7.5){\includegraphics[width = 0.45\textwidth]{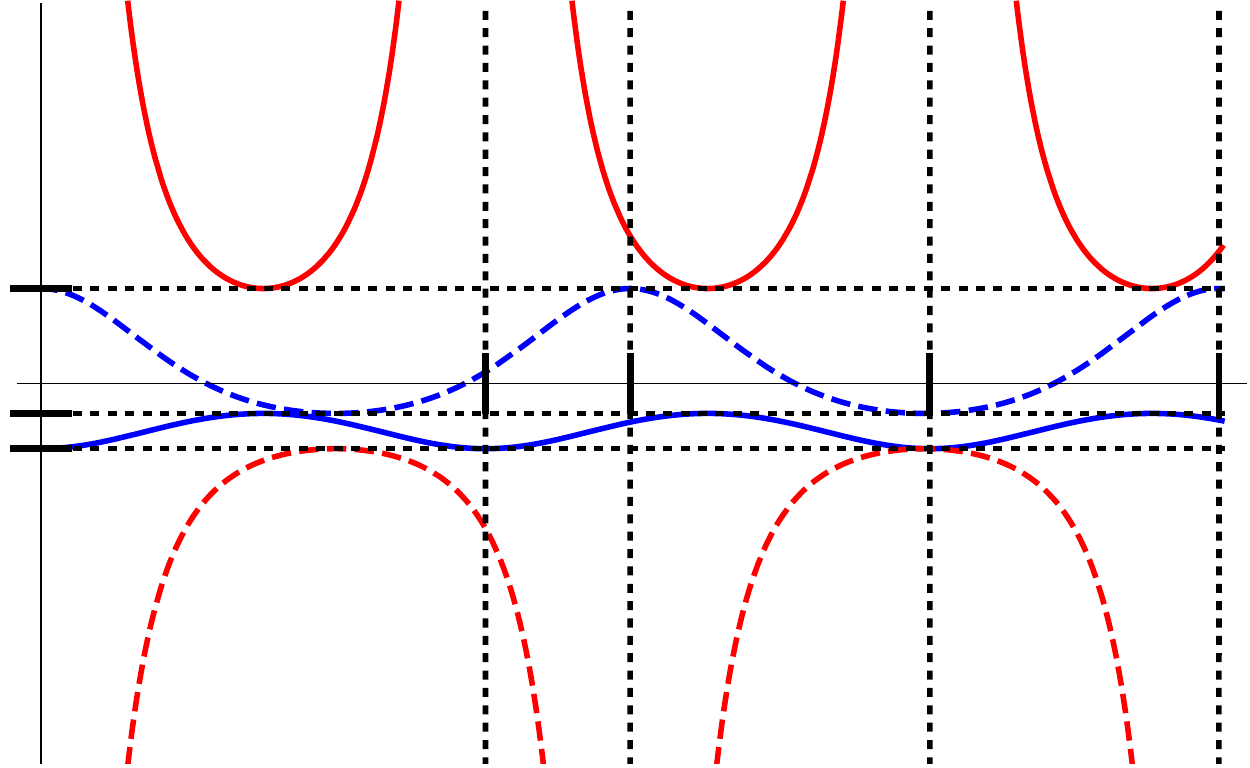}}
\put(52.5,7.5){\includegraphics[width = 0.45\textwidth]{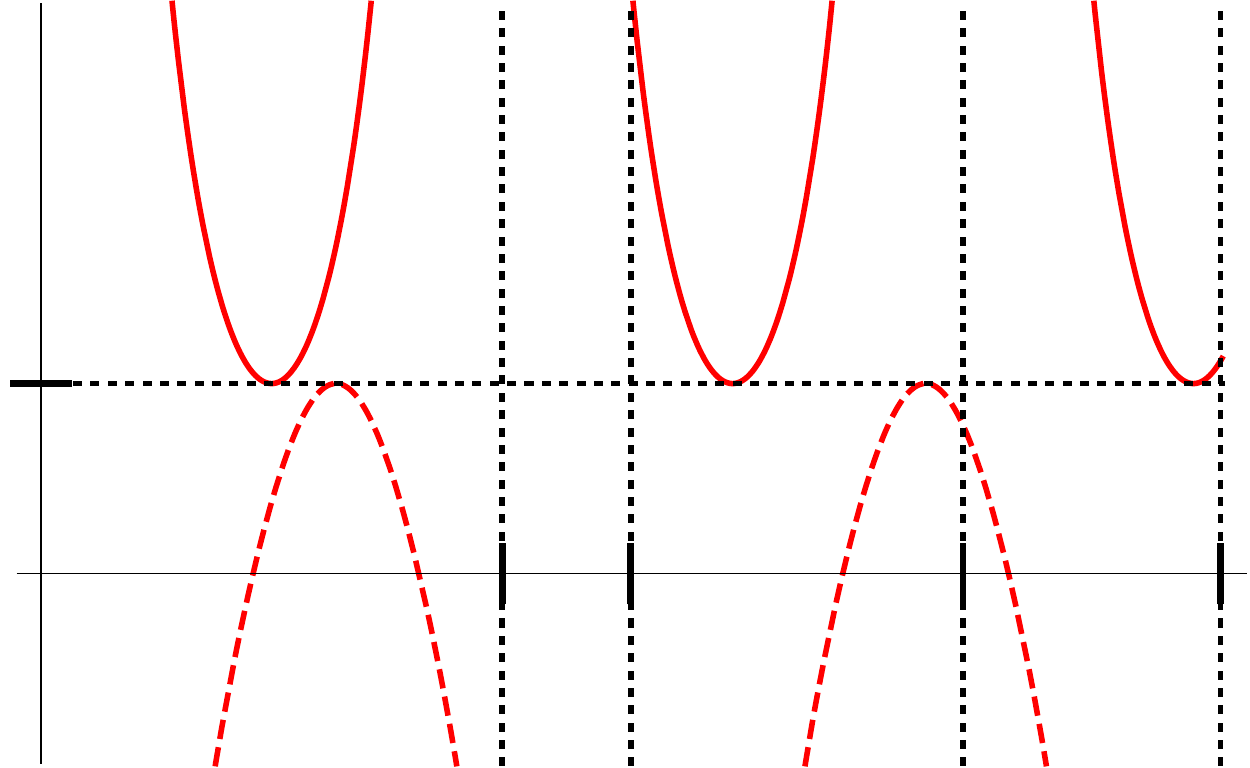}}
\put(35,-6){\includegraphics[height = 0.08\textwidth]{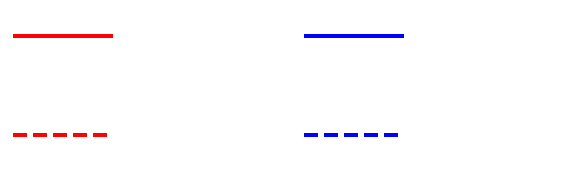}}
\put(0.5,24.5){$e_1$}
\put(0.5,20.25){$e_2$}
\put(0.5,18.25){$e_3$}
\put(18,5.5){$2\omega_1$}
\put(23.25,5.5){$2\omega_2$}
\put(34,5.5){$4\omega_1$}
\put(44.5,5.5){$4\omega_2$}
\put(50.5,20.75){$e_2$}
\put(68.75,5.5){$2\omega_1$}
\put(73.25,5.5){$2\omega_2$}
\put(85.25,5.5){$4\omega_1$}
\put(94.5,5.5){$4\omega_2$}
\put(3,35.5){$\wp$}
\put(53,35.5){$\wp$}
\put(20,0){$\Delta > 0$}
\put(73.5,0){$\Delta < 0$}
\put(40.5,0){$\wp \left( x \right)$}
\put(53,0){$\wp \left( x + \omega_2 \right)$}
\put(40.5,-4.25){$\wp \left( ix \right)$}
\put(53,-4.25){$\wp \left( ix + \omega_1 \right)$}
\put(34.5,-5.75){\line(0,1){8.5}}
\put(34.5,-5.75){\line(1,0){31.75}}
\put(66.25,-5.75){\line(0,1){8.5}}
\put(34.5,2.75){\line(1,0){31.75}}
\end{picture}
\end{center}
\vspace{-5pt}
\caption{The values of $\wp \left( z \right)$ on the real and imaginary axes}
\vspace{5pt}
\label{fig:wpreal}
\end{figure}

For reasons that will become apparent in section \ref{subsec:lame}, it is also interesting to study the behaviour of $\wp$ on the purely imaginary axis $z = i x$. Then, the Weierstrass equation \eqref{eq:elliptic_wp_equation} becomes
\begin{equation}
{\left( {\frac{{d\wp \left(i x\right) }}{{d x}}} \right)^2} = -\left(4{\wp ^3 \left(ix\right)} - {g_2}\wp \left(ix\right) - {g_3} \right) .
\end{equation}
Thus, the Weierstrass function behaves on the imaginary axis in exactly the same way as it behaves on the real axis, but with inverted cubic potential \eqref{eq:elliptic_wp_cubic_polynomial}. The function $\wp$ is still real on the imaginary axis, taking values in intervals complementary to those defined on the real axis. By the same token $\wp \left( i x + \omega_1 \right)$ is also real when $\Delta > 0$. The above are depicted in figure \ref{fig:wpreal}.

In later sections we will be interested in knowing the position of a number $z$ in the complex plane obeying $\wp \left( z \right) = y$ for given real number $y$. Based on the previous analysis we are led to consider the following cases:
\pagebreak
\begin{itemize}
\item If $\Delta > 0$, all real numbers will appear on the rectangle of the complex plane with corners at $0$, $\omega_1$, $\omega_2$ and $\omega_3$. Then,
\begin{itemize}
\item for any $y \geq e_1$, there is an $x \in \left( 0 , \omega_1 \right]$ such that $\wp \left( x \right) = y$
\item for any $e_2 \leq y \leq e_1$, there is an $x \in \left[ 0 , -i \omega_2 \right]$ such that $\wp \left( ix + \omega_1 \right) = y$
\item for any $e_3 \leq y \leq e_2$, there is an $x \in \left[ 0 , \omega_1 \right]$ such that $\wp \left( x + \omega_2 \right) = y$
\item for any $y \leq e_3$, there is an $x \in \left( 0 , -i \omega_2 \right]$ such that $\wp \left( i x \right) = y$.
\end{itemize}
\item If $\Delta < 0$, all real numbers will appear on two segments of the complex plane, one on the real axis with endpoints at $0$ and $\omega_1 + \omega_2 $ and one in the imaginary axis with endpoints at $0$ and $\omega_1 - \omega_2 $. Then,
\begin{itemize}
\item for any $y \geq e_2$, there is an $x \in \left( 0 , \omega_1 + \omega_2 \right]$ such that $\wp \left( x \right) = y$
\item for any $y \leq e_2$, there is an $x \in \left( 0 , -i \left( \omega_1 - \omega_2 \right) \right]$ such that $\wp \left( i x \right) = y$.
\end{itemize}
\end{itemize}

\subsection{The Elliptic Solutions of the Sinh- and Cosh-Gordon Equations}
\label{subsec:elliptic_solutions}
Next we focus on the equation \eqref{eq:elliptic_p_equation} in order to obtain solutions of the sinh- and cosh-Gordon equations that depend only on one of the two coordinates. This equation is of the Weierstrass form \eqref{eq:elliptic_wp_equation} with specific constants $g_2$ and $g_3$. Equation \eqref{eq:elliptic_p_equation} is solved by
\begin{align}
y &= \wp \left( {{\xi_1} ;{g_2}\left( E, t \right),{g_3}\left( E, t \right)} \right),\\
y &= \wp \left( {{\xi_1} +\omega_2 ;{g_2}\left( E, t \right),{g_3}\left( E, t \right)} \right),
\end{align}
bearing in mind that the second solution is valid only when there are three real roots. The coefficients $g_2$ and $g_3$ are given by
\begin{equation}
{g_2}\left( E, t \right) = \frac{1}{3}{E^2} + t\frac{{{m^4}}}{4},\quad {g_3}\left( E, t \right) =  - \frac{E}{3}\left( {\frac{1}{9}{E^2} + t\frac{{{m^4}}}{8}} \right) 
\label{eq:elliptic_moduli_energy}
\end{equation}
and the related cubic polynomial is
\begin{equation}
Q \left( x \right) = 4{x^3} - \left( {\frac{1}{3}{E^2} + t\frac{{{m^4}}}{4}} \right)x + \frac{E}{3}\left( {\frac{1}{9}{E^2} + t\frac{{{m^4}}}{8}} \right) .
\label{eq:elliptic_cubic_equation}
\end{equation}

\begin{figure}[ht]
\vspace{10pt}
\begin{center}
\begin{picture}(100,35)
\put(0,5){\includegraphics[width = 0.4\textwidth]{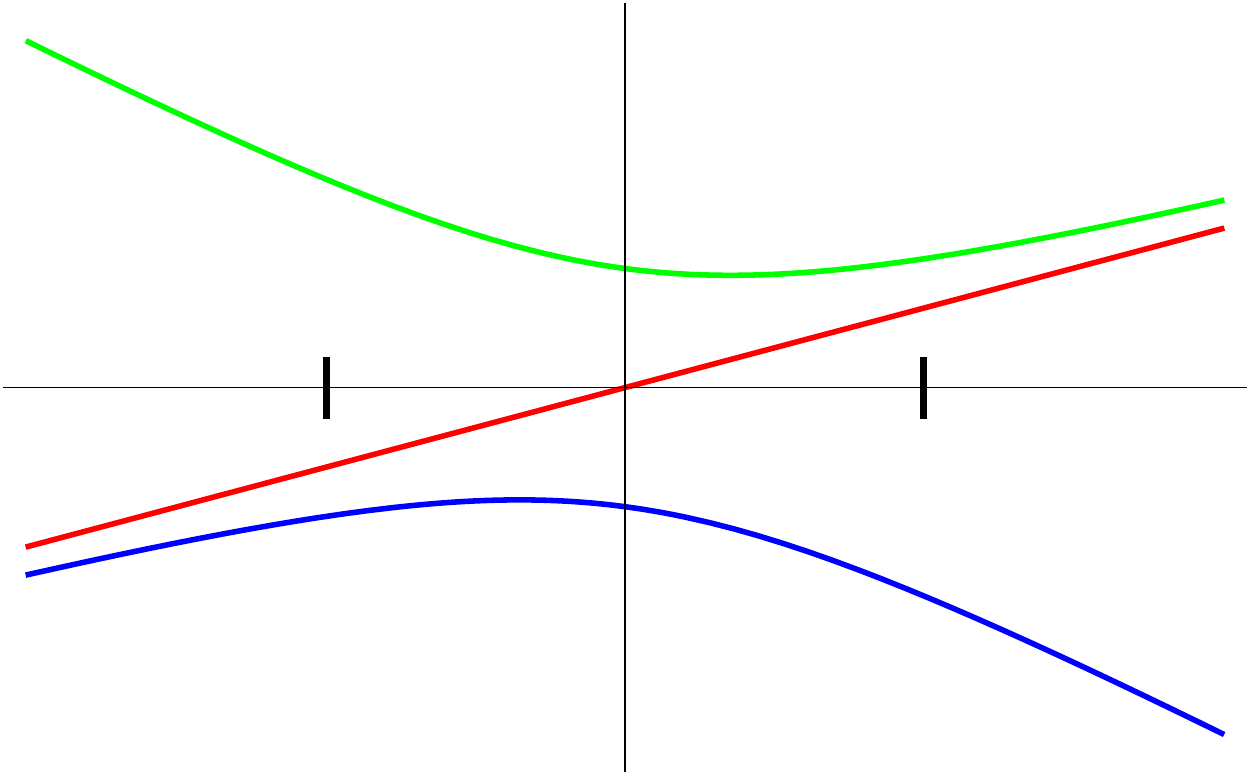}}
\put(55,5){\includegraphics[width = 0.4\textwidth]{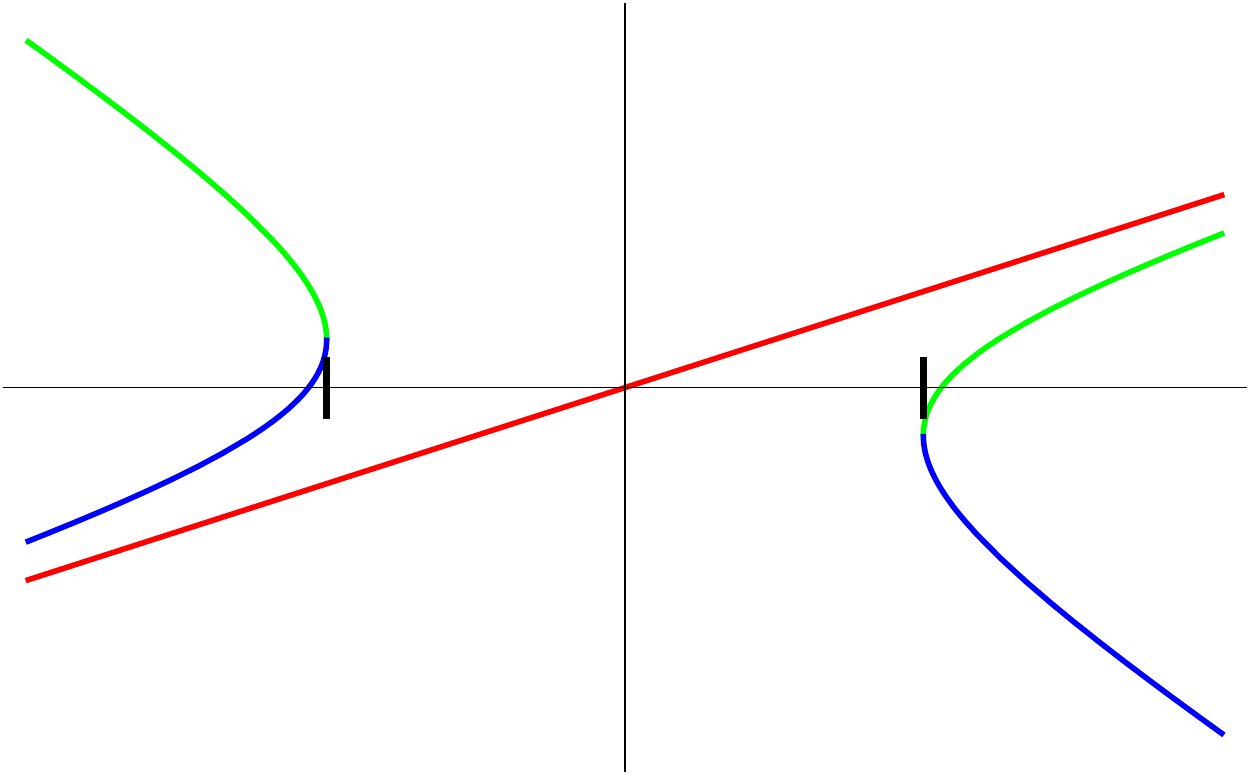}}
\put(44.5,9.5){\includegraphics[height = 0.15\textwidth]{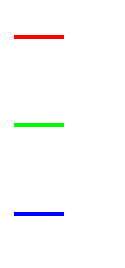}}
\put(7.25,19){$-m^2$}
\put(28.25,14.5){$m^2$}
\put(59.5,18){$-m^2$}
\put(85,15){$m^2$}
\put(49,21.75){$x_1$}
\put(49,16.75){$x_2$}
\put(49,11.75){$x_3$}
\put(19,31){$x_i$}
\put(74,31){$x_i$}
\put(40,16.75){$E$}
\put(95,16.75){$E$}
\put(13.5,0){cosh-Gordon}
\put(68.5,0){sinh-Gordon}
\put(44.5,10.5){\line(0,1){14}}
\put(44.5,10.5){\line(1,0){7.75}}
\put(52.25,10.5){\line(0,1){14}}
\put(44.5,24.5){\line(1,0){7.75}}
\end{picture}
\end{center}
\vspace{-5pt}
\caption{The roots of the cubic polynomial \eqref{eq:elliptic_cubic_equation} as function of the energy constant $E$}
\vspace{5pt}
\label{fig:roots}
\end{figure}
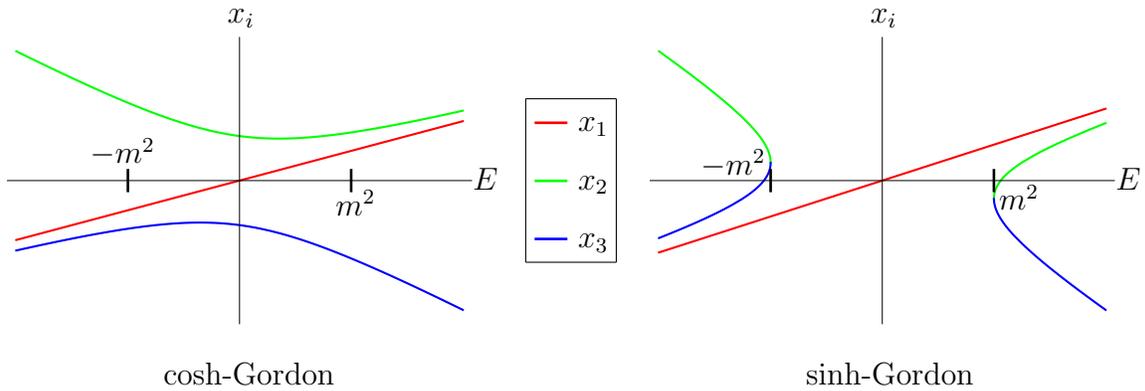

The roots of the cubic polynomial are easy to obtain noting that $x = E / 6$ is one of them. Thus, all three roots of $Q \left( x \right)$ are
\begin{equation}
{x_1} = \frac{E}{6},\quad {x_{2,3}} =  - \frac{E}{{12}} \pm \frac{1}{4}\sqrt {{E^2} + t{m^4}} .
\label{eq:elliptic_xroots}
\end{equation}
In the following we use the notation $x_i$ for the roots of $Q \left( x \right)$ as written in equations \eqref{eq:elliptic_xroots} and deserve the notation $e_i$ for the ordered roots of $Q \left( x \right)$:  we let $e_1 > e_2 > e_3$ when all roots are real, whereas if only one root is real this will be $e_2$ and $e_1$ will be the complex root with positive imaginary part. The roots $x_i$ are plotted as functions of the energy constant $E$ in figure \ref{fig:roots}.

The advantage of using the Weierstrass elliptic function, instead of the Jacobi elliptic functions is clear. The Weierstrass function allows for a unifying description of the elliptic solutions of both sinh- and cosh-Gordon equations. Different classes of solutions simply correspond to different ordering of the roots $x_i$. Figure \ref{fig:roots} suggests that there are four distinct cases for the ordering of the three roots $x_i$, which are summarized in table \ref{tb:root_assignments}.
\begin{table}[ht]
\vspace{10pt}
\begin{center}
\begin{tabular}{ | c || c | c | }
\hline
 & reality of roots & ordering of roots \\
\hline\hline		
$t = + 1$ & 3 real roots & $e_1 = x_2$, $e_2 = x_1$, $e_3 = x_3$ \\
\hline
$t = - 1$, $E > m^2$ & 3 real roots & $e_1 = x_1$, $e_2 = x_2$, $e_3 = x_3$ \\
\hline
$t = - 1$, $E < - m^2$ & 3 real roots & $e_1 = x_2$, $e_2 = x_3$, $e_3 = x_1$ \\
\hline
$t = - 1$, $\left| E \right| < m^2$ & 1 real, 2 complex roots & $e_1 = x_2$, $e_2 = x_1$, $e_3 = x_3$ \\
\hline
\end{tabular}
\vspace{3pt}
\caption{The possible orderings of the roots $x_i$}
\label{tb:root_assignments}
\end{center}
\end{table}

The unbounded solution ranges from $e_1$ to infinity when $\Delta > 0$ and from $e_2$ to infinity when $\Delta < 0$, whereas the bounded solution ranges from $e_3$ to $e_2$. Then, using equation \eqref{eq:elliptic_y_definition}, $V_1 = 2 \left( y - x_1 \right)$, we can explore the range of $V_1$ in all cases. The results are summarized in table \ref{tb:v1_range}.
\begin{table}[ht]
\vspace{10pt}
\begin{center}
\begin{tabular}{ | c || c | c | c | }
\hline
 & $V_1$ & unbounded range & bounded range \\
\hline\hline		
$t = + 1$ & $2 \left( y - e_2 \right)$ & $\left[ {2 \left( {e_1 - e_2} \right) , + \infty} \right)$ & $\left[ { - 2 \left( {e_2 - e_3} \right) , 0} \right]$\\
\hline
$t = - 1$, $E > m^2$ & $2 \left( y - e_1 \right)$ & $\left[ {0 , + \infty} \right)$ & $\left[ { - 2 \left( {e_1 - e_3} \right) , - 2 \left( {e_1 - e_2} \right)} \right]$ \\
\hline
$t = - 1$, $E < - m^2$ & $2 \left( y - e_3 \right)$ & $\left[ {2 \left( {e_1 - e_3} \right) , + \infty} \right)$ & $\left[ { 0 , 2 \left( {e_2 - e_3} \right)} \right]$ \\
\hline
$t = - 1$, $\left| E \right| < m^2$ & $2 \left( y - e_2 \right)$ & $\left[ {0 , + \infty} \right)$ & -- \\
\hline
\end{tabular}
\vspace{3pt}
\caption{The solutions for $V_1$ and their range}
\label{tb:v1_range}
\end{center}
\end{table}
In all cases, the sign of $V_1$ does not alternate within its range. Using $V_1 = - s{m^2}{e^{\varphi_1} } / 2$, we obtain the corresponding real static elliptic solutions of the sinh- or cosh-Gordon equation for the appropriate choice of the sign $s$. For the translationally invariant solutions, $V_1$ is replaced by $V_0$, but since $V_0 = s{m^2}{e^{\varphi_0} } / 2$ the corresponding real translationally invariant elliptic solutions of the sinh- or cosh-Gordon equation arise for the complementary choice of the sign $s$.

Coming back to the static solutions, we examine the range of the field $\varphi_1 \left( \xi_1 \right)$. The solutions have the behaviour expected by the effective one-dimensional mechanical problem, as analysed before and they are summarized in table \ref{tb:phi_range}. We denote the real period of $y$ by $2\omega$, which is equal to $2 \omega_1$ for $\Delta > 0$ and $2 \left( \omega_1 + \omega_2 \right)$ for $\Delta < 0$.
\begin{table}[ht]
\vspace{10pt}
\begin{center}
\begin{tabular}{ | c || c | c | c || c | c | c | }
\hline
\multirow{2}{*}{} & \multicolumn{3}{|c||}{unbounded} & \multicolumn{3}{|c|}{bounded} \\
\cline{2-7}
{}& $\varphi_1 \left( 0 \right)$ & $\varphi_1 \left( \omega \right)$ & $\varphi_1 \left( 2\omega \right)$ & $\varphi_1 \left( 0 \right)$ & $\varphi_1 \left( \omega \right)$ & $\varphi_1 \left( 2\omega \right)$ \\
\hline\hline
\multicolumn{1}{|c||}{} & \multicolumn{6}{|c|}{$s = - 1$}\\
\hline		
$t = + 1$ & $+ \infty$ & $\ln {\frac{4\left( {e_1 - e_2} \right)}{m^2}} $ & $+ \infty$ & \multicolumn{3}{|c|}{--}\\
\hline
$t = - 1$, $E > m^2$ & $+ \infty$ & $- \infty$ & $+ \infty$ & \multicolumn{3}{|c|}{--} \\
\hline
$t = - 1$, $E < - m^2$ & $+ \infty$ & $\ln {\frac{4\left( {e_1 - e_3} \right)}{m^2}} $ & $+ \infty$ & $- \infty$ & $\ln {\frac{4\left( {e_2 - e_3} \right)}{m^2}} $ & $- \infty$ \\
\hline
$t = - 1$, $\left| E \right| < m^2$ & $+ \infty$ & $- \infty$ & $+ \infty$ & \multicolumn{3}{|c|}{--} \\
\hline\hline
\multicolumn{1}{|c||}{} & \multicolumn{6}{|c|}{$s = + 1$}\\
\hline	
$t = + 1$ & \multicolumn{3}{|c||}{--} & $\ln {\frac{4\left( {e_2 - e_3} \right)}{m^2}} $ & $- \infty$ & $\ln {\frac{4\left( {e_2 - e_3} \right)}{m^2}} $ \\
\hline
$t = - 1$, $E > m^2$ & \multicolumn{3}{|c||}{--} & $\ln {\frac{4\left( {e_1 - e_2} \right)}{m^2}} $ & $\ln {\frac{4\left( {e_1 - e_3} \right)}{m^2}} $ & $\ln {\frac{4\left( {e_1 - e_2} \right)}{m^2}} $ \\
\hline
$t = - 1$, $E < - m^2$ & \multicolumn{3}{|c||}{--} & \multicolumn{3}{|c|}{--} \\
\hline
$t = - 1$, $\left| E \right| < m^2$ & \multicolumn{3}{|c||}{--} & \multicolumn{3}{|c|}{--} \\
\hline
\end{tabular}
\vspace{3pt}
\caption{The range of the elliptic solutions of the sinh- or cosh-Gordon equation}
\label{tb:phi_range}
\end{center}
\end{table}

All elliptic solutions of the sinh- and cosh-Gordon equations take the following form
\begin{align}
V_1 \left( {\xi_1 ;E} \right) &= 2\wp \left( {{\xi _1} + \delta {\xi _1};{g_2}\left( E \right),{g_3}\left( E \right)} \right) - \frac{E}{3} ,\label{eq:elliptic_general_V1}\\
\varphi_1 \left( {\xi_1 ;E} \right) &= \ln \left[ { - s\frac{2}{{{m^2}}}\left( {2\wp \left( {{\xi _1} + \delta {\xi _1};{g_2}\left( E \right),{g_3}\left( E \right)} \right) - \frac{E}{3}} \right)} \right] ,\label{eq:elliptic_general_phi1}
\end{align}
for all choices of the overall sign $s$. In particular, we choose
\begin{itemize}
\item $\delta {\xi _1} = 0$ for the reflecting solutions of the cosh-Gordon equation as well as for the right incoming reflecting and transmitting solutions of the sinh-Gordon equation, having $s = -1$ in both cases
\item $\delta {\xi _1} = \omega_1$ for the left incoming transmitting solutions of the sinh-Gordon equation, having $s = -1$
\item $\delta {\xi _1} = \omega_2$ for the left incoming reflecting solutions of the sinh-Gordon equation with $s = -1$ as well as for the oscillating solutions of the sinh-Gordon equation with $s = +1$
\item $\delta {\xi _1} = \omega_1 + \omega_2$ for the reflecting solutions of the cosh-Gordon equation with $s = +1$.
\end{itemize}

\pagebreak
It is worth commenting on the behaviour exhibited by the static elliptic solutions of the sinh-Gordon equation with $s = - 1$. In that case, the effective mechanical problem describes the scattering of a particle coming from the left or the right and it is reflected by the potential barrier or else it overcomes it, depending on the energy $E$. The parity of the effective mechanical problem corresponds to the shift of the argument of the Weierstrass function by a half period -- if the particle passes the barrier the shift will be in the real axis, but if it gets reflected it will be in the imaginary axis.

Finally, we note that for every solution of the equations, there is yet another solution obtained by shifting $\xi_1$ by $\omega_x$, which is the half-period corresponding to the root $x_1$, 
\begin{equation}
\wp \left( \omega_x \right) = x_1.
\end{equation}
This symmetry connects the left and right incoming transmitting solutions of the sinh-Gordon equation, the left and right incoming reflecting solutions of the sinh-Gordon equation, the solutions of the cosh-Gordon equations with opposite signs $s$ as well as the oscillating solutions of the sinh-Gordon equation with $s = + 1$ differing by half-period.

\subsection{Double Root Limits of the Solutions}
\label{subsec_elliptic_kink}
The Weierstrass function degenerates to trigonometric or hyperbolic functions when two roots coincide (for the details see appendix \ref{sec:Weierstrass_functions}). At these degenerate limits, $\wp$ is not a doubly periodic function, but rather one of the two periods tends to infinity; it is the real one when the two larger roots coincide and it is the imaginary when the two smaller roots coincide. Only in the former case the ``time of flight'' of the effective point particle becomes infinite.

Figure \ref{fig:roots} suggests that there are exactly two possibilities for having such a degeneracy.  They arise in the sinh-Gordon equation for $E = - m^2$ and for $E = m^2$. In the first case, $E = - m^2$, the roots are
\begin{equation}
e_1 = e_2 = \frac{m^2}{12}, \quad e_3 = - \frac{m^2}{6} ,
\end{equation}
the real half-period diverges and the imaginary half-period becomes
\begin{equation}
\omega_2 = i \frac{\pi}{m}.
\end{equation}
Using formula \eqref{eq:Weierstrass_wp_e1e2} we obtain
\begin{equation}
V_1 \left( \xi_1 ; - \frac{m^2}{2} \right) = \frac{m^2}{2} \coth^2\frac{m \xi_1}{2}
\end{equation}
for the unbounded solution and
\begin{equation}
V_1 \left( \xi_1 ; - \frac{m^2}{2} \right) = \frac{m^2}{2} \tanh^2\frac{m \xi_1}{2} 
\end{equation}
for the bounded solution. They both correspond to solutions of the sinh-Gordon solution solutions with $s = - 1$. In terms of the field $\varphi$ the two solutions are
\begin{equation}
\varphi_1 \left( \xi_1 ; - \frac{m^2}{2} \right) = 2 \ln \coth\frac{m \xi_1}{2} 
\end{equation}
and
\begin{equation}
\varphi_1 \left( \xi_1 ; - \frac{m^2}{2} \right) = - 2 \ln \coth\frac{m \xi_1}{2} ,
\end{equation}
respectively. Thus, in this limit, we recover the kink and anti-kink solutions of the sinh-Gordon equation on the real line.

In the second case, $E = m^2$, the roots are
\begin{equation}
e_1 = \frac{m^2}{6}, \quad  e_2 = e_3 = - \frac{m^2}{12}
\end{equation}
and the imaginary half period diverges. Then, formula \eqref{eq:Weierstrass_wp_e2e3} implies
\begin{equation}
V_1 \left( \xi_1 ; - \frac{m^2}{2} \right) = \frac{m^2}{2} \cot^2\frac{m \xi_1}{2} 
\end{equation}
for the unbounded solution and
\begin{equation}
V_1 \left( \xi_1 ; - \frac{m^2}{2} \right) = - \frac{m^2}{2} 
\end{equation}
for the bounded one. The former solves the sinh-Gordon equation with $s = - 1$ and the latter solves the same equation with $s = + 1$. In terms of the field variable $\varphi_1$ we have, respectively,
\begin{equation}
\varphi_1 \left( \xi_1 ; - \frac{m^2}{2} \right) = 2 \ln \cot\frac{m \xi_1}{2} 
\end{equation}
and
\begin{equation}
\varphi_1 \left( \xi_1 ; - \frac{m^2}{2} \right) = 0 .
\end{equation}
Thus, in this limit, the bounded solution becomes the vacuum of the sinh-Gordon equation.

\subsection{Modular Transformations}
\label{subsec:modular}
The various solutions are interrelated by exchanging the roots $x_i$, which corresponds to modular transformations.

The details are provided in terms of the complex modulus
\begin{equation}
\tau  = \frac{{{\omega _2}}}{{{\omega _1}}} = \frac{{iK\left( {k'} \right)}}{{K\left( k \right)}} ,
\end{equation}
where $k$ and $k'$ are given by equations \eqref{eq:elliptic_moluli_D_pos} and \eqref{eq:elliptic_moluli_D_neg}. The action of the basic modular transformations $T$ and $S$ on $\tau$ are
\begin{align}
T:\tau  &\to \tau  + 1 ,\\
S:\tau  &\to  - \frac{1}{\tau } .
\end{align}

Using the following properties of the complete elliptic integral of the first kind
\begin{equation}
K\left( {ik} \right) = \frac{1}{{\sqrt {{k^2} + 1} }}K\left( {\sqrt {\frac{{{k^2}}}{{{k^2} + 1}}} } \right),\quad K\left( {\frac{1}{k}} \right) = k\left( {K\left( k \right) \mp iK\left( {k'} \right)} \right) ,
\label{eq:modular_K_properties}
\end{equation}
where in the last relation the minus sign holds when ${\mathop{\rm Im}\nolimits} \left( {{k^2}} \right) \ge 0$ and the plus sign holds when ${\mathop{\rm Im}\nolimits} \left( {{k^2}} \right) < 0$,
we find that the permutation $p_{ij}$ of any two roots $e_i$ and $e_j$ are described as follows:
\begin{align}
{p_{13}}:\tau  &\to \frac{{iK\left( k \right)}}{{K\left( {k'} \right)}} =  - \frac{1}{\tau } ,\\
{p_{12}}:\tau  &\to \frac{{iK\left( {i\frac{{k'}}{k}} \right)}}{{K\left( {\frac{1}{k}} \right)}} = \frac{{ikK\left( {k'} \right)}}{{k\left( {K\left( k \right) - iK\left( {k'} \right)} \right)}} = \frac{\tau }{{\tau - 1}} ,\\
{p_{23}}:\tau  &\to \frac{{iK\left( {\frac{1}{{k'}}} \right)}}{{K\left( {i\frac{k}{{k'}}} \right)}} = \frac{{ik'\left( {K\left( {k'} \right) - iK\left( k \right)} \right)}}{{k'K\left( k \right)}} = \tau  + 1 .
\end{align}
Thus, the permutation of roots can be identified with the following elements of the modular group
\begin{align}
{p_{13}} &= S ,\\
{p_{12}} &= T^{ - 1}ST^{ - 1} ,\\
{p_{23}} &= T .
\end{align}

Using the explicit formulas for the roots $x_i$ in the case of the sinh-Gordon equation we find that the complex modulus of the elliptic solution with $E <  - {m^2}$ is
\begin{equation}
{\tau _{E <  - {m^2}}} = \frac{{iK\left( {\sqrt {\frac{{2\sqrt {{E^2} - {m^4}} }}{{ - E + \sqrt {{E^2} - {m^4}} }}} } \right)}}{{K\left( {\sqrt {\frac{{ - E - \sqrt {{E^2} - {m^4}} }}{{ - E + \sqrt {{E^2} - {m^4}} }}} } \right)}} = \frac{{2iK\left( {\sqrt {\frac{{ - E - {m^2}}}{{ - E + {m^2}}}} } \right)}}{{K\left( {\sqrt {\frac{{2{m^2}}}{{ - E + {m^2}}}} } \right)}} ,
\end{equation}
the complex modulus of the elliptic solution with $\left| E \right| < {m^2}$ is
\begin{equation}
{\tau _{\left| E \right| < {m^2}}} = \frac{{K\left( {\sqrt {\frac{{{m^2} - E}}{{2{m^2}}}} } \right) + iK\left( {\sqrt {\frac{{{m^2} + E}}{{2{m^2}}}} } \right)}}{{K\left( {\sqrt {\frac{{{m^2} - E}}{{2{m^2}}}} } \right) - iK\left( {\sqrt {\frac{{{m^2} + E}}{{2{m^2}}}} } \right)}} 
\end{equation}
and the complex modulus of the elliptic solution with ${E > {m^2}}$ is
\begin{equation}
{\tau _{E > {m^2}}} = \frac{{iK\left( {\sqrt {\frac{{E - \sqrt {{E^2} - {m^4}} }}{{E + \sqrt {{E^2} - {m^4}} }}} } \right)}}{{K\left( {\sqrt {\frac{{2\sqrt {{E^2} - {m^4}} }}{{E + \sqrt {{E^2} - {m^4}} }}} } \right)}} = \frac{{iK\left( {\sqrt {\frac{{2{m^2}}}{{E + {m^2}}}} } \right)}}{{2K\left( {\sqrt {\frac{{E - {m^2}}}{{E + {m^2}}}} } \right)}} .
\end{equation}
In the calculations above, we made use of the following properties of the complete elliptic integral of the first kind,
\begin{equation}
K\left( {\frac{{1 - k}}{{1 + k}}} \right) = \frac{{1 + k}}{2}K\left( {k'} \right),\quad K\left( {\frac{{2\sqrt k }}{{1 + k}}} \right) = \left( {1 + k} \right)K\left( k \right) .
\end{equation}
Using the properties \eqref{eq:modular_K_properties}, we can show that the various classes of solutions of the sinh-Gordon are connected with modular transformations corresponding to the appropriate permutations of the roots. For example,
\begin{align}
S{T^{ - 1}}:{\tau _{E <  - {m^2}}} &\to \frac{1}{{1 - {\tau _{E <  - {m^2}}}}} = {\tau _{\left| E \right| < {m^2}}}, \\
{T^{ - 1}}S:{\tau _{E > {m^2}}} &\to - \frac{1}{{{\tau _{E > {m^2}}}}} - 1 = {\tau _{\left| E \right| < {m^2}}} .
\end{align}


The modular transformations do not trivialize the different classes of solutions of the sinh-Gordon equation. They correspond to topologically distinct sectors of the theory (see for example \cite{Bakas:2002qi}).


\section{The Spiky String in AdS$_3$ and its Pohlmeyer Reduction}
\label{sec:Spiky_Strings} 

In this section, we revisit the spiky string solutions in AdS$_3$, following the literature \cite{Kruczenski:2004wg} with some variations, which are more appropriate for the Pohlmeyer reduction and the use of the Weierstrass elliptic functions. This study serves as a benchmark for the construction of string solutions corresponding to more general elliptic solutions of the sinh- and cosh-Gordon equations following the methodology that we discuss later.

\subsection{The Spiky Strings in AdS$_3$}
\label{subsec:spiky_solution}

We consider the AdS$_3$ space in global coordinates. The metric is given by
\begin{equation}
d{s^2} =  \Lambda^2 \left(- {\cosh ^2} \rho d{t^2} + d{r^2} + {\sinh ^2} \rho d{\varphi ^2} \right).
\label{eq:spiky_AdS_metric}
\end{equation}
The coordinates $Y^\mu$ of the four-dimensional enhanced space are expressed in terms of the global coordinates as
\begin{equation}
\begin{split}
Y = \Lambda \left( {\begin{array}{*{20}{c}}
{\cosh \rho \cos t}\\
{\cosh \rho \sin t}\\
{\sinh \rho \cos \varphi }\\
{\sinh \rho \sin \varphi }
\end{array}} \right) .
\end{split}
\end{equation}

We search for rigid-body rotating string solutions in conformal parametrization. The appropriate ansatz is
\begin{equation}
\begin{split}
t &= {\xi_0}  + f\left( {\xi_1}  \right) ,\\
\varphi  &= \omega {\xi_0}  + g\left( {\xi_1}  \right) ,\\
\rho &= \rho\left( {\xi_1}  \right) .
\end{split}
\label{eq:ansatz}
\end{equation}
We need to specify the functions $f\left( {\xi_1}  \right)$, $g\left( {\xi_1}  \right)$ and $\rho\left( {\xi_1}  \right)$ that satisfy the Virasoro constraints \eqref{eq:Pohlmeyer_Virasoro} and equations of motion \eqref{eq:eom}.

It is matter of simple algebra to show that the Virasoro constraints \eqref{eq:Pohlmeyer_Virasoro} in the ansatz \eqref{eq:ansatz} are written as
\begin{align}
f'{\cosh ^2}\rho &= \omega g'{\sinh ^2}\rho , \label{eq:Virasoro1}\\
\left( {1 + f{'^2} - \rho{'^2}} \right){\cosh ^2}\rho &= \left( {{\omega ^2} + g{'^2} - \rho{'^2}} \right){\sinh ^2}\rho .\label{eq:Virasoro2}
\end{align}

Similarly, we can express the equations of motion in terms of the unknown functions appearing in the ansatz \eqref{eq:ansatz}. All components of the equations of motion contain the quantity ${\partial _ + }{Y} \cdot {\partial _ - }{Y}$, which equals
\begin{equation}
\begin{split}
{\partial _ + }{Y} \cdot {\partial _ - }{Y} &= \Lambda^2 \left( {1 - f{'^2} + \rho{'^2}} \right){{\cosh }^2}\rho - \Lambda^2 \left( {{\omega ^2} - g{'^2} + \rho{'^2}} \right){{\sinh }^2}\rho\\
 &= 2 \Lambda^2 \left( {{{\cosh }^2}\rho - {\omega ^2}{{\sinh }^2}\rho} \right) ,
\end{split}
\label{eq:Pohlmeyer_field_ansatz}
\end{equation}
where in the last step we made use of the equation \eqref{eq:Virasoro2}.

Linear combinations of the $Y^{-1}$ and $Y^0$ components of the equations of motion are used to eliminate the dependence upon $t$ and $\varphi$ coordinates giving rise to the following system of ordinary differential equations
\begin{align}
\left( {1 - f{'^2} + \rho{'^2}} \right)\cosh \rho + \rho''\sinh \rho &= 2 \left( {{{\cosh }^2}\rho - {\omega ^2}{{\sinh }^2}\rho} \right)\cosh \rho , \label{eq:eomt1}\\
f''\cosh r + 2f'\rho'\sinh \rho &= 0 .\label{eq:eomt2}
\end{align}
The last one can be easily integrated to yield
\begin{equation}
f' = \frac{{{C_f}}}{{{{\cosh }^2}\rho}} \, .
\label{eq:fprime}
\end{equation}
Similarly, appropriate linear combinations of the $Y^1$ and $Y^2$ components of the equations of motion yield the system of ordinary differential equations
\begin{align}
\left( {{\omega ^2} - g{'^2} + \rho{'^2}} \right)\sinh \rho + \rho''\cosh \rho &= 2 \left( {{{\cosh }^2}\rho - {\omega ^2}{{\sinh }^2}\rho} \right)\sinh \rho . \label{eq:eomphi1}\\
g''\sinh \rho + 2g'\rho'\cosh \rho &= 0 .\label{eq:eomphi2}
\end{align}
As before, the last equation can be easily integrated leading to the relation
\begin{equation}
g' = \frac{{{C_g}}}{{{{\sinh }^2}\rho}} \, .
\label{eq:gprime}
\end{equation}

Next, substituting the form of $f'$ and $g'$ into the first Virasoro constraint \eqref{eq:Virasoro1} gives rise to following relation among the different integration constants,
\begin{equation}
{C_f} = \omega {C_g} \equiv \omega C ,
\end{equation}
while the second Virasoro constraint \eqref{eq:Virasoro2} implies
\begin{equation}
\rho{'^2} = \frac{{\left( {4{C^2} - {{\sinh }^2}2\rho} \right)\left( {{\omega ^2}{{\sinh }^2}\rho - {{\cosh }^2}\rho} \right)}}{{{{\sinh }^2}2\rho}} \, .
\label{eq:rhoprime}
\end{equation}
Note at this point that if equations \eqref{eq:fprime}, \eqref{eq:gprime} and \eqref{eq:rhoprime} are satisfied, then the remaining equations of motion \eqref{eq:eomt1} and \eqref{eq:eomphi1} will also be satisfied. Thus, complete treatment of the problem requires solving equation \eqref{eq:rhoprime} for the coordinate function $\rho$ and substituting it to equations \eqref{eq:fprime} and \eqref{eq:gprime} to determine $f$ and $g$.

It is convenient to parametrize $C$ and $\omega$ as
\begin{equation}
C \equiv \frac{1}{2}\sinh 2{\rho_0} ,\quad \omega \equiv \coth {\rho_1} .
\end{equation}
Here, we implicitly assume that $\left| \omega \right| > 1$, but we will also discuss the case $\left| \omega \right| < 1$ later. We also introduce the change of variables
\begin{equation}
u = \cosh 2\rho ,\quad {u_0} = \cosh 2{\rho_0} ,\quad {u_1} = \cosh 2{\rho_1} .
\label{eq:spiky_u_def}
\end{equation}
Then, equation \eqref{eq:rhoprime} is written as
\begin{equation}
{\left( {\frac{{du}}{{d{\xi_1} }}} \right)^2} = 4\frac{{\left( {{u^2} - u_0^2} \right)\left( {{u_1} - u} \right)}}{{{u_1} - 1}} .
\label{eq:ueq}
\end{equation}
It is obvious from its form that real solutions require the bounds on $u$
\begin{equation}
\min \left( {{u_0},{u_1}} \right) < u < \max \left( {{u_0},{u_1}} \right) .
\label{eq:spiky_ubounds}
\end{equation}

the right hand side of equation \eqref{eq:ueq} is cubic in $u$, and, thus, with appropriate rescaling and shifting of the variable $u$ it can be brought into the Weierstrass form. Indeed, letting
\begin{equation}
u = - \left( u_1 - 1 \right) w + \frac{u_1}{3} \, ,
\end{equation}
equation \eqref{eq:ueq} takes the final form
\begin{equation}
{\left( {\frac{{dw}}{{d{\xi_1} }}} \right)^2} = 4 w^3 - \frac{4}{3} \frac{u_1^2 + 3 u_0^2}{ \left( u_1 - 1 \right) ^2} w + \frac{8}{27} \frac{u_1 \left( u_1^2 - 9 u_0^2 \right)}{ \left( u_1 - 1 \right) ^3} 
\label{eq:spiky_Weierstrall_final}
\end{equation}
and its solution is
\begin{equation}
u = - \left( u_1 - 1 \right) \wp \left( \xi_1 + \delta \xi_1 ; \frac{4}{3} \frac{u_1^2 + 3 u_0^2}{ \left( u_1 - 1 \right) ^2} , - \frac{8}{27} \frac{u_1 \left( u_1^2 - 9 u_0^2 \right)}{ \left( u_1 - 1 \right) ^3}\right) + \frac{u_1}{3},
\label{eq:spiky_solution_u}
\end{equation}
where $\delta \xi_1$ can be either $0$ or $\omega_2$ for reality.

The roots of the cubic polynomial appearing in equation \eqref{eq:spiky_Weierstrall_final} are real. They are
\begin{equation}
r_1 = - \frac{2 u_1}{3 \left( u_1 - 1 \right)}, \quad r_{2,3} = \frac{u_1 \pm 3 u_0}{3 \left( u_1 - 1 \right)} 
\label{eq:spiky_r_roots}
\end{equation}
and they are all distinct unless $\rho_0 = \rho_1$ in which case $r_1 = r_3$. It can be easily seen that 
the elliptic solution \eqref{eq:spiky_solution_u} takes values within the interval $\left( - \infty , - u_0 \right]$ when $\delta \xi_1 = 0$, but since $u=\cosh 2\rho$ and $u_0 = \cosh 2 \rho_0$ it is ruled out. On the other hand, when $\delta \xi_1 = \omega_2$, the solution $u$ takes values within the interval $\left[ \min \left( {{u_0},{u_1}} \right) , \max \left( {{u_0},{u_1}} \right) \right]$, agreeing with equation \eqref{eq:spiky_ubounds} and it is acceptable. Thus, the physical solution is uniquely determined to be
\begin{equation}
u = - \left( u_1 - 1 \right) \wp \left( \xi_1 + \omega_2 ; \frac{4}{3} \frac{u_1^2 + 3 u_0^2}{ \left( u_1 - 1 \right) ^2} , - \frac{8}{27} \frac{u_1 \left( u_1^2 - 9 u_0^2 \right)}{ \left( u_1 - 1 \right) ^3}\right) + \frac{u_1}{3} \, .
\label{eq:spiky_u_solution}
\end{equation}

The unknown functions $f$ and $g$ can be calculated using the integral formula \eqref{eq:Weierstrass_integral}. We find,
\begin{align}
f &=  - \sqrt {\frac{{\left( {u_0^2 - 1} \right)\left( {{u_1} + 1} \right)}}{{{{\left( {{u_1} - 1} \right)}^3}}}} \frac{1}{{\wp '\left( {{a_1}} \right)}}\left( {2\zeta \left( {{a_1}} \right)\left( {{\xi _1} + {\omega _2}} \right) + \ln \frac{{\sigma \left( {{\xi _1} - {a_1}} \right)}}{{\sigma \left( {{\xi _1} + {a_1}} \right)}}} \right) ,\\
g &=  - \sqrt {\frac{{\left( {u_0^2 - 1} \right)}}{{{{\left( {{u_1} - 1} \right)}^2}}}} \frac{1}{{\wp '\left( {{a_2}} \right)}}\left( {2\zeta \left( {{a_2}} \right)\left( {{\xi _1} + {\omega _2}} \right) + \ln \frac{{\sigma \left( {{\xi _1} - {a_2}} \right)}}{{\sigma \left( {{\xi _1} + {a_2}} \right)}}} \right) ,
\end{align}
where $a_1$ and $a_2$ are defined to be
\begin{equation}
\wp \left( {{a_1}} \right) = \frac{{{u_1} + 3}}{{3\left( {{u_1} - 1} \right)}}\, ,\quad \wp \left( {{a_2}} \right) = \frac{{{u_1} - 3}}{{3\left( {{u_1} - 1} \right)}} \, .
\end{equation}

We study the shape of the rigidly rotating string by freezing time. We find that $d \rho / d \varphi$ vanishes as $\rho \to \rho_0$ and it diverges as $\rho \to \rho_1$. Thus, at radius $\rho = \rho_1$ the solution has singular behaviour, hence the name ``spiky strings'' characterizing this family of solutions. A reasonable question is why the tension of the string does not change its shape evolving it towards a smoother configuration. The answer to this question turns out to be kinematic. The form of the AdS$_3$ metric \eqref{eq:spiky_AdS_metric} implies that the velocity of a given point with radial coordinate $\rho$ of a rigidly rotating configuration with constant angular velocity $\omega$ equals
\begin{equation}
v = \omega \tanh \rho .
\end{equation}
Consequently, if $\left| \omega \right| > 1$, there will be a finite radius $\rho_c = \coth^{-1} \left| \omega \right|$ where the velocity $v$ reaches the speed of light. Actually, it has been implicitly assumed that $\left| \omega \right| > 1$ when we let $\omega = \coth \rho_1$, and, thus, $\rho_1$ turns out to be equal to $\rho_c$. The physical explanation for the existence of the spikes is that these points of the string move with the speed of light. If we were in the range $\left| \omega \right| < 1$, no spikes would occur at finite radius and consequently the corresponding solutions would extend up to infinite radius.

\subsection{Pohlmeyer Reduction of Spiky Strings}
\label{subsec:Spiky_Pohlmeyer}

Let us now describe the spiky string solutions in the Pohlmeyer reduced theory. For this purpose, we compute the inner product ${\partial _ + }Y \cdot {\partial _ - }Y$, which provides the exponential of the Pohlmeyer reduced field $a$. After some algebra, we find
\begin{equation}
{\partial _ + }Y \cdot {\partial _ - }Y = 2{\Lambda ^2}\left( {{{\cosh }^2}\rho  - {\omega ^2}{{\sinh }^2}\rho } \right) = 2{\Lambda ^2}\frac{{{u_1} - u}}{{{u_1} - 1}} \, .
\end{equation}
Thus, if $u_1 > u_0$, the inner product is positive definite and the Pohlmeyer field is
\begin{equation}
a = \ln \left[ {2{\Lambda ^2}\left( {\wp \left( {{\xi _1} + {\omega _2};\frac{4}{3}\frac{{u_1^2 + 3u_0^2}}{{{{\left( {{u_1} - 1} \right)}^2}}}, - \frac{8}{{27}}\frac{{{u_1}\left( {u_1^2 - 9u_0^2} \right)}}{{{{\left( {{u_1} - 1} \right)}^3}}}} \right) + \frac{{2{u_1}}}{{3\left( {{u_1} - 1} \right)}}} \right)} \right] .
\end{equation}
If $u_1 < u_0$, as in reference \cite{Mosaffa:2007ty}, the inner product will be negative, and the surface will be not time-like, but space-like. The spikes lie at radius $\rho_1$ and they move with the speed of light. Since $\min \left( {{u_0},{u_1}} \right) < u < \max \left( {{u_0},{u_1}} \right) $, it is immediately clear that if $u_1 < u_0$ all other points of the string will move with velocity larger than the speed of light. Thus we will not pursue those solutions any further.

Following the details of the Pohlmeyer reduction, as explained in section \ref{subsec:Pohlmeyer_reduction}, we will specify the functions ${a_4^{\left( \pm \right)}}$, associated to the spiky strings. They are given by equation
\begin{equation}
{a_4^{\left( \pm \right)}}{v_4} = \partial _ \pm ^2 Y - {\partial _ \pm }a{\partial _ \pm }Y ,
\end{equation}
and, thus, we arrive at the quadratic relations
\begin{align}
\left( {a_4^{\left( \pm \right)}} \right)^2 = \left( \partial _ \pm ^2 Y - {\partial _ \pm }a{\partial _ \pm }Y \right) \cdot \left( \partial _ \pm ^2 Y - {\partial _ \pm }a{\partial _ \pm }Y \right) ,\\
{a_4^{\left( + \right)}} {a_4^{\left( - \right)}} = \left( \partial _ + ^2 Y - {\partial _ + }a{\partial _ + }Y \right) \cdot \left( \partial _ - ^2 Y - {\partial _ - }a{\partial _ - }Y \right) .
\end{align}
There is an ambiguity in the signs of ${a_4^{\left( \pm \right)}}$, but there is no ambiguity in the relative sign. After some algebra we obtain
\begin{align}
{\left( {a_4^{\left(  \pm  \right)}} \right)^2} &= {\left( {\Lambda \frac{{\sinh 2{\rho _0} \mp \sinh 2{\rho _1}}}{{{{\sinh }^2}{\rho _1}}}} \right)^2} = {\left( {\frac{{2\Lambda }}{{{u_1} - 1}}} \right)^2}{\left( {\sqrt {u_0^2 - 1}  \mp \sqrt {u_1^2 - 1} } \right)^2} ,\\
a_4^{\left(  +  \right)}a_4^{\left(  -  \right)} &= {\Lambda ^2}\frac{{{{\sinh }^2}2{\rho _0} - {{\sinh }^2}2{\rho _1}}}{{{{\sinh }^4}{\rho _1}}} = {\left( {\frac{{2\Lambda }}{{{u_1} - 1}}} \right)^2}\left( {u_0^2 - u_1^2} \right) ,
\end{align}
showing that ${a_4^{\left( \pm \right)}}$ are both constants. The product $a_4^{\left(  +  \right)}a_4^{\left(  -  \right)}$ is negative, since $u_1 > u_0$ for the spiky strings. Thus, the spiky strings in AdS$_3$ are naturally related to solutions of the sinh-Gordon equation.

Once the coefficients ${a_4^{\left(  \pm  \right)}}$ are known, we rescale the world-sheet coordinate $\xi_1$ according to equation \eqref{eq:coordinate_rescaling},
\begin{equation}
{\xi_1} = \frac{{\sqrt {{u_1} - 1} \left( {\gamma {\xi_1} ' + \gamma \beta {\xi_0} '} \right)}}{{\sqrt 2 \Lambda \sqrt[4]{{ {u_1^2 - u_0^2} }}}} = \frac{{\sinh {\rho _1}\left( {\gamma {\xi_1} ' + \gamma \beta {\xi_0} '} \right)}}{{\Lambda \sqrt[4]{{ {{{\sinh }^2}2{\rho _1} - {{\sinh }^2}2{\rho _0}} }}}},
\end{equation}
where
\begin{equation}
\beta = \frac{{\sqrt {u_1^2 - 1} - \sqrt {u_1^2 - u_0^2} }}{{\sqrt {u_0^2 - 1} }} = \frac{{\sinh 2{\rho _1} - \sqrt {{{\sinh }^2}2{\rho _1} - {{\sinh }^2}2{\rho _0}} }}{{\sinh 2{\rho _0}}}
\label{eq:rapidity}
\end{equation}
and $\gamma = 1 / \sqrt{1 - \beta^2}$. We also redefine the Pohlmeyer field according to equation \eqref{eq:Pohlmeyer_field_redef}. At first sight, the solution appears to depend on both world-sheet coordinates $\xi_0$ and $\xi_1$. However, one can reparametrize those coordinates using a boost to set $\gamma {\xi_1} ' + \gamma \beta {\xi_0} ' \to {\xi_1}$ within the conformal gauge. Finally, using the homogeneity relation \eqref{eq:Weierstras_homogeneity_wp}, we can perform a rescaling of the Weierstrass function argument with scaling parameter
\begin{equation}
\mu = \frac{{\sqrt {{u_1} - 1}}}{{\sqrt 2 \Lambda \sqrt[4]{{ {u_1^2 - u_0^2} }}}}
\end{equation}
to obtain the following solution for the sinh-Gordon field
\begin{equation}
\varphi = \ln \left[ {2{\Lambda ^2}\left( {\wp \left( {{\xi _1} + {\omega _2};\frac{{u_1^2 + 3u_0^2}}{{3{\Lambda ^4}\left( {u_1^2 - u_0^2} \right)}}, - \frac{{{u_1}\left( {u_1^2 - 9u_0^2} \right)}}{{27{\Lambda ^6}{{\left( {u_1^2 - u_0^2} \right)}^{\frac{3}{2}}}}}} \right) + \frac{{{u_1}}}{{3{\Lambda ^2}\sqrt {u_1^2 - u_0^2} }}} \right)} \right] .
\label{eq:spiky_final_solution}
\end{equation}

The roots of the characteristic cubic polynomial of the Weierstrass equation are
\begin{equation}
r_1 = - \frac{{{u_1}}}{{3{\Lambda ^2}\sqrt {u_1^2 - u_0^2} }} , \quad r_{2,3} = - \frac{{{u_1 \pm 3 u_0}}}{{6{\Lambda ^2}\sqrt {u_1^2 - u_0^2} }} 
\end{equation}
after rescaling by $\mu$.
Thus, comparing the solution \eqref{eq:spiky_final_solution} to the general form \eqref{eq:elliptic_general_phi1}, we arrive at the identification $x_1 = r_1$. Since $u_1 > u_0$, $r_1$ is the smallest root and the spiky strings correspond to static solutions of the sinh-Gordon equation with $s = -1$. Their interpretation in terms of the effective one-dimensional mechanical system is that of incoming particles from the left that get reflected by the potential barrier, as described in section \ref{subsec:elliptic_solutions}. The energy of the effective particle follows from $x_1$,
\begin{equation}
E = - \frac{2}{{{\Lambda ^2}}}\frac{{u_1}}{{\sqrt {u_1^2 - u_0^2} }} \, .
\end{equation}
The energy is indeed smaller than $- 2 / \Lambda^2$, as required for reflecting solutions.

\section{The Building Blocks of the Elliptic String Solutions}
\label{sec:SUSYQM}

Given a classical string configuration, it is straightforward to find the corresponding solution of the Pohlmeyer reduced system, as for the spiky strings in AdS$_3$ space. The inverse problem is highly non-trivial due to the non-local nature of the transformation relating the embedding functions $Y^\mu$ with the reduced field $\varphi$ and because the Pohlmeyer reduction is a many-to-one mapping. Such a construction requires the solution of the equations of motion for the embedding functions,
\begin{equation}
\frac{{{\partial^2 Y^\mu}}}{{\partial{\xi_1 ^2}}} - \frac{{{\partial^2 Y^\mu}}}{{\partial{\xi_0 ^2}}} = - s \frac{1}{{{\Lambda ^2}}}{e^\varphi}{Y^\mu } ,
\label{eq:susy_reconstruction}
\end{equation}
supplemented with the geometric constraint as well as the Virasoro constraints of the embedding problem,
\begin{align}
Y \cdot Y &= s {\Lambda ^2} ,\label{eq:susy_constraint}\\
{\partial _ \pm }Y \cdot {\partial _ \pm }Y &= 0 .\label{eq:susy_Virasoro}
\end{align}

We will construct the classical string solutions whose Pohlmeyer counterparts are the elliptic solutions presented before. Here, we focus on the general solutions of equation \eqref{eq:susy_reconstruction}, which will be used as building blocks for the construction of string solutions.

\subsection{The Effective \Schrodinger Problem}
\label{subsec:eff_Schr_sigma}

Consider the special case of a static solution of the reduced system $\varphi \left( \xi_0, \xi_1 \right) = \varphi_1 \left( \xi_1 \right)$. We define 
\begin{equation}
V_1\left( \xi_1  \right) := - s {\frac{1}{\Lambda^2}} {e^{\varphi_1}} ,
\label{eq:SUSY_potential_1}
\end{equation}
as in equation \eqref{eq:elliptic_V_definition_1} introduced earlier. Then, the equations of motion \eqref{eq:susy_reconstruction} can be rewritten as
\begin{equation}
\frac{{{d^2 Y^\mu}}}{{d{\xi_1 ^2}}} - \frac{{{d^2 Y^\mu}}}{{d{\xi_0 ^2}}} = V_1\left( \xi_1  \right){Y^\mu } .
\end{equation}

Since $V_1$ depends solely on $\xi_1$, it is possible to separate the variables letting
\begin{equation}
{Y^\mu }\left( \xi_0 , \xi_1  \right) := {\Sigma ^\mu }\left( \xi_1  \right){{\rm T}^\mu }\left( \xi_0  \right) .
\label{eq:susy_separation}
\end{equation}
We arrive at a pair of ordinary differential equations,
\begin{align}
 - \frac{{{d^2}{\rm T^\mu}}}{{d{\xi_0 ^2}}} &= {\kappa^\mu}{\rm T^\mu} , \label{eq:Schr_tau_1}\\
 - \frac{{{d^2}\Sigma^\mu}}{{d{\xi_1 ^2}}} + V_1\left( \xi_1 \right) \Sigma^\mu  &={\kappa^\mu} \Sigma^\mu ,\label{eq:Schr_sigma_1}
\end{align}
which can be viewed as two effective \Schrodinger problems with common eigenvalues. One of them has flat potential and the other has a potential that can be derived from the solution of the Pohlmeyer reduced system. This pair of \Schrodinger problems does not require any normalization condition for the effective wavefunction.

Translationally invariant solutions of the reduced system can be related to classical string configurations in a similar manner. The only difference is that $V_1 \left( \xi_1 \right)$ is replaced by 
\begin{equation}
V_0\left( \xi_0  \right) := s\frac{1}{{{\Lambda ^2}}}{e^{\varphi_0}}.
\label{eq:susy_tau_potential}
\end{equation}
Using the separation of variables \eqref{eq:susy_separation}, as before, the classical equations of motion \eqref{eq:susy_reconstruction} reduce to following pair of effective \Schrodinger equations
\begin{align}
 - \frac{{{d^2}{\Sigma ^\mu }}}{{d{\xi_1 ^2}}} &= \kappa^\mu {\Sigma ^\mu } ,\label{eq:susy:inv_schr_sigma_0}\\
 - \frac{{{d^2}{{\rm T}^\mu }}}{{d{\xi_0 ^2}}} + V_0\left( \xi_0  \right) {{\rm T}^\mu } &= \kappa^\mu  {{\rm T}^\mu } \label{eq:susy:inv_schr_tau_0}
\end{align}
with common eigenvalues. Thus, the construction of string configurations from translationally invariant solutions of the Pohlmeyer reduced system is similar to the previous construction, taking into account the necessary inversion of the sign $s$.

In either case, the solution of the effective \Schrodinger problems have to be compatible to the constraints \eqref{eq:susy_constraint} and \eqref{eq:susy_Virasoro}.

\subsection{The Supersymmetric Effective \Schrodinger Problem}
\label{subsec:special_solutions}

The \Schrodinger problem with the flat potential is trivial and so we focus on the other \Schrodinger problem with potential $V_1 \left( \xi_1 \right)$ by considering static solutions. In this context, methods of supersymmetric quantum mechanics will prove useful.

Recall that the sinh-Gordon or the cosh-Gordon equation yields an energy relation for an effective point particle,
\begin{equation}
\frac{1}{2}{\left( {\frac{{d\varphi_1}}{{d\xi_1 }}} \right)^2} + s \frac{1}{{{\Lambda ^2}}}\left( {{e^{\varphi_1}} - t{e^{ - {\varphi_1}}}} \right) = E ,
\label{eq:SUSY_energy_cons}
\end{equation}
which can be used to show the following functional relation for the potential $V_1$,
\begin{equation}
V_1 = \frac{1}{2}\left( {\frac{{V_1''}}{V_1} - \frac{{V_1{'^2}}}{{2{V_1^2}}} - E} \right) .
\end{equation}
Here, prime denotes differentiation with respect to $\xi_1$. At this point, we introduce
\begin{equation}
W\left( \xi_1  \right) := \frac{{V_1'\left( \xi_1  \right)}}{{2V_1\left( \xi_1  \right)}} = \frac{{\varphi_1'\left( \xi_1  \right)}}{2} \, ,
\label{eq:SUSY_superpotential_1}
\end{equation}
in terms of which the \Schrodinger potential $V_1$ takes the form
\begin{equation}
V_1 = {W^2} + W' - \frac{E}{2} .
\end{equation}
Thus, $W \left( \xi_1 \right)$ serves as superpotential and the pair of effective \Schrodinger problems takes the convenient form
\begin{align}
 - \frac{{{d^2}{\rm T^\mu}}}{{d{\xi_0 ^2}}} &= {\kappa^\mu}{\rm T^\mu} , \label{eq:susy:Schr_tau_1}\\
 - \frac{{{d^2}\Sigma^\mu}}{{d{\xi_1 ^2}}} + \left({W^2} + \frac{d W}{d \xi_1} \right)\Sigma^\mu  &= \left({\kappa^\mu} + \frac{E}{2} \right) \Sigma^\mu .\label{eq:susy:Schr_sigma_1}
\end{align}

Based on the techniques of supersymmetric quantum mechanics, we introduce the creation and annihilation operators
\begin{equation}
A = \frac{d}{{d\xi_1 }} + W,\quad {A^\dag } = - \frac{d}{{d\xi_1 }} + W ,
\label{eq:susy_a}
\end{equation}
which recast the \Schrodinger equation \eqref{eq:susy:Schr_sigma_1} into the form
\begin{equation}
A{A^\dag }\Sigma = \left({\kappa} + \frac{E}{2} \right)\Sigma .
\label{eq:susy_Schr_sigma_a}
\end{equation}
The same equation arises for all components labelled by $\mu$ although $\kappa^\mu$ are not necessarily the same. Here, we suppress the index $\mu$ for convenience.

It is easy to find a solution of this problem, by constructing a state $\Sigma_0$ that is annihilated by ${A^\dag }$, which we call the ``roof'' state. It satisfies the first order equation 
\begin{equation}
{\Sigma _0}' = W{\Sigma _0} = \frac{{\varphi_1'}}{2}{\Sigma _0} \, ,
\label{eq:susy_ground}
\end{equation}
which can be easily integrated as
\begin{equation}
\Sigma _0^2 = {e^{\varphi_1}} = - s{\Lambda ^2} V_1 ,
\end{equation}
up to a normalization constant. The corresponding eigenvalue is zero so that
\begin{equation}
\kappa _0 = - \frac{E}{2} .
\end{equation}

There is another interesting class of wavefunctions having the following form,
\begin{equation}
\Sigma  = \sqrt {\Sigma _0^2 + c} ,
\end{equation}
which, as will be seen later, provide particular examples of classical string configurations. Such solutions can be solely found by supersymmetric quantum mechanics. For any state of the form $f\left( {{\Sigma _0}} \right)$ we have the identities,
\begin{align}
{A^\dag }f\left( {{\Sigma _0}} \right) &=  W\left( { - f'\left( {{\Sigma _0}} \right){\Sigma _0} + f\left( {{\Sigma _0}} \right)} \right) ,\\
Af\left( {{\Sigma _0}} \right) &= W\left( {f'\left( {{\Sigma _0}} \right){\Sigma _0} + f\left( {{\Sigma _0}} \right)} \right) ,
\end{align}
which, in turn, imply that
\begin{equation}
A{A^\dag }\sqrt {\Sigma _0^2 + c}  = - s c\frac{1}{{2{\Lambda ^2}}}\left[ {\sqrt {\Sigma _0^2 + c}  - \frac{{{c^2} + sE{\Lambda ^2}c - t}}{{{{\left( {\sqrt {\Sigma _0^2 + c} } \right)}^3}}}} \right] .
\end{equation}
$\sqrt {\Sigma _0^2 + c}$ is an eigenstate not only when $c=0$, corresponding to the roof state itself, but also when $c$ satisfies the quadratic equation ${c^2} + sE{\Lambda ^2}c - t = 0$, corresponding to the choices
\begin{equation}
{c_ \pm } = \frac{1}{2}\left( {- s E{\Lambda ^2} \mp \sqrt {{E^2}{\Lambda ^4} + 4t} } \right) .
\end{equation}
The respective eigenvalues yield
\begin{equation}
\kappa_ \pm = - s c\frac{1}{{2{\Lambda ^2}}} - \frac{E}{2} = \frac{1}{4}\left( { - E \pm s\sqrt {{E^2} + \frac{{4t}}{{{\Lambda ^4}}}} } \right) 
\label{eq:special_solutions_eigenvalues}
\end{equation}
and, as it turns out, $c_\pm$ is proportional to $\kappa_\mp$,
\begin{equation}
{c_ \pm } = 2s{\Lambda ^2}\kappa_ \mp .
\end{equation}

Summarizing, there are two additional eigenfunctions of the effective supersymmetric \Schrodinger problem
\begin{equation}
{\Sigma _ \pm } = \sqrt {{\Sigma _0^2} + \frac{1}{2}\left( {- s E{\Lambda ^2} \mp \sqrt {{E^2}{\Lambda ^4} + 4t} } \right)} = \sqrt {{\Sigma _0^2} + 2s{\Lambda ^2}{\kappa _ \mp }},
\label{eq:special_solutions}
\end{equation}
up to normalization constant, which can be appended to $\Sigma_0$.

Using the Weierstrass form \eqref{eq:elliptic_general_V1} for the potential $V_1$ and noting that $c_\pm$ are related to the roots $x_i$ as follows,
\begin{equation}
{c_ \pm } =  - 2s{\Lambda ^2}\left( {{x_1} - {x_{2,3}}} \right) ,
\end{equation}
the three eigenstates we have just constructed can be rewritten as
\begin{equation}
{\Sigma _0} = N\sqrt {\wp \left( {{\xi _1} + \delta {\xi _1}} \right) - {x_1}} ,\quad {\Sigma _ \pm } = N\sqrt {\wp \left( {{\xi _1} + \delta {\xi _1}} \right) - {x_{2,3}}} .
\end{equation}
The superpotential can also be written in terms of the Weierstrass function,
\begin{equation}
W \left( \xi_1 \right) = \frac{\wp ' \left( \xi_1 + \delta \xi_1 \right)}{2 \left( \wp \left( \xi_1 + \delta \xi_1 \right) - x_1 \right)}
\end{equation}
and
\begin{equation}
V_+ := W^2 + W' = 2 {\wp \left( \xi_1 + \delta \xi_1 \right)} + x_1 .
\end{equation}
As a side remark, the supersymmetric partner potential takes similar form,
\begin{equation}
\begin{split}
V_- :=& W^2 - W' = \frac{2 \left( {\wp \left( \xi_1 + \delta \xi_1 \right) - x_2} \right) \left( {\wp \left( \xi_1 + \delta \xi_1 \right) - x_3} \right)}{\wp \left( \xi_1 + \delta \xi_1 \right) - x_1} - 2 \wp \left( \xi_1 + \delta \xi_1 \right) - x_1 \nonumber \\
=& 2 \wp \left( \xi_1 + \delta \xi_1 + \omega_x \right) + x_1 ,
\end{split}
\end{equation}
where $\omega_x$ is the half-period corresponding to the root $x_1$. In the last step we made use the addition formula \eqref{eq:Weierstrass_addition_wp} for the Weierstrass function. This is in accordance to the fact noted in section \ref{subsec:elliptic_solutions} that for every elliptic solution, there is another one obtained by shifting $\xi_1$ by $\omega_x$.

\subsection{The \Lame Potential}
\label{subsec:lame}

According to the previous discussion, the effective potential $V_1$ takes the special form
\begin{equation}
{V_1} = 2 {\wp \left( {{\xi _1} + \delta \xi_1} \right) - \frac{E}{3}} ,\label{eq:lame_pot}
\end{equation}
where $\delta \xi_1$ is either $0$ or $\omega_2$ depending on the use of the unbounded or the bounded real solution of the Weierstrass equation.

The special class of periodic potentials
\begin{equation}
V \left( x \right) = n\left( {n + 1} \right)\wp \left( x \right)
\end{equation}
are called \Lame potentials and they are known to have magnificent structure for any integer $n$. In fact, it has been shown that the spectrum of the corresponding \Schrodinger problem contains $n$ finite allowed bands, some of which may become degenerate, plus one more continuous band extending to infinite energy. Our case corresponds to the $n = 1$ \Lame problem,
\begin{equation}
- \frac{{{d^2}y}}{{d{x^2}}} + 2\wp \left( x \right)y = \lambda y ,
\label{eq:lame_n1_problem}
\end{equation}
whose solutions are given in general by
\begin{equation}
{y_ \pm } \left( {x ; a} \right) = \frac{{\sigma \left( {x \pm a} \right)}}{{\sigma \left( x \right) \sigma \left( \pm a \right)}}{e^{ - \zeta \left( \pm \alpha  \right)x}} 
\label{eq:lame_eigenstates}
\end{equation}
with corresponding eigenvalues
\begin{equation}
\lambda = - \wp \left( a \right) .
\label{eq:lame:eigenvalues}
\end{equation}

The functions $\zeta \left( z \right)$ and $\sigma \left( z \right)$ are the Weierstrass zeta and sigma functions, respectively, which are defined as
\begin{equation}
\frac{{d\zeta }}{{dz}} = - \wp , \quad \frac{1}{\sigma }\frac{{d\sigma }}{{dz}}= \zeta .
\label{eq:lame_zeta_sigma_defs}
\end{equation}
Unlike the Weierstrass function $\wp$, the Weierstrass functions $\zeta$ and $\sigma$ are not doubly periodic functions. Specifically, $\zeta$ is a doubly quasi-periodic function, while $\sigma$ transforms in an even more complicated way under shifts of the variable $z$ by the periods $2\omega_1$ and $2\omega_2$. More information about them is provided in appendix \ref{sec:Weierstrass_functions}.

The assertion made above can be verified by direct computation. Using the defining relations \eqref{eq:lame_zeta_sigma_defs}, it can be shown that
\begin{align}
\frac{{d{y_ \pm }}}{{dx}} &= \left( {\zeta \left( {x \pm a} \right) - \zeta \left( x \right) \mp \zeta \left( \alpha  \right)} \right){y_ \pm } ,\\
\frac{{{d^2}{y_ \pm }}}{{d{x^2}}} &= \left[ {{{\left( {\zeta \left( {x \pm a} \right) - \zeta \left( x \right) \mp \zeta \left( \alpha  \right)} \right)}^2} - \left( {\wp \left( {x \pm a} \right) - \wp \left( x \right)} \right)} \right]{y_ \pm } .
\end{align}
Then, applying the addition theorems \eqref{eq:Weierstrass_addition_wp} and \eqref{eq:Weierstrass_addition_zeta}, we obtain
\begin{align}
\frac{{d{y_ \pm }}}{{dx}} &= \frac{1}{2}\frac{{\wp '\left( x \right) \mp \wp '\left( a \right)}}{{\wp \left( x \right) - \wp \left( a \right)}}{y_ \pm } ,\\
\frac{{{d^2}{y_ \pm }}}{{d{x^2}}} &= \left( {2\wp \left( x \right) + \wp \left( a \right)} \right){y_ \pm } .
\end{align}
The latter relation implies that indeed $y_ \pm$ are eigenstates of the \Lame problem \eqref{eq:lame_n1_problem}. In either case the corresponding eigenvalue is provided by equation \eqref{eq:lame:eigenvalues}.

As long as $y_\pm$ are linearly independent, they provide the general solution of the \Lame problem \eqref{eq:lame_n1_problem}. However, as it turns out, $y_\pm$ are not linearly independent when $ - \lambda$ is any of the three roots $e_i$. In those special cases, equation \eqref{eq:Weierstrass_addition_sigma_special} implies that the two solutions coincide and they can be written as
\begin{equation}
{y_ \pm }\left( {x;{\omega _{1,2,3}}} \right) = \sqrt {\wp \left( x \right) - {e_{1,3,2}}} \, ,
\end{equation}
since $e_i$ are given by the values of the Weierstrass function at half-periods. These are precisely
the eigenstates found in section \ref{subsec:special_solutions}, using supersymmetric quantum mechanics. For those eigenvalues, the second independent solution is given by
\begin{equation}
\tilde y\left( {x;{\omega _{1,2,3}}} \right) = \sqrt {\wp \left( x \right) - {e_{1,3,2}}} \left( {\zeta \left( {x + {\omega _{1,2,3}}} \right) + {e_{1,3,2}}x} \right) . 
\end{equation}

Returning back to $y_\pm \left( x \right)$ for other values of $\lambda$, we want to examine their reality and study how they transform by shifting $x$ by the period $2\omega_1$. There are two basic cases to be considered, $\Delta > 0$ and $\Delta < 0$. For $\Delta > 0$, there are three real roots, and so, according to the analysis of section \ref{subsec:elliptic_wp}, we may distinguish four subcases ($b$ below is taken to be real and $\bar y$ denotes the complex conjugate of $y$):
\begin{enumerate}
\item If $\wp \left( a \right) > e_1$, $a$ will be real. Setting $a = b$, we have
\begin{align}
{{\bar y}_ \pm }\left( {x;b} \right) &= {y_ \pm }\left( {x;b} \right) ,\\
{y_ \pm }\left( {x + 2{\omega _1};b} \right) &= {e^{ \pm 2\left( {b\zeta \left( {{\omega _1}} \right) - {\omega _1}\zeta \left( b \right)} \right)}}{y_ \pm }\left( {x;b} \right) .
\end{align}
\item If $e_2 < \wp \left( a \right) < e_1$, $a$ will be complex number of the form $a = \omega_1 + i b$ and
\begin{align}
{{\bar y}_ \pm }\left( {x;{\omega _1} + ib} \right) &= {y_ \mp }\left( {x;{\omega _1} + ib} \right) ,\\
{y_ \pm }\left( {x + 2{\omega _1};{\omega _1} + ib} \right) &= {e^{ \pm 2\left( {ib\zeta \left( {{\omega _1}} \right) - {\omega _1}\zeta \left( {ib} \right) - \frac{1}{2}\frac{{{\omega _1}\wp '\left( {ib} \right)}}{{\wp \left( {ib} \right) - {e_1}}}} \right)}}{y_ \pm }\left( {x;{\omega _1} + ib} \right) .
\end{align}
\item If $e_3 < \wp \left( a \right) < e_2$, $a$ will be complex number of the form $a = \omega_2 + b$ and
\begin{align}
{{\bar y}_ \pm }\left( {x;{\omega _2} + b} \right) &= {y_ \pm }\left( {x;{\omega _2} + b} \right) ,\\
{y_ \pm }\left( {x + 2{\omega _1};{\omega _2} + b} \right) &=  - {e^{ \pm 2\left( {b\zeta \left( {{\omega _1}} \right) - {\omega _1}\zeta \left( b \right) - \frac{1}{2}\frac{{{\omega _1}\wp '\left( b \right)}}{{\wp \left( b \right) - {e_3}}}} \right)}}{y_ \pm }\left( {x;{\omega _2} + b} \right) .
\end{align}
\item If $\wp \left( a \right) < e_3$, $a$ will be purely imaginary, $a = i b$, and
\begin{align}
{{\bar y}_ \pm }\left( {x;ib} \right) &= {y_ \mp }\left( {x;ib} \right) ,\\
{y_ \pm }\left( {x + 2{\omega _1};ib} \right) &= {e^{ \pm 2\left( {ib\zeta \left( {{\omega _1}} \right) - {\omega _1}\zeta \left( {ib} \right)} \right)}}{y_ \pm }\left( {x;ib} \right) .
\end{align}
\end{enumerate}
The derivation of these equations is based on the transformation properties of the Weierstrass functions $\zeta$ and $\sigma$ under translations in the complex plane by the periods $2\omega_1$ and $2\omega_2$, using equations \eqref{eq:Weierstrass_period_zeta} and \eqref{eq:Weierstrass_period_sigma}, as well as the formula \eqref{eq:Weierstrass_zeta_special} found in the appendix.

The case $\Delta < 0$ is much simpler. In fact, there are only two subcases to be considered (using the same notation as for $\Delta > 0$):
\begin{enumerate}
\item If $\wp \left( a \right) > e_2$, $a$ will real. Setting $a = b$, we have
\begin{align}
{{\bar y}_ \pm }\left( {x;b} \right) &= {y_ \pm }\left( {x;b} \right) ,\\
{y_ \pm }\left( {x + 2{\omega _1} + 2{\omega _2};b} \right) &= {e^{ \pm 2\left[ {b\left( {\zeta \left( {{\omega _1}} \right) + \zeta \left( {{\omega _2}} \right)} \right) - \zeta \left( b \right)\left( {{\omega _1} + {\omega _2}} \right)} \right]}}{y_ \pm }\left( {x;b} \right) .
\end{align}
\item If $\wp \left( a \right) < e_2$, $a$ will be purely imaginary, $a = i b$, and
\begin{align}
{{\bar y}_ \pm }\left( {x;ib} \right) &= {y_ \mp }\left( {x;ib} \right) ,\\
{y_ \pm }\left( {x + 2{\omega _1} + 2{\omega _2};ib} \right) &= {e^{ \pm 2\left[ {ib\left( {\zeta \left( {{\omega _1}} \right) + \zeta \left( {{\omega _2}} \right)} \right) - \zeta \left( ib \right)\left( {{\omega _1} + {\omega _2}} \right)} \right]}}{y_ \pm }\left( {x;ib} \right) .
\end{align}
\end{enumerate}

At this point we make the following comments which are applicable to either $\Delta > 0$ or $\Delta < 0$:
\begin{itemize}
\item In the subcases 1 and 3, the eigenstates $y_\pm \left( x \right)$ are real and if they are shifted by a period $2\omega_1$ they will get multiplied by a real number. In those cases, the eigenfunctions diverge exponentially as $x \to \pm \infty$.
\item In the subcases 2 and 4, the eigenstates $y_\pm \left( x \right)$ are complex conjugate to each other and if they are shifted by a period $2\omega_1$ they will acquire a complex phase. These states are the familiar Bloch waves of periodic potentials.
\end{itemize}
Thus, the band structure of the $n = 1$ \Lame potential contains a finite ``valence'' band between the energies $-e_1$ and $-e_2$ an infinite ``conduction'' band above $-e_3$ in the case $\Delta > 0$. On the other hand, there is only one infinite ``conduction'' band at energies higher than $-e_2$ in the case $\Delta < 0$.

The addition theorem for the Weierstrass $\sigma$ function \eqref{eq:Weierstrass_addition_sigma} implies two useful properties of the eigenstates $y_\pm$,
\begin{align}
{y_ + }{y_ - } &= \wp \left( x \right) - \wp \left( a \right) , \label{eq:lame_eigen_property_1}\\
{y_ + }'{y_ - } - {y_ + }{y_ - }' &= - \wp '\left( a \right) . \label{eq:lame_eigen_property_2}
\end{align}

\begin{figure}[ht]
\vspace{10pt}
\begin{center}
\begin{picture}(100,40)
\put(4.5,10){\includegraphics[width = 0.43\textwidth]{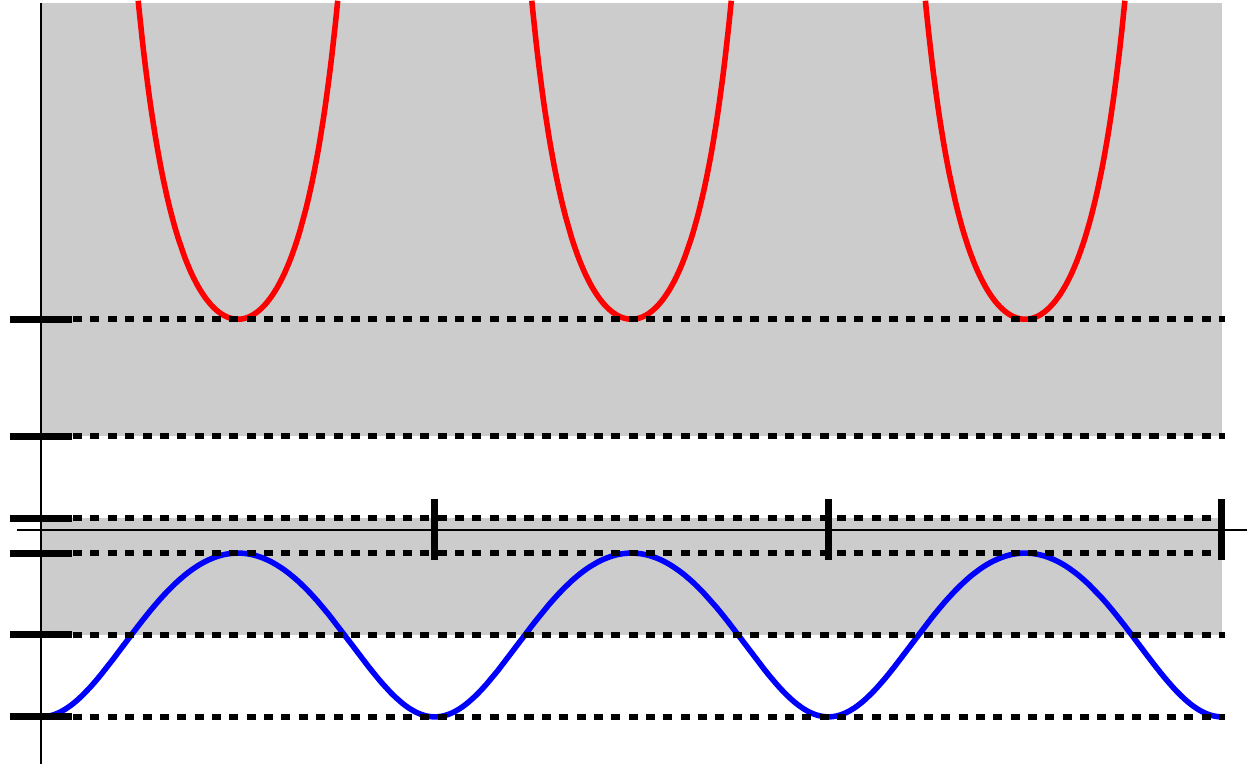}}
\put(52.5,10){\includegraphics[width = 0.43\textwidth]{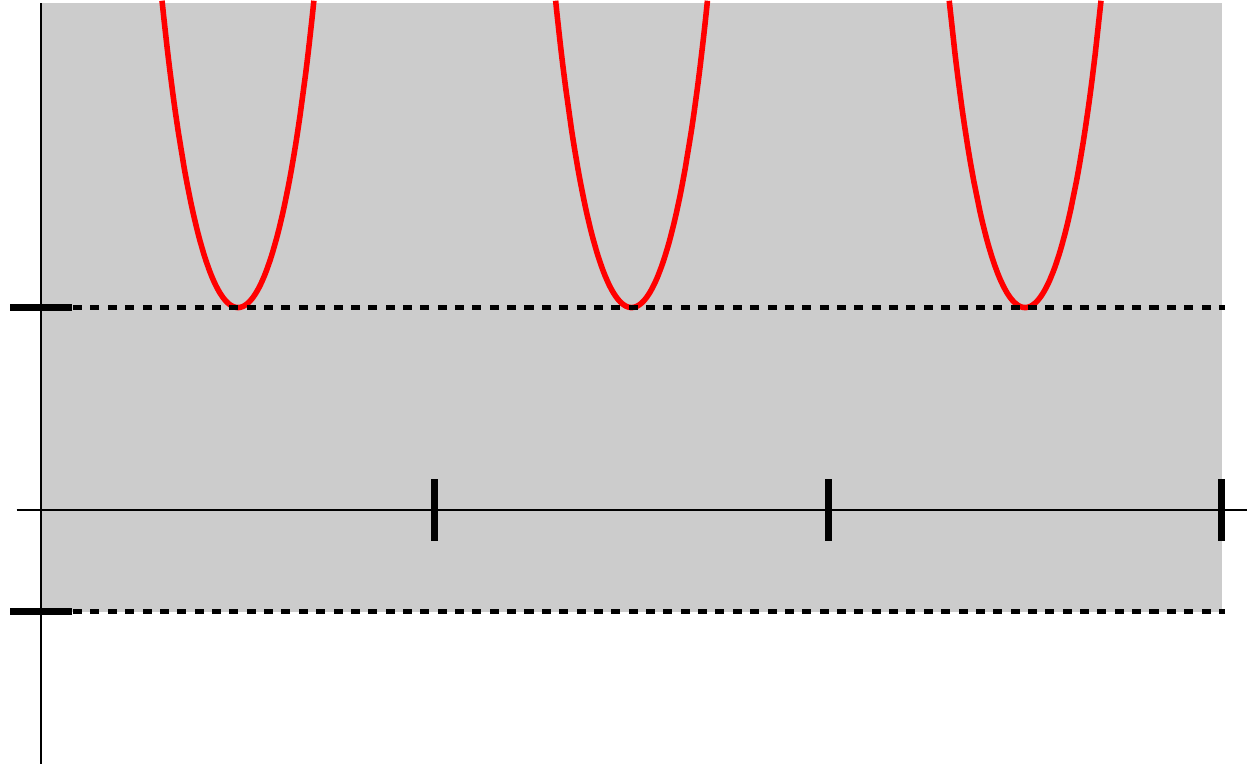}}
\put(41.5,0){\includegraphics[height = 0.1\textwidth]{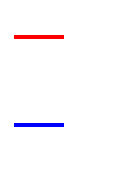}}
\put(1,25){$2e_1$}
\put(1,16.5){$2e_2$}
\put(1,11.25){$2e_3$}
\put(0.25,14.25){$-e_1$}
\put(0.25,18.5){$-e_2$}
\put(0.25,21){$-e_3$}
\put(17.5,19.5){$2\omega_1$}
\put(31,19.5){$4\omega_1$}
\put(44.5,19.5){$6\omega_1$}
\put(49,25){$2e_2$}
\put(48.35,14.75){$-e_2$}
\put(65.5,20.25){$2\omega_1$}
\put(79,20.25){$4\omega_1$}
\put(92.5,20.25){$6\omega_1$}
\put(3,37.5){$V \left( x \right)$}
\put(51,37.5){$V \left( x \right)$}
\put(48,17.5){$x$}
\put(96,18){$x$}
\put(23,7.5){$\Delta > 0$}
\put(71,7.5){$\Delta < 0$}
\put(45.5,7.25){$2\wp \left( x \right)$}
\put(45.5,2.25){$2\wp \left( x + \omega_2 \right)$}
\put(41.5,1){\line(0,1){9}}
\put(41.5,1){\line(1,0){17.5}}
\put(59,1){\line(0,1){9}}
\put(41.5,10){\line(1,0){17.5}}
\end{picture}
\end{center}
\vspace{-10pt}
\caption{The band structure of the \Lame potential $2\wp$}
\vspace{5pt}
\label{fig:bands}
\end{figure}
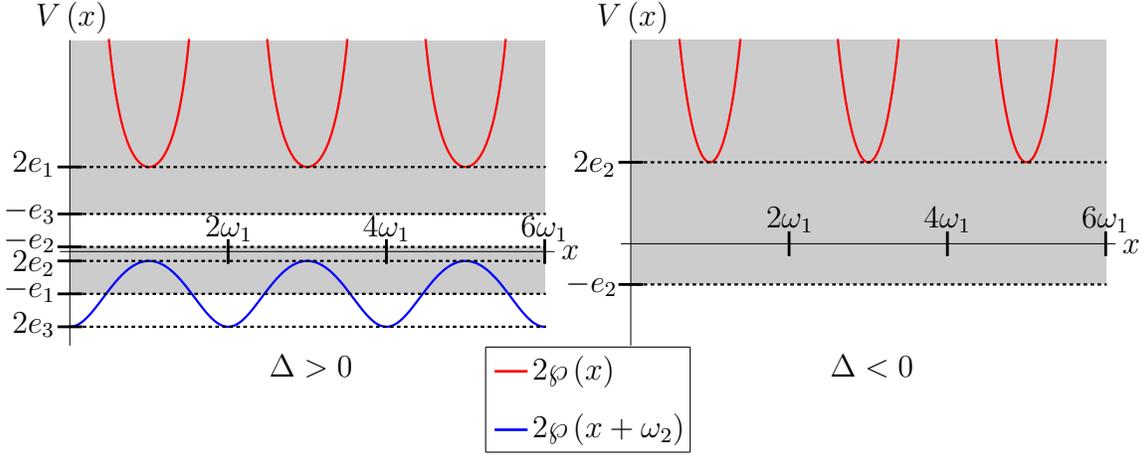

The whole process of finding the eigenstates and the band structure can be repeated for the potential $V = 2 \wp \left( x + \omega_2 \right)$. The results are the same apart from making a shift by $\omega_2$ in the definition of the eigenfunctions and an appropriate choice of the normalization constant in order to absorb the complex phases,
\begin{equation}
{y_ \pm }\left( {x;a} \right) = \frac{{\sigma \left( {x + {\omega _2} \pm a} \right)\sigma \left( {{\omega _2}} \right)}}{{\sigma \left( {x + {\omega _2}} \right)\sigma \left( {{\omega _2} \pm a} \right)}}{e^{ - \zeta \left( { \pm a} \right)x}} .
\label{eq:lame_eigenstates_shifted}
\end{equation}
As a result, the potentials $V = 2 \wp \left( x \right)$ and $V = 2 \wp \left( x + \omega_2 \right)$ have the same band structure. The two potentials are quite dissimilar functions, the first one having poles and the other being smooth and bounded function, as shown in figure \ref{fig:bands}.

There is a small but important detail that needs to be taken into account for $V = 2 \wp \left( x + \omega_2 \right)$. The analog property \eqref{eq:lame_eigen_property_1} for this potential is
\begin{equation}
{y_ + }{y_ - } = \frac{{\wp \left( {x + {\omega _2}} \right) - \wp \left( a \right)}}{{{e_3} - \wp \left( a \right)}} 
\label{eq:lame_eigen_property_1_corrected}
\end{equation}
and it will turn out to be very useful in the next section.
The absolute value of the denominator can be absorbed in the definition of $y_\pm$, but not its sign, as it will alter the reality properties of $y_\pm$. As a consequence, there is a sign difference for ${y_ + }{y_ - }$ appearing in equations \eqref{eq:lame_eigen_property_1} and \eqref{eq:lame_eigen_property_1_corrected} with the exception of states that belong to the infinite conduction band of the corresponding potential.

The $n = 1$ \Lame potential has an interesting limit when the two larger roots $e_1$ and $e_2$ coincide, in which case $k = 1$. The potential is expressed in terms of hyperbolic functions and it is not periodic, since $2\omega_1$ becomes infinite. This limit gives rises to the \Poschl -Teller potential having one bound state, which is the degeneration of the valence band, plus a continuous spectrum extending to infinity. This limit corresponds to the kink solution of the sinh-Gordon equation.

\section{Construction of Classical String Solutions}
\label{sec:construction}
In this section, we use the solutions of the effective \Schrodinger problems \eqref{eq:Schr_tau_1} and \eqref{eq:Schr_sigma_1} together with the geometric and Virasoro constraints \eqref{eq:susy_constraint} and \eqref{eq:susy_Virasoro} in order to construct classical string solutions in AdS$_3$ space. Similar constructions are applicable and will be discussed for the case of dS$_3$ space. For this purpose, it is convenient to write the Virasoro constraints in the form
\begin{align}
\left( {\frac{\partial^2 Y}{\partial\xi_0^2} + \frac{\partial^2 Y}{\partial\xi_1^2}} \right) \cdot {Y} &= 0 ,\label{eq:susy_Virasoro1}\\
\frac{\partial^2 Y}{\partial\xi_0 \partial\xi_1} \cdot {Y} &= 0 ,\label{eq:susy_Virasoro2}
\end{align}
making use of the geometric constraint \eqref{eq:susy_constraint}.

We proceed starting with the \Schrodinger problem for the flat potential \eqref{eq:Schr_tau_1}. The eigenfunctions are $T^\mu \left( \xi_0 \right)$ and the corresponding eigenvalues are $\kappa^\mu$. Actually, it turns out that if all eigenvalues $\kappa^\mu$ are equal there will be no string solution that is compatible with the constraints. The simplest solution to obtain is provided by two distinct eigenvalues, as will be seen shortly. The form of the target space metrics suggests that AdS$_3$ favours the selection of eigenvalues of the same sign, which can be either positive or negative, whereas dS$_3$ favours the selection of eigenvalues of opposite sign.

For notational convenience we drop the index $\mu$ but this does not mean that all components are the same. For positive eigenvalues $\kappa = \ell^2$, the solution of the flat \Schrodinger problem is
\begin{equation}
{\rm T}\left( \xi_0  \right) = {c_1}\cos \left( \ell\xi_0 \right) + {c_2}\sin \left( \ell\xi_0 \right) ,
\label{eq:construction_flat_solutions_positive}
\end{equation}
while for negative eigenvalues $\kappa = - \ell^2$, the corresponding solution is
\begin{equation}
{\rm T}\left( \xi_0  \right) = {c_1}\cosh \left( \ell\xi_0 \right) + {c_2}\sinh \left( \ell\xi_0 \right) .
\label{eq:construction_flat_solutions_negative}
\end{equation}
Any of these solutions should be combined with the eigenfunctions $\Sigma \left( \xi_1 \right)$ of the effective \Schrodinger problem \eqref{eq:Schr_sigma_1} which is of \Lame type. Thus, according to the discussion in the previous section, the eigenvalues $\kappa$ should be $\kappa = - \wp \left( a \right) - 2 x_1 $.

\subsection{AdS$_3$ and Positive Eigenvalues}
\label{subsec:construction_static_positive}
Let us consider string solutions with embedding coordinates associated with two distinct positive eigenvalues $\kappa = \ell_{1,2}^2$ given by the ansatz
\begin{equation}
Y = \left( {\begin{array}{*{20}{c}}
{c_1^ + \Sigma _1^ + \left( {{\xi _1}} \right)\cos \left( {\ell _1}{\xi _0} \right) + c_1^ - \Sigma _1^ - \left( {{\xi _1}} \right)\sin \left( {\ell _1}{\xi _0}\right)}\\
{c_1^ + \Sigma _1^ + \left( {{\xi _1}} \right)\sin \left( {\ell _1}{\xi _0} \right) - c_1^ - \Sigma _1^ - \left( {{\xi _1}} \right)\cos \left( {\ell _1}{\xi _0}\right)}\\
{c_2^ + \Sigma _2^ + \left( {{\xi _1}} \right)\cos \left( {\ell _2}{\xi _0} \right) + c_2^ - \Sigma _2^ - \left( {{\xi _1}} \right)\sin \left( {\ell _2}{\xi _0}\right)}\\
{c_2^ + \Sigma _2^ + \left( {{\xi _1}} \right)\sin \left( {\ell _2}{\xi _0} \right) - c_2^ - \Sigma _2^ - \left( {{\xi _1}} \right)\cos \left( {\ell _2}{\xi _0}\right)}
\end{array}} \right) .
\label{eq:construction_refl_sinh_ansatz}
\end{equation}
The functions $\Sigma _{1,2}^ \pm \left( {{\xi _1}} \right)$ are in general linear combinations of the functions $y_\pm \left( \xi_1 \right)$ given by \eqref{eq:lame_eigenstates} with moduli equal to $a_{1,2}$, respectively. The connection between the eigenvalues of the $\xi_0$ problem and the $\xi_1$ problem implies that
\begin{equation}
\ell _{1,2}^2 = - \wp \left( {{a_{1,2}}} \right) -2 x_1 .
\label{eq:construction_refl_sinh_eigenvalues}
\end{equation}

We will demonstrate that this ansatz is compatible with the geometric and Virasoro constraints for strings in AdS$_3$ space. Indeed, the constraint \eqref{eq:susy_constraint} implies
\begin{equation}
{\left( {c_1^ + \Sigma _1^ + } \right)^2} + {\left( {c_1^ - \Sigma _1^ - } \right)^2} - {\left( {c_2^ + \Sigma _2^ + } \right)^2} - {\left( {c_2^ - \Sigma _2^ - } \right)^2} = {\Lambda ^2} .
\label{eq:construction_refl_sinh_constraint_step1}
\end{equation}
The Virasoro constraint \eqref{eq:susy_Virasoro2} implies
\begin{equation}
{\ell _1}c_1^ + c_1^ - \left( {{\Sigma _1^ +} '{\Sigma _1^ -}  - {\Sigma _1^ -} '{\Sigma _1^ +} } \right) = {\ell _2}c_2^ + c_2^ - \left( {{\Sigma _2^ +} '{\Sigma _2^ -}  - {\Sigma _2^ -} '{\Sigma _2^ +} } \right) ,
\label{eq:construction_refl_sinh_Virasoro2_step1}
\end{equation}
and the Virasoro constraint \eqref{eq:susy_Virasoro1} implies
\begin{multline}
{\left( {c_1^ + } \right)^2}{\Sigma _1^ +} ''{\Sigma _1^ +}  + {\left( {c_1^ - } \right)^2}{\Sigma _1^ -} ''{\Sigma _1^ -}  - {\left( {c_2^ + } \right)^2}{\Sigma _2^ +} ''{\Sigma _2^ +}  - {\left( {c_2^ - } \right)^2}{\Sigma _2^ -} ''{\Sigma _2^ -}  = \\
\ell _1^2{\left( {c_1^ + \Sigma _1^ + } \right)^2} + \ell _1^2{\left( {c_1^ - \Sigma _1^ - } \right)^2} - \ell _2^2{\left( {c_2^ + \Sigma _2^ + } \right)^2} - \ell _2^2{\left( {c_2^ - \Sigma _2^ - } \right)^2} .
\label{eq:construction_refl_sinh_Virasoro1_step0}
\end{multline}
Making use of the \Lame equation to eliminate the second derivatives of $\Sigma _{1,2}^ \pm$, the Virasoro constraint \eqref{eq:construction_refl_sinh_Virasoro1_step0} can be rewritten as follows,
\begin{equation}
\left[ {{{\left( {c_1^ + \Sigma _1^ + } \right)}^2} + {{\left( {c_1^ - \Sigma _1^ - } \right)}^2}} \right]\left( {\wp \left( {{\xi _1}} \right) - x_1 - \ell _1^2} \right) = \left[ {{{\left( {c_2^ + \Sigma _2^ + } \right)}^2} + {{\left( {c_2^ - \Sigma _2^ - } \right)}^2}} \right]\left( {\wp \left( {{\xi _1}} \right) - x_1 - \ell _2^2} \right) .
\end{equation}
Using the geometric constraint \eqref{eq:construction_refl_sinh_constraint_step1}, this can be further simplified to
\begin{equation}
\left[ {{{\left( {c_1^ + \Sigma _1^ + } \right)}^2} + {{\left( {c_1^ - \Sigma _1^ - } \right)}^2}} \right]\ell _1^2 - \left[ {{{\left( {c_2^ + \Sigma _2^ + } \right)}^2} + {{\left( {c_2^ - \Sigma _2^ - } \right)}^2}} \right]\ell _2^2 = {\Lambda ^2} \left( {\wp \left( {{\xi _1}} \right) - x_1} \right) .
\label{eq:construction_refl_sinh_Virasoro1_step1}
\end{equation}

Combining the form of the geometric constraint \eqref{eq:construction_refl_sinh_constraint_step1} with the property \eqref{eq:lame_eigen_property_1} of the eigenfunctions of the \Lame potential, suggests the following choice,
\begin{align}
c_1^ +  = c_1^ -  \equiv {c_1},&\quad c_2^ +  = c_2^ -  \equiv {c_2} ,\\
\Sigma _{1,2}^ +  = \frac{1}{2}\left( {y_{1,2}^ +  + y_{1,2}^ - } \right),&\quad \Sigma _{1,2}^ -  = \frac{1}{{2i}}\left( {y_{1,2}^ +  - y_{1,2}^ - } \right) . \label{eq:construction_refl_sinh_linear_combinations}
\end{align}
Since the embedding functions $Y^\mu$ have to be real, this selection can be performed only if $y^ \pm$ are complex conjugate to each other. Thus, $y_{1,2}^ \pm$ should be Bloch waves of the \Lame potential, in which case the moduli $a_{1,2}$ should obey the relations
\begin{equation}
\wp \left( a_{1,2} \right) < e_3 \quad \textrm{or}\quad e_2 < \wp \left( a_{1,2} \right) < e_1 .
\label{eq:construction_refl_sinh_a_bounds}
\end{equation}

With this choice, the various combinations appearing in the constraints simplify as
\begin{align}
&{\left( {c_{1,2}^ + \Sigma _{1,2}^ + } \right)^2} + {\left( {c_{1,2}^ - \Sigma _{1,2}^ - } \right)^2} = c_{1,2}^2y_{1,2}^ + y_{1,2}^ -  = c_{1,2}^2 \left( {\wp \left( {{\xi _1}} \right) - \wp \left( {{a_{1,2}}} \right)} \right) ,\\
&2i \left({ {\Sigma _{1,2}^ +} '{\Sigma _{1,2}^ -}  - {\Sigma _{1,2}^ -} '{\Sigma _{1,2}^ +} }\right) = {y_{1,2}^ -} '{y_{1,2}^ +}  - {y_{1,2}^ +} '{y_{1,2}^ -} = \wp '\left( {{a_{1,2}}} \right) .
\end{align}
Substituting these relations in the constraints \eqref{eq:construction_refl_sinh_constraint_step1}, \eqref{eq:construction_refl_sinh_Virasoro2_step1} and \eqref{eq:construction_refl_sinh_Virasoro1_step1}, we arrive at the very simple relations
\begin{align}
& c_1^2 \left( {\wp \left( {{\xi _1}} \right) - \wp \left( {{a_1}} \right)} \right) - c_2^2 \left( {\wp \left( {{\xi _1}} \right) - \wp \left( {{a_2}} \right)} \right) = {\Lambda ^2} , \label{eq:construction_refl_sinh_constraint_step2}\\
&{\ell _1}c_1^2 \wp '\left( {{a_1}} \right) = {\ell _2}c_2^2 \wp '\left( {{a_1}} \right) , \label{eq:construction_refl_sinh_Virasoro2_step2}\\
& c_1^2\ell _1^2 \left( {\wp \left( {{\xi _1}} \right) - \wp \left( {{a_1}} \right)} \right) - c_2^2\ell _2^2 \left( {\wp \left( {{\xi _1}} \right) - \wp \left( {{a_2}} \right)} \right) = {\Lambda ^2} \left( {\wp \left( {{\xi _1}} \right) - x_1} \right) . \label{eq:construction_refl_sinh_Virasoro1_step2}
\end{align}

Equation \eqref{eq:construction_refl_sinh_constraint_step2} specifies the coefficients $c_{1,2}$ in terms of $a_{1,2}$,
\begin{align}
&c_1^2 = c_2^2 \equiv {c^2} ,\label{eq:construction_refl_sinh_constraint_step3a}\\
&{c^2}\left( {\wp \left( {{a_2}} \right) - \wp \left( {{a_1}} \right)} \right) = {\Lambda ^2} , \label{eq:construction_refl_sinh_constraint_step3b}
\end{align}
since, otherwise, the left hand side should not be independent of the variable $\xi_1$. Notice that \eqref{eq:construction_refl_sinh_constraint_step3b} implies that $ \wp \left( {{a_1}} \right) < \wp \left( {{a_2}} \right)$. It also hints why it is not possible to build a string solution using only one eigenvalue instead of two. Also, making use of equations \eqref{eq:construction_refl_sinh_constraint_step3a} and \eqref{eq:construction_refl_sinh_constraint_step3b} as well as the fact that the eigenvalues $\ell_{1,2}^2$ are determined by $a_{1,2}$ via equations \eqref{eq:construction_refl_sinh_eigenvalues}, the Virasoro constraint \eqref{eq:construction_refl_sinh_Virasoro1_step2} becomes
\begin{equation}
\wp \left( {{a_1}} \right) + \wp \left( {{a_2}} \right) =  - x_1 = {x_2} + {x_3} .
\label{eq:construction_refl_sinh_a_solution}
\end{equation}

The only relation that remains to be verified is the Virasoro constraint \eqref{eq:construction_refl_sinh_Virasoro2_step2}. Making use of equation \eqref{eq:construction_refl_sinh_constraint_step3a}, it takes the form
\begin{equation}
{\ell _1}\wp '\left( {{a_1}} \right) = {\ell _2}\wp '\left( {{a_2}} \right) .
\end{equation}
This is satisfied for all appropriate choices of $a_{1,2}$, since the values of the Weierstrass function $\wp \left( a_{1,2} \right)$ obey the relation
\begin{equation}
\ell _{1,2}^2\wp {'^2}\left( {{a_{1,2}}} \right) =  - 4\left( {\wp \left( {{a_{1,2}}} \right) + 2{x_1}} \right)\left( {\wp \left( {{a_{1,2}}} \right) - {x_1}} \right)\left( {\wp \left( {{a_{1,2}}} \right) - {x_2}} \right)\left( {\wp \left( {{a_{1,2}}} \right) - {x_3}} \right)
\end{equation}
and \eqref{eq:construction_refl_sinh_a_solution} can be written as
\begin{equation}
\left( {\wp \left( {{a_{1,2}}} \right) - {x_2}} \right) + \left( {\wp \left( {{a_{2,1}}} \right) - {x_3}} \right) = 0, \quad \left( {\wp \left( {{a_{1,2}}} \right) + 2{x_1}} \right) + \left( {\wp \left( {{a_{2,1}}} \right) - {x_1}} \right) = 0.
\end{equation}

Summarizing, the ansatz \eqref{eq:construction_refl_sinh_ansatz} provides a classical string solution in AdS$_3$ as long as $\wp \left( {{a_1}} \right) + \wp \left( {{a_2}} \right) =  - x_1 = {x_2} + {x_3} $, where $\wp \left( {{a_1}} \right) $ and $ \wp \left( {{a_2}} \right)$ correspond to Bloch waves of the \Lame potential and the eigenvalues $\kappa_{1,2} = \ell_{1,2}^2 = - \wp \left( {{a_{1,2}}} \right) - 2 x_1$ are both positive. We have also shown the inequality $ \wp \left( {{a_1}} \right) < \wp \left( {{a_2}} \right)$. The classical string solutions that we constructed above correspond to the unbounded configurations of the reduced Pohlmeyer equation. There is yet another, equally important sector of solutions corresponding to the bounded configurations, which can be constructed in a similar manner substituting the \Lame potential $2\wp \left( \xi_1 \right) -2 x_1$ with the potential $2\wp \left( \xi_1 + \omega_2 \right) -2 x_1$. There is no need to repeat all the steps of the construction for this class of string configurations, as it turns out that the requirements for $ \wp \left( a_{1,2} \right)$ are identical to those before, except for the inequality $ \wp \left( {{a_1}} \right) < \wp \left( {{a_2}} \right)$ that gets inverted to $ \wp \left( {{a_1}} \right) > \wp \left( {{a_2}} \right)$ when $a_{1,2}$ correspond to the valence band of the \Lame potential. This is due to the fact that the product $y_+ y_-$ of eigenstates of the shifted potential acquires an extra minus sign for all eigenstates, except for those in the infinite conduction band, as shown at the end of section \ref{subsec:lame}.

There are four classes of possible string solutions depending on the ordering of the three roots $x_1$, $x_2$ and $x_3$.

Case 1: Assume that there is only one real root. Then $x_1 = e_2$, and the allowed band of the corresponding \Lame potential is $\wp \left( a \right) < e_2$. The constraints of the problem cannot be simultaneously satisfied. When $e_2 < 0$, $\wp \left( a_{1,2} \right)$ are equidistant from $- e_2 / 2$ and they cannot both lie in the allowed band of the \Lame potential. If $e_2 > 0$, $\wp \left( a_{1,2} \right)$ are still equidistant from $- e_2 / 2$, but this time can both lie in the allowed band of the potential, as long as they are larger than $- 2 e_2$: in the latter case both eigenvalues are negative. String solutions do not exist in this case.

Case 2: Assume that there are three real roots and $x_1$ is the largest, and, thus, positive, in which case $x_{1,2,3} = e_{1,2,3}$. $\wp \left( a_{1,2} \right)$ are equidistant from $- e_1 / 2$, which is larger than $ - 2 e_1$, and, therefore, the corresponding eigenvalues cannot be both positive. In this case too, string solutions do not exist.

Case 3: Assume that there are three real roots and $x_1$ is the intermediate root, in which case $x_{1,2,3} = e_{2,1,3}$. When $e_2 > 0 $, $\wp \left( a_{1,2} \right)$ cannot sum up to $- e_2 / 2$ and be both smaller than $- 2 e_2$. When $e_2 < 0$, there is an allowed region for $\wp \left( a_{1,2} \right)$, as shown in figure \ref{fig:const_pos_cosh}. 
\afterpage{
\begin{figure}[p]
\centering
\begin{picture}(100,60)
\put(10,5){\includegraphics[width = 0.5\textwidth]{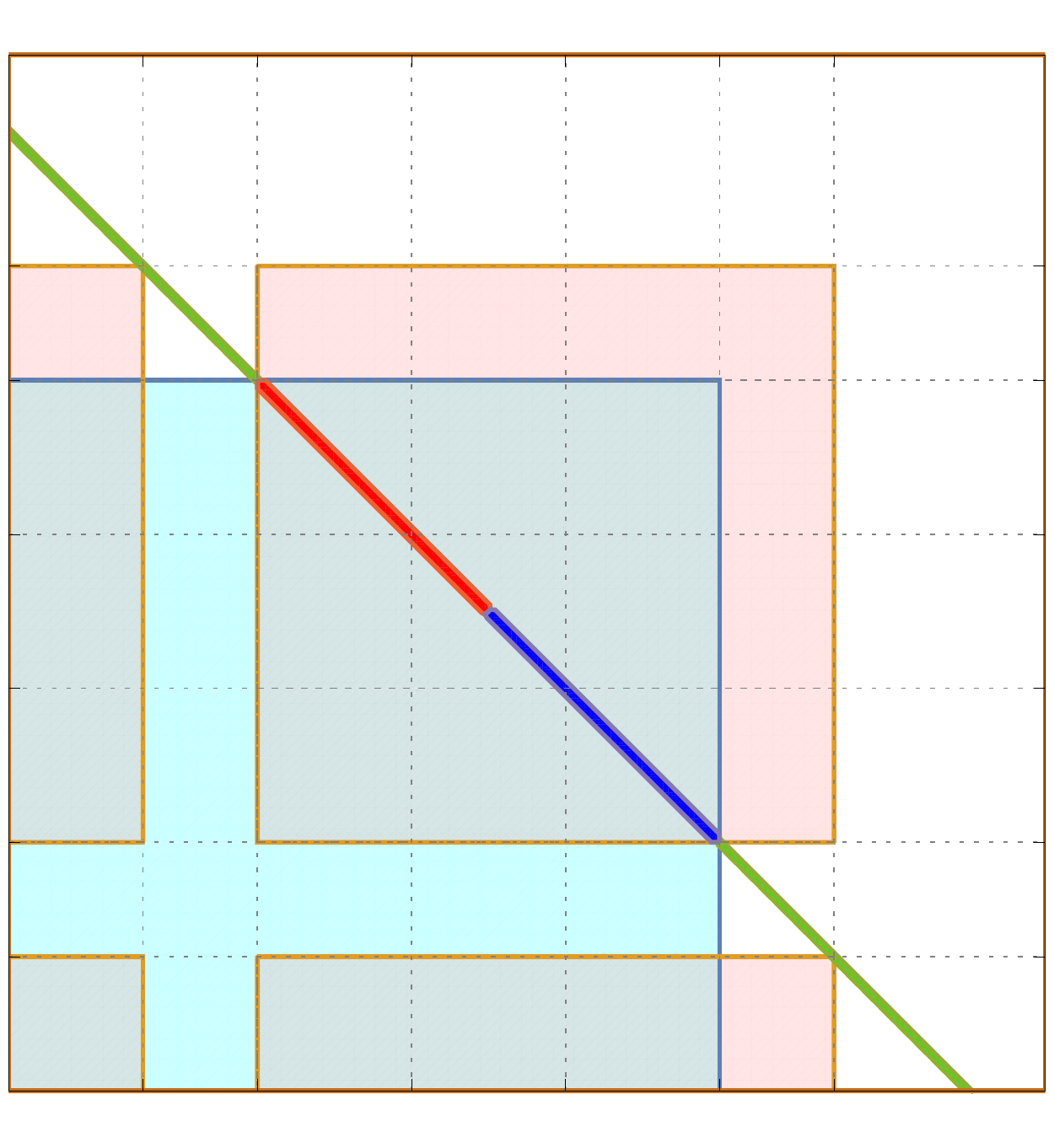}}
\put(62,16){\includegraphics[width = 0.1\textwidth]{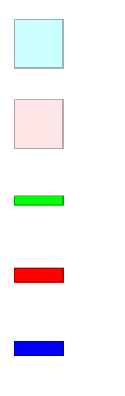}}
\put(7.75,46.5){$e_1$}
\put(4.25,41){$-2e_2$}
\put(5.5,33.5){$-e_2$}
\put(8.5,26){$0$}
\put(7.75,19.25){$e_2$}
\put(7.75,13.5){$e_3$}
\put(15.75,5.25){$e_3$}
\put(21.25,5.25){$e_2$}
\put(29,5.25){$0$}
\put(33.75,5.25){$-e_2$}
\put(40.75,5.25){$-2e_2$}
\put(48.75,5.25){$e_1$}
\put(32,2){$\wp \left( a_1 \right)$}
\put(2,28){\rotatebox{90}{$\wp \left( a_2 \right)$}}
\put(68,43.5){positive eigenvalues}
\put(68,36.75){Bloch states}
\put(68,31.25){$\wp \left( a_1 \right)+\wp \left( a_2 \right) = -e_2$}
\put(68,24.75){$V = 2 \wp \left( x \right)$ solutions}
\put(68,18.75){$V = 2 \wp \left( x +\omega_2 \right)$ solutions}
\end{picture}
\vspace{-20pt}
\caption{The allowed $\wp\left( a_{1,2} \right)$ for classical string solutions when $x_1 = e_2$, $E<0$}
\label{fig:const_pos_cosh}
\vspace{10pt}
\begin{picture}(100,60)
\put(10,5){\includegraphics[width = 0.5\textwidth]{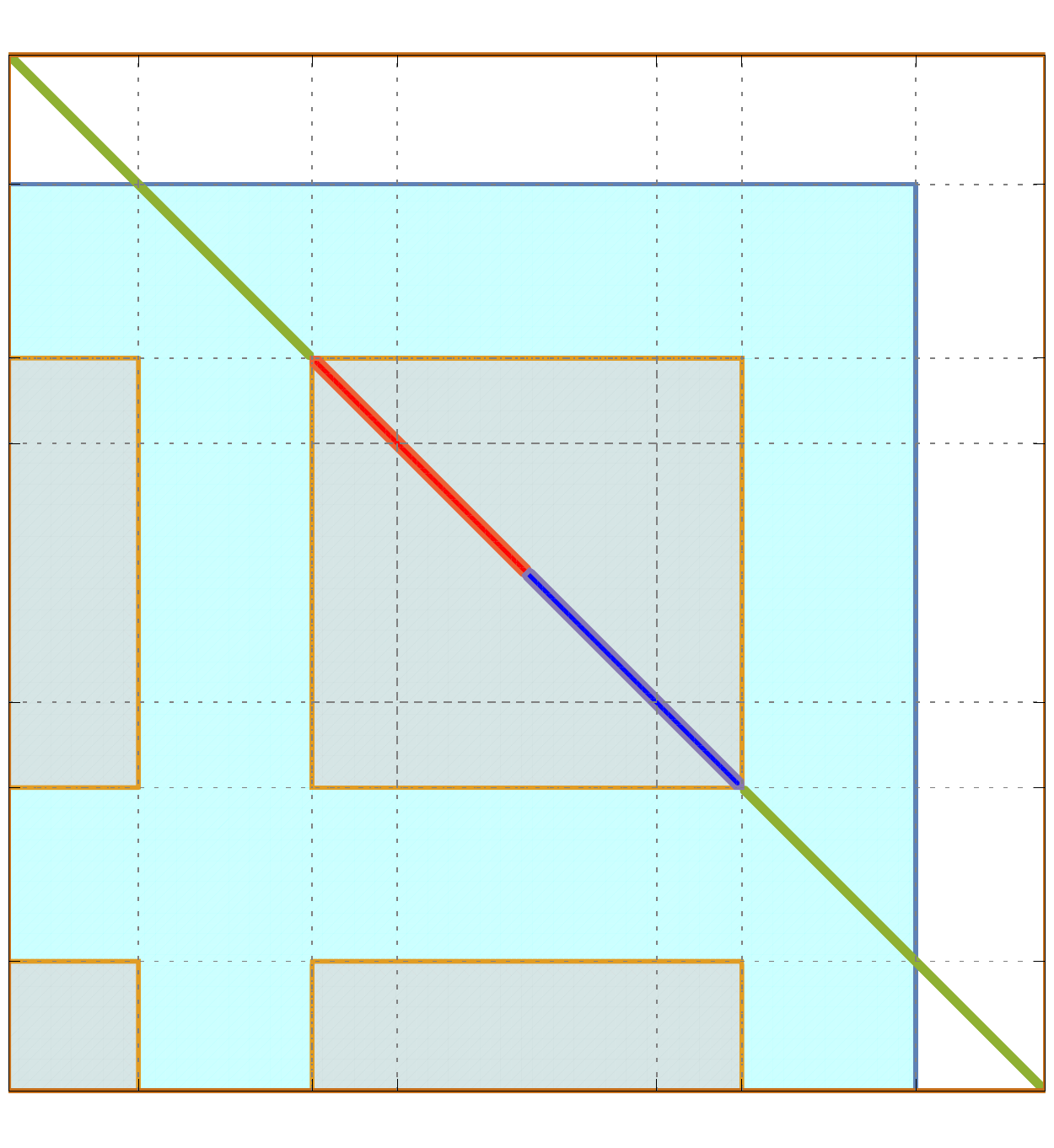}}
\put(62,16){\includegraphics[width = 0.1\textwidth]{regions_legend.pdf}}
\put(4.25,50){$-2e_3$}
\put(7.75,42){$e_1$}
\put(5.5,37.75){$-e_3$}
\put(8.5,25.25){$0$}
\put(7.75,21.5){$e_2$}
\put(7.75,13.5){$e_3$}
\put(15.5,5.25){$e_3$}
\put(23.5,5.25){$e_2$}
\put(28.25,5.25){$0$}
\put(38,5.25){$-e_3$}
\put(44.25,5.25){$e_1$}
\put(49.75,5.25){$-2e_3$}
\put(32,2){$\wp \left( a_1 \right)$}
\put(2,28){\rotatebox{90}{$\wp \left( a_2 \right)$}}
\put(68,43.5){positive eigenvalues}
\put(68,36.75){Bloch states}
\put(68,31.25){$\wp \left( a_1 \right)+\wp \left( a_2 \right) = -e_3$}
\put(68,24.75){$V = 2 \wp \left( x \right)$ solutions}
\put(68,18.75){$V = 2 \wp \left( x +\omega_2 \right)$ solutions}
\end{picture}
\vspace{-20pt}
\caption{The allowed $\wp\left( a_{1,2} \right)$ for classical string solutions when $x_1 = e_3$}
\label{fig:const_pos_sinh}
\end{figure}
\clearpage
}
In reading the figures, here and later, the red line denotes the range of parameters for the existence of classical string solutions corresponding to unbounded configurations and the blue line for the bounded case. Going back to table \ref{tb:phi_range} that summarizes all elliptic solutions of sinh- and cosh-Gordon equations, we observe that the region $\wp \left( a_1 \right) < \wp \left( a_2 \right)$ corresponds to static solutions of the cosh-Gordon equation, while the region $\wp \left( a_1 \right) > \wp \left( a_2 \right)$ corresponds to translationally invariant solutions of the cosh-Gordon.

Case 4: Assume that there are three real roots and $x_1$ is the smallest root, in which case $x_{1,2,3} = e_{3,1,2}$. $\wp \left( a_{1,2} \right)$ are equidistant from the center $(e_1 + e_2) / 2$ of the valence band of the corresponding \Lame problem. Thus, we may select any $\wp \left( a_1 \right)$ and $\wp \left( a_2 \right)$ in the valence band lying symmetrically from its center, as shown in figure \ref{fig:const_pos_sinh}. According to table \ref{tb:phi_range}, we conclude that these solutions correspond to static reflecting solutions of the sinh-Gordon equation: they come from the right when $\wp \left( a_1 \right) < \wp \left( a_2 \right)$ and from the left when $\wp \left( a_1 \right) > \wp \left( a_2 \right)$.

To visualize the form of the solutions, we convert to global coordinates
\begin{equation}
Y = \Lambda \left( {\begin{array}{*{20}{c}}
{\sqrt{{1 + {r^2}}} \cos t}\\
{\sqrt{{1 + {r^2}}} \sin t}\\
{r \cos \varphi }\\
{r \sin \varphi }
\end{array}} \right) ,
\label{eq:global_coordinates}
\end{equation}
in which the AdS$_3$ metric takes the usual form
\begin{equation}
d{s^2} =  - \left( {1 + {r^2}} \right)d{t^2} + \frac{1}{{1 + {r^2}}}d{r^2} + {r^2}d{\varphi ^2} .
\end{equation}
The string solution associated to the unbounded configurations takes the form
\begin{equation}
\begin{split}
r &= \sqrt {\frac{{\wp \left( \xi _1 \right) - \wp \left( {{a_2}} \right)}}{{\wp \left( {{a_2}} \right) - \wp \left( {{a_1}} \right)}}} , \\
t &= {\ell _1}{\xi _0} - \arg \frac{{\sigma \left( {{\xi _1} + {a_1}} \right)}}{{\sigma \left( {{\xi _1}} \right)\sigma \left( {{a_1}} \right)}}{e^{ - \zeta \left( {{a_1}} \right){\xi _1}}} ,\\
\varphi  &= {\ell _2}{\xi _0} - \arg \frac{{\sigma \left( {{\xi _1} + {a_2}} \right)}}{{\sigma \left( {{\xi _1}} \right)\sigma \left( {{a_2}} \right)}}{e^{ - \zeta \left( {{a_2}} \right){\xi _1}}} .
\end{split}
\end{equation}
Likewise, for the bounded configurations, the corresponding string solution is
\begin{equation}
\begin{split}
r &= \sqrt {\frac{{\wp \left( {{a_2}} \right) - \wp \left( {\xi _1 + {\omega _2}} \right)}}{{\wp \left( {{a_1}} \right) - \wp \left( {{a_2}} \right)}}} , \\
t &= {\ell _1}{\xi _0} - \arg \frac{{\sigma \left( {{\xi _1} + {\omega _2} + {a_1}} \right) \sigma \left( {{\omega _2}} \right)}}{{\sigma \left( {{\xi _1} + {\omega _2}} \right)\sigma \left( {{a_1 + {\omega _2}}} \right)}}{e^{ - \zeta \left( {{a_1}} \right){\xi _1}}} ,\\
\varphi &= {\ell _2}{\xi _0} - \arg \frac{{\sigma \left( {{\xi _1} + {\omega _2} + {a_2}} \right) \sigma \left( {\omega _2} \right)}}{{\sigma \left( {{\xi _1} + {\omega _2}} \right)\sigma \left( {{a_2}} + {\omega _2} \right)}}{e^{ - \zeta \left( {{a_2}} \right){\xi _1} }} .
\end{split}
\end{equation}

In both cases, the solution corresponds to a rigidly rotating spiky string with constant angular velocity $\omega = \ell_2 / \ell_1$, which
\begin{equation}
\begin{split}
&\omega < 1, \quad \mathrm{when} \quad \wp \left( a_1 \right) < \wp \left( a_2 \right) , \\
&\omega > 1, \quad \mathrm{when} \quad \wp \left( a_1 \right) > \wp \left( a_2 \right) .
\end{split}
\end{equation}
In the spirit of section \ref{sec:Spiky_Strings}, $\omega$ is smaller than one for the unbounded solution and larger than one for the bounded one, since the radial coordinate $r$ is also unbounded or bounded, respectively depending on the form of the solution. Only for $\omega > 1$ there are spikes, i.e., they are special points of the string that move with the speed of light.

The periodic sinh-Gordon configurations exhibit an interesting limit as
\begin{equation}
\wp \left( a_{1,2} \right) \to e_{1,2} \quad \textrm{or} \quad \wp \left( a_{1,2} \right) \to e_{2,1} .
\end{equation}
In this limit, the functions $y_\pm \left( \xi_1 ; a_{1,2} \right)$ both tend to $\sqrt{\wp \left( \xi_1  \right) - e_{1,2}}$. The construction of the string solutions is still valid, even though half of the eigenfunctions of the \Lame problem are used. Note, however that these eigenfunctions are real, and, thus, we have the relation $\phi - \omega t = 0$ and the solution degenerates to a straight string rotating like a rigid rod around its center. The string has finite size for bounded configurations and infinite size for unbounded configurations. In either case the limit gives rise to the Gubser-Klebanov-Polyakov solution \cite{Gubser:2002tv}, which in fact arises as degenerate limit of a spiky string with two spikes.

If one considers the translationally invariant solutions of the cosh-Gordon equation, $\xi_0$ and $\xi_1$ will be interchanged and the solution will be written as
\begin{equation}
\begin{split}
r &= \sqrt {\frac{{\wp \left( {{a_2}} \right) - \wp \left( {{\xi _0} + {\omega _2}} \right)}}{{\wp \left( {{a_1}} \right) - \wp \left( {{a_2}} \right)}}} ,\\
t &= {\ell _1}{\xi _1} - \arg \frac{{\sigma \left( {{\xi _0} + {\omega _2} + {a_1}} \right) \sigma \left( {{\omega _2}} \right)}}{{\sigma \left( {{\xi _0} + {\omega _2}} \right)\sigma \left( {{a_1} + {\omega _2}} \right)}}{e^{ - \zeta \left( {{a_1}} \right){\xi _0} }} ,\\
\varphi  &= {\ell _2}{\xi _1} - \arg \frac{{\sigma \left( {{\xi _0} + {\omega _2} + {a_2}} \right)  \sigma \left( {{\omega _2}} \right)}}{{\sigma \left( {{\xi _0} + {\omega _2}} \right)\sigma \left( {{a_2} + {\omega _2}} \right)}}{e^{ - \zeta \left( {{a_2}} \right){\xi _0} }} .
\end{split}
\end{equation}
It describes the space-time ``dual'' picture of a finite spiky string. The variables $r$ and $\phi - \omega t$ are independent of $\xi_1$, thus this solution is a circular string that rotates with angular velocity and radius that vary periodically in time. In this solutions, the radius of the string oscillates between two extremes. When it reaches the maximum value the string moves with the speed of light. Then, it is reflected towards smaller radii and starts shrinking until it reaches the minimum and it keeps oscillating.

From the point of view of the enhanced space, the coordinates $Y^{-1}$ and $Y^0$ have a periodic dependence on the global coordinate $t$ with period equal to $2\pi$. Thus, demanding that these solutions are single valued in the enhanced space enforces the oscillatory behaviour of the circular strings to have period equal to $2 \pi / n$, where $n \in \mathbb{N}$. This condition is analogous to the condition imposed on the angular opening of two consecutive spikes in the spiky string solutions.

\subsection{AdS$_3$ and Negative Eigenvalues}
We repeat the procedure to construct string solutions associated to two negative eigenvalues $\kappa_{1,2} = -\ell_{1,2}^2$ of the effective \Schrodinger problems. According to the opening remarks of this section, the ansatz for the coordinates $Y^\mu$ takes the form
\begin{equation}
Y = \left( {\begin{array}{*{20}{c}}
{c_1^ + \Sigma _1^ + \left( {{\xi _1}} \right)\cosh \left({\ell _1}{\xi _0}\right) + c_1^ - \Sigma _1^ - \left( {{\xi _1}} \right)\sinh \left({\ell _1}{\xi _0}\right)}\\
{c_2^ + \Sigma _2^ + \left( {{\xi _1}} \right)\cosh \left({\ell _2}{\xi _0}\right) + c_2^ - \Sigma _2^ - \left( {{\xi _1}} \right)\sinh \left({\ell _2}{\xi _0}\right)}\\
{c_1^ + \Sigma _1^ + \left( {{\xi _1}} \right)\sinh \left({\ell _1}{\xi _0}\right) + c_1^ - \Sigma _1^ - \left( {{\xi _1}} \right)\cosh \left({\ell _1}{\xi _0}\right)}\\
{c_2^ + \Sigma _2^ + \left( {{\xi _1}} \right)\sinh \left({\ell _2}{\xi _0}\right) + c_2^ - \Sigma _2^ - \left( {{\xi _1}} \right)\cosh \left({\ell _2}{\xi _0}\right)}
\end{array}} \right).
\label{eq:construction_ansatz_negative}
\end{equation}
In this case, the eigenvalues are given by the relations
\begin{equation}
- \ell _{1,2}^2 = - \wp \left( a _{1,2} \right) - 2{x_1} .
\end{equation}

Using this ansatz, the geometric constraint equation \eqref{eq:susy_constraint} as well as the Virasoro constraints \eqref{eq:susy_Virasoro1} and \eqref{eq:susy_Virasoro2} take the form
\begin{align}
&{\left( {c_1^ + \Sigma _1^ + } \right)^2} - {\left( {c_1^ - \Sigma _1^ - } \right)^2} + {\left( {c_2^ + \Sigma _2^ + } \right)^2} - {\left( {c_2^ - \Sigma _2^ - } \right)^2} = {\Lambda ^2} ,\\
&{\ell _1}c_1^ + c_1^ - \left( {{\Sigma _1^ +} '\Sigma _1^ -  - {\Sigma _1^ -} '\Sigma _1^\_} \right) =  - {\ell _2}c_2^ + c_2^ - \left( {{\Sigma _2^ +} '\Sigma _2^ -  - {\Sigma _2^ -} '\Sigma _2^\_} \right) ,\\
&\left[ {{{\left( {c_1^ + \Sigma _1^ + } \right)}^2} - {{\left( {c_1^ - \Sigma _1^ - } \right)}^2}} \right]\ell _1^2 + \left[ {{{\left( {c_2^ + \Sigma _2^ + } \right)}^2} - {{\left( {c_2^ - \Sigma _2^ - } \right)}^2}} \right]\ell _2^2 =  - {\Lambda ^2}\left( {\wp \left( {{\xi _1}} \right) - {x_1}} \right) .
\end{align}
In this case, we are led to the following choice,
\begin{align}
c_1^ +  = c_1^ -  = {c_1},&\quad c_2^ +  = c_2^ -  = {c_2} ,\\
\Sigma _1^ \pm  = \frac{1}{2}\left( {y_1^ +  \pm y_1^ - } \right),&\quad \Sigma _2^ \pm  = \frac{1}{2}\left( {y_2^ +  \mp y_2^ - } \right) . \label{eq:construction_negative_sigma_initial}
\end{align}
Reality of the solution implies that the states $y_{1,2}^\pm$ are not Bloch states but they should be non-normalizable states with eigenvalues lying in the gaps of the \Lame spectrum. Thus, $a_{1,2}$ should obey the relations
\begin{equation}
\wp \left( {{a_{1,2}}} \right) > {e_1}\quad \mathrm{or} \quad {e_3} < \wp \left( {{a_{1,2}}} \right) < {e_2} .
\end{equation}

Further manipulation of the constraints leads to the following equations,
\begin{align}
&c_1^2 = c_2^2 \equiv {c^2} ,\\
&{c^2}\left( {\wp \left( {{a_2}} \right) - \wp \left( {{a_1}} \right)} \right) = {\Lambda ^2} ,\label{eq:construction_negative_sign}\\
&\wp \left( {{a_1}} \right) + \wp \left( {{a_2}} \right) =  - x_1 = {x_2} + {x_3} .
\end{align}
which are identical with the case of positive eigenvalues. Equation \eqref{eq:construction_negative_sign} implies that $\wp \left( {{a_1}} \right) < \wp \left( {{a_2}} \right)$, which is valid for string solutions associated to unbounded configurations. For string solutions corresponding to the bounded configurations, the inequality is reversed to $\wp \left( {{a_1}} \right) > \wp \left( {{a_2}} \right)$.

Summarizing, the ansatz \eqref{eq:construction_ansatz_negative} provides a classical string solution, as long as $\wp \left( {{a_1}} \right) + \wp \left( {{a_2}} \right) =  - x_1 = {x_2} + {x_3} $, where $\wp \left( {{a_1}} \right) $ and $ \wp \left( {{a_2}} \right)$ correspond to eigenvalues in the gaps of the \Lame spectrum and the eigenvalues $\kappa_{1,2} = - \ell_{1,2}^2 = - \wp \left( {{a_{1,2}}} \right) - 2 x_1$ are both negative. A case by case analysis can be performed as for positive eigenvalues. Briefly, as it turns out, there are valid string solutions when there are three real roots and $x_1$ is either the intermediate root and $E > 0$ or the largest root. The former case corresponds to specific configurations of the cosh-Gordon equation, whereas the latter case corresponds to transmitting solutions of the sinh-Gordon equation with $s = - 1$ and $E > m^2$ or oscillating solutions of the sinh-Gordon equation with $s = + 1$. In all cases, both eigenvalues belong to the finite gap between the valence and conduction bands of the \Lame spectrum. The pairs of $\wp \left( {{a_1}} \right)$ and $\wp \left( {{a_2}} \right)$ that give rise to classical string solutions are depicted in figures \ref{fig:const_neg_cosh} and \ref{fig:const_neg_sinh}.
\afterpage{
\begin{figure}[p]
\centering
\begin{picture}(100,60)
\put(10,5){\includegraphics[width = 0.5\textwidth]{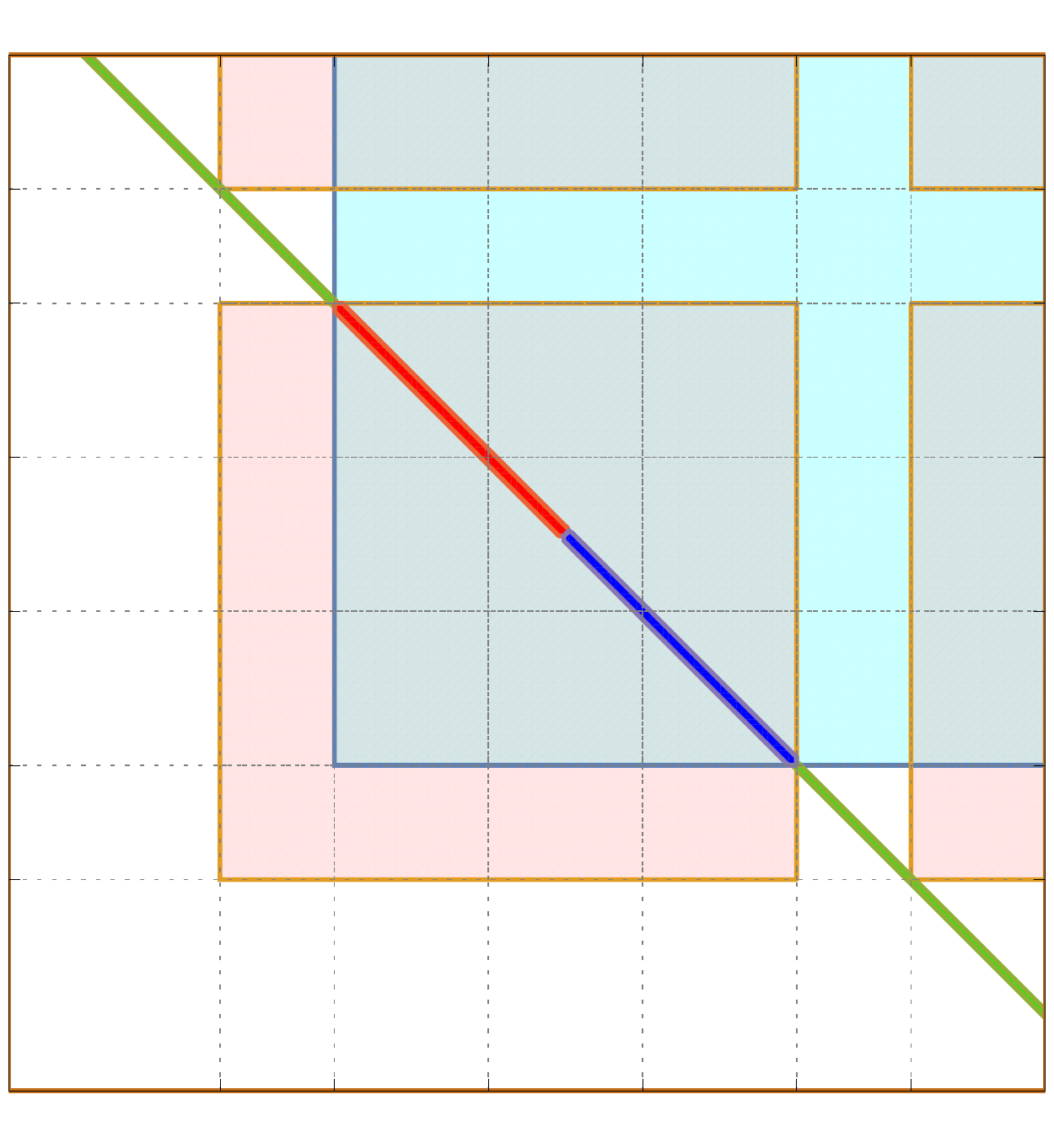}}
\put(62,16){\includegraphics[width = 0.1\textwidth]{regions_legend.pdf}}
\put(7.75,50.5){$e_1$}
\put(7.75,45){$e_2$}
\put(8.5,37){$0$}
\put(5.5,30.25){$-e_2$}
\put(4.25,22.75){$-2e_2$}
\put(7.75,17.5){$e_3$}
\put(19.25,5.25){$e_3$}
\put(22.25,5.25){$-2e_2$}
\put(30.25,5.25){$-e_2$}
\put(39.9,5.25){$0$}
\put(46.75,5.25){$e_2$}
\put(52.5,5.25){$e_1$}
\put(32,2){$\wp \left( a_1 \right)$}
\put(2,28){\rotatebox{90}{$\wp \left( a_2 \right)$}}
\put(68,43.5){negative eigenvalues}
\put(68,36.75){gap states}
\put(68,31.25){$\wp \left( a_1 \right)+\wp \left( a_2 \right) = -e_2$}
\put(68,24.75){$V = 2 \wp \left( x \right)$ solutions}
\put(68,18.75){$V = 2 \wp \left( x +\omega_2 \right)$ solutions}
\end{picture}
\vspace{-20pt}
\caption{The allowed $\wp\left( a_{1,2} \right)$ for classical string solutions when $x_1 = e_2$, $E>0$}
\label{fig:const_neg_cosh}
\vspace{10pt}
\begin{picture}(100,60)
\put(10,5){\includegraphics[width = 0.5\textwidth]{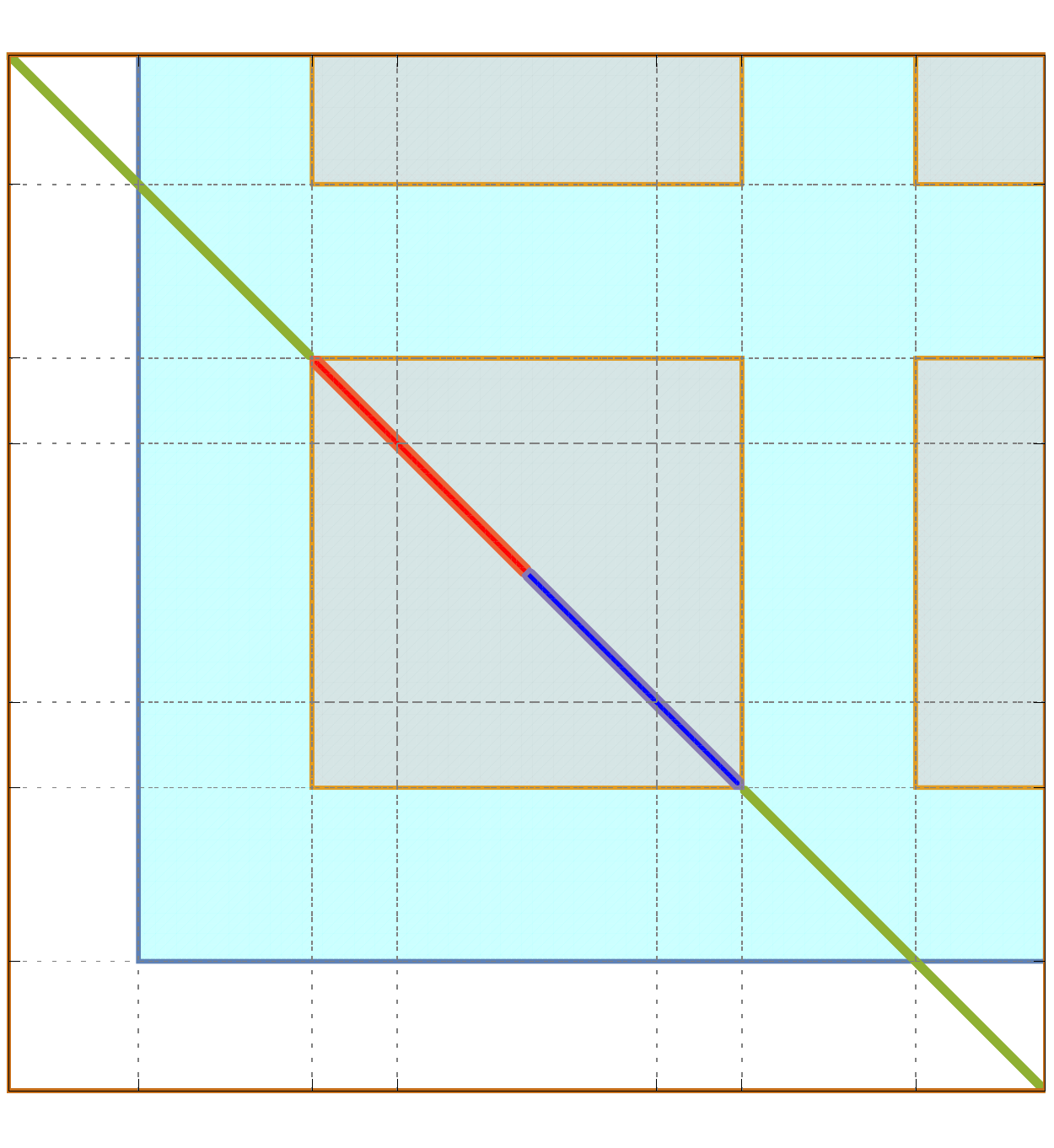}}
\put(62,16){\includegraphics[width = 0.1\textwidth]{regions_legend.pdf}}
\put(7.75,50){$e_1$}
\put(7.75,42){$e_2$}
\put(8.5,37.75){$0$}
\put(5.5,25.25){$-e_1$}
\put(7.75,21.5){$e_3$}
\put(4.25,13.5){$-2e_1$}
\put(13,5.25){$-2e_1$}
\put(23.75,5.25){$e_3$}
\put(26.25,5.25){$-e_1$}
\put(40.5,5.25){$0$}
\put(44.25,5.25){$e_2$}
\put(52.5,5.25){$e_1$}
\put(32,2){$\wp \left( a_1 \right)$}
\put(2,28){\rotatebox{90}{$\wp \left( a_2 \right)$}}
\put(68,43.5){negative eigenvalues}
\put(68,36.75){gap states}
\put(68,31.25){$\wp \left( a_1 \right)+\wp \left( a_2 \right) = -e_3$}
\put(68,24.75){$V = 2 \wp \left( x \right)$ solutions}
\put(68,18.75){$V = 2 \wp \left( x +\omega_2 \right)$ solutions}
\end{picture}
\vspace{-20pt}
\caption{The allowed $\wp\left( a_{1,2} \right)$ for classical string solutions when $x_1 = e_1$}
\label{fig:const_neg_sinh}
\end{figure}
\clearpage
}

The string solutions associated to two negative eigenvalues can also be formulated in global coordinates. For the ones originating from unbounded configurations, which are static, we define
\begin{equation}
\theta_{1,2} \left( {\xi_1} \right) = - \zeta \left( {{a_{1,2}}} \right){\xi _1} + \ln \frac{{\sigma \left( {{\xi _1} + {a_{1,2}}} \right)}}{{\sigma \left( {{\xi _1}} \right)\sigma \left( {{a_{1,2}}} \right)}} - \frac{1}{2} \ln \left( { \wp \left( {\xi_1} \right) - \wp \left( {a_{1,2}} \right) } \right) 
\end{equation}
and obtain
\begin{equation}
\begin{split}
\tan t &= \sqrt {\frac{{\wp \left( {{\xi _1}} \right) - \wp \left( {{a_2}} \right)}}{{\wp \left( {{\xi _1}} \right) - \wp \left( {{a_1}} \right)}}} \frac{{\sinh \left( {{\ell _2}{\xi _0} + \theta_2 \left( {\xi_1} \right)} \right)}}{{\cosh \left( {{\ell _1}{\xi _0} + \theta_1 \left( {\xi_1} \right)} \right)}} , \\
\tan \varphi &= \sqrt {\frac{{\wp \left( {{\xi _1}} \right) - \wp \left( {{a_2}} \right)}}{{\wp \left( {{\xi _1}} \right) - \wp \left( {{a_1}} \right)}}} \frac{{\cosh \left( {{\ell _2}{\xi _0} + \theta_2 \left( {\xi_1} \right)} \right)}}{{\sinh \left( {{\ell _1}{\xi _0} + \theta_1 \left( {\xi_1} \right)} \right)}} ,\\
{r^2} &= \frac{{\wp \left( {{\xi _1}} \right) - \wp \left( {{a_1}} \right)}}{{\wp \left( {{a_2}} \right) - \wp \left( {{a_1}} \right)}}{\sinh ^2}\left( {{\ell _1}{\xi _0} +\theta_1 \left( {\xi_1} \right)} \right) \\
&+ \frac{{\wp \left( {{\xi _1}} \right) - \wp \left( {{a_2}} \right)}}{{\wp \left( {{a_2}} \right) - \wp \left( {{a_1}} \right)}}{\cosh ^2}\left( {{\ell _2}{\xi _0} + \theta_2 \left( {\xi_1} \right)} \right) . 
\end{split}
\end{equation}

As for the string solutions originating from bounded configurations, which are translationally invariant, we define
\begin{equation}
\theta_{1,2} \left( {\xi_0} \right) = - \zeta \left( {{a_{1,2}}} \right){\xi _0} + \ln \frac{{\sigma \left( {{\xi _0} + {a_{1,2}} + {\omega _2}} \right)\sigma \left( {{\omega _2}} \right)}}{{\sigma \left( {{\xi _0}} \right)\sigma \left( {{a_{1,2}} + {\omega _2}} \right)}} - \frac{1}{2} \ln \left( { \wp \left( {a_{1,2}} \right) - \wp \left( {\xi_0 + \omega_2} \right) } \right)
\end{equation}
and obtain
\begin{equation}
\begin{split}
\tan t &= \sqrt {\frac{{\wp \left( {{a_2}} \right) - \wp \left( {{\xi _0} + {\omega _2}} \right)}}{{\wp \left( {{a_1}} \right) - \wp \left( {{\xi _0} + {\omega _2}} \right)}}} \frac{{\sinh \left( {{\ell _2}{\xi _1} + \theta_2 \left( {\xi_0} \right)} \right)}}{{\cosh \left( {{\ell _1}{\xi _1} + \theta_1 \left( {\xi_0} \right)} \right)}} , \\
\tan \varphi &= \sqrt {\frac{{\wp \left( {{a_2}} \right) - \wp \left( {{\xi _0} + {\omega _2}} \right)}}{{\wp \left( {{a_1}} \right) - \wp \left( {{\xi _0} + {\omega _2}} \right)}}} \frac{{\cosh \left( {{\ell _2}{\xi _1} + \theta_2 \left( {\xi_0} \right)} \right)}}{{\sinh \left( {{\ell _1}{\xi _1} + \theta_1 \left( {\xi_0} \right)} \right)}} ,\\
{r^2} &= \frac{{\wp \left( {{a_1}} \right) - \wp \left( {{\xi _0} + {\omega _2}} \right)}}{{\wp \left( {{a_1}} \right) - \wp \left( {{a_2}} \right)}}{\sinh ^2}\left( {{\ell _1}{\xi _1} + \theta_1 \left( {\xi_0} \right)} \right) \\
&+ \frac{{\wp \left( {{a_2}} \right) - \wp \left( {{\xi _0} + {\omega _2}} \right)}}{{\wp \left( {{a_1}} \right) - \wp \left( {{a_2}} \right)}}{\cosh ^2}\left( {{\ell _2}{\xi _1} + \theta_2 \left( {\xi_0} \right)} \right) .
\end{split}
\end{equation}

Finally, we note for completeness that the solutions take a much simpler form using the hyperbolic slicing of AdS$_3$
\begin{equation}
Y = \Lambda \left( {\begin{array}{*{20}{c}}
{r\cosh x}\\
{\sqrt {{r^2} - 1} \sinh t}\\
{r\sinh x}\\
{\sqrt {{r^2} - 1} \cosh t}
\end{array}} \right) .
\end{equation}
In these coordinates, the AdS$_3$ metric takes the usual form
\begin{equation}
d{s^2} =  - \left( {{r^2} - 1} \right)d{t^2} + \frac{1}{{{r^2} - 1}}d{r^2} + {r^2}d{x^2} .
\end{equation}

Then, for the unbounded configurations we get
\begin{equation}
\begin{split}
x &= {\ell _1}{\xi _0} + \theta_1 \left( {\xi_1} \right) ,\\
t &= {\ell _2}{\xi _0} + \theta_2 \left( {\xi_1} \right) ,\\
r &= \sqrt {\frac{{\wp \left( {{\xi _1}} \right) - \wp \left( {{a_1}} \right)}}{{\wp \left( {{a_2}} \right) - \wp \left( {{a_1}} \right)}}} \, ,
\end{split}
\end{equation}
whereas for the bounded ones we get
\begin{equation}
\begin{split}
x &= {\ell _1}{\xi _1} + \theta_1 \left( {\xi_0} \right) ,\\
t &= {\ell _2}{\xi _1} + \theta_2 \left( {\xi_0} \right) ,\\
r &= \sqrt {\frac{{\wp \left( {{a_1}} \right) - \wp \left( {{\xi _0} + {\omega _2}} \right)}}{{\wp \left( {{a_1}} \right) - \wp \left( {{a_2}} \right)}}} \, ,
\end{split}
\end{equation}
using the respective parameters $\theta_{1,2}$. The solutions can be visualized as a periodic spiky structure moving along the $x$ direction. In the unbounded case, the string extends to infinite $r$ and for the bounded case the string can reach a maximum value of $r$ where there are spikes moving with the speed of light.

\subsection{The Case of dS$_3$}
For dS$_3$, we use a pair of eigenvalues with opposite sign suggested by the form of the metric. Thus, the two eigenvalues $\kappa$ are $-\ell_1^2$ and $\ell_2^2$, where 
\begin{equation}
- \ell _1^2 = - \wp \left( {{a_1}} \right) - 2 x_1, \quad \ell _2^2 = - \wp \left( {{a_2}} \right) - 2 x_1 .
\end{equation}
It follows immediately that the following inequality is always satisfied,
\begin{equation}
\wp \left( {{a_1}} \right) > \wp \left( {{a_2}} \right) .
\end{equation}
We also make the ansatz for the coordinates of the enhanced space
\begin{equation}
Y = \left( {\begin{array}{*{20}{c}}
{c_1^ + \Sigma _1^ + \left( {{\xi _1}} \right)\cosh \left({\ell _1}{\xi _0}\right) + c_1^ - \Sigma _1^ - \left( {{\xi _1}} \right)\sinh \left({\ell _1}{\xi _0}\right)}\\
{c_1^ + \Sigma _1^ + \left( {{\xi _1}} \right)\sinh \left({\ell _1}{\xi _0}\right) + c_1^ - \Sigma _1^ - \left( {{\xi _1}} \right)\cosh \left({\ell _1}{\xi _0}\right)}\\
{c_2^ + \Sigma _2^ + \left( {{\xi _1}} \right)\cos \left({\ell _2}{\xi _0}\right) + c_2^ - \Sigma _2^ - \left( {{\xi _1}} \right)\sin \left({\ell _2}{\xi _0}\right)}\\
{c_2^ + \Sigma _2^ + \left( {{\xi _1}} \right)\sin \left({\ell _2}{\xi _0}\right) - c_2^ - \Sigma _2^ - \left( {{\xi _1}} \right)\cos \left({\ell _2}{\xi _0}\right)}
\end{array}} \right) .
\end{equation}

Substituting the above to the constraint equations, we obtain the relations
\begin{align}
& - {\left( {c_1^ + \Sigma _1^ + } \right)^2} + {\left( {c_1^ - \Sigma _1^ - } \right)^2} + {\left( {c_2^ + \Sigma _2^ + } \right)^2} + {\left( {c_2^ - \Sigma _2^ - } \right)^2} = {\Lambda ^2} ,\\
&{\ell _1}c_1^ + c_1^ - \left( {{\Sigma _1^ +} '\Sigma _1^ -  - {\Sigma _1^ -} '\Sigma _1^ + } \right) = {\ell _2}c_2^ + c_2^ - \left( {{\Sigma _2^ +} '\Sigma _2^ -  - {\Sigma _2^ -} '\Sigma _2^ + } \right) ,\\
& - \left[ { - {{\left( {c_1^ + \Sigma _1^ + } \right)}^2} + {{\left( {c_1^ - \Sigma _1^ - } \right)}^2}} \right]\ell _1^2 + \left[ {{{\left( {c_2^ + \Sigma _2^ + } \right)}^2} + {{\left( {c_2^ - \Sigma _2^ - } \right)}^2}} \right]\ell _2^2 = \left( {\wp \left( {{\xi _1}} \right) - x_1} \right){\Lambda ^2} ,
\end{align}
which lead to the choice
\begin{align}
c_1^ +  = c_1^ -  \equiv {c_1},&\quad c_2^ +  = c_2^ -  \equiv {c_2} ,\\
\Sigma _1^ +  = \frac{1}{2}\left( {y_1^ +  + y_1^ - } \right),&\quad \Sigma _1^ -  = \frac{1}{2}\left( {y_1^ +  - y_1^ - } \right) ,\\
\Sigma _2^ +  = \frac{1}{2}\left( {y_2^ +  + y_2^ - } \right),&\quad \Sigma _2^ -  = \frac{1}{{2i}}\left( {y_2^ +  - y_2^ - } \right) .
\end{align}
As in AdS$_3$, positive eigenvalues are enforced by the constraints to lie in the valence or conduction bands of the spectrum, whereas negative eigenvalues are enforced to lie in the gaps,
\begin{align}
\wp \left( {{a_1}} \right) > {e_1}&\quad \mathrm{or} \quad {e_3} < \wp \left( {{a_1}} \right) < {e_2} ,\\
\wp \left( {{a_2}} \right) < {e_3}&\quad \mathrm{or} \quad {e_2} < \wp \left( {{a_2}} \right) < {e_1} .
\end{align}

Further manipulation of the constraints yields the equations,
\begin{align}
&c_1^2 = c_2^2 \equiv {c^2} ,\\
&{c^2}\left( {\wp \left( {{a_1}} \right) - \wp \left( {{a_2}} \right)} \right) = {\Lambda ^2} ,\\
&\wp \left( {{a_1}} \right) + \wp \left( {{a_2}} \right) =  - x_1 = {x_2} + {x_3} ,
\end{align}
which are identical to the ones arising in AdS$_3$ space. The existence of string solutions is independent of the type of roots and their ordering. The positive eigenvalue lies in the infinite conduction band and the negative eigenvalue in the infinite gap.

When the potential $2 \wp \left( \xi_1 +\omega_2 \right) - 2 x_1$ is considered instead of $2 \wp \left( \xi_1 \right) -2 x_1$ for the bounded instead of the unbounded solutions, the normalization of the product $y_+ y_-$ of the conduction band states does not change sign, unlike the normalization of the product of the infinite gap states. Thus, in this case, the parameters $a_{1,2}$ remain the same, but $\Sigma _1^ \pm$ should be taken to be
\begin{equation}
\Sigma _1^ +  = \frac{1}{2}\left( {y_1^ +  - y_1^ - } \right),\quad \Sigma _1^ -  = \frac{1}{2}\left( {y_1^ +  + y_1^ - } \right) .
\end{equation}

The pairs of $\wp \left( {{a_1}} \right)$ and $\wp \left( {{a_2}} \right)$ that lead to classical string solutions in dS$_3$ are depicted in figures \ref{fig:const_ds_sinh_trans}, \ref{fig:const_ds_cosh}, \ref{fig:const_ds_sinh_refl} and \ref{fig:const_ds_sinh_trans_complex}.
\afterpage{
\begin{figure}[p]
\centering
\begin{picture}(100,60)
\put(10,5){\includegraphics[width = 0.5\textwidth]{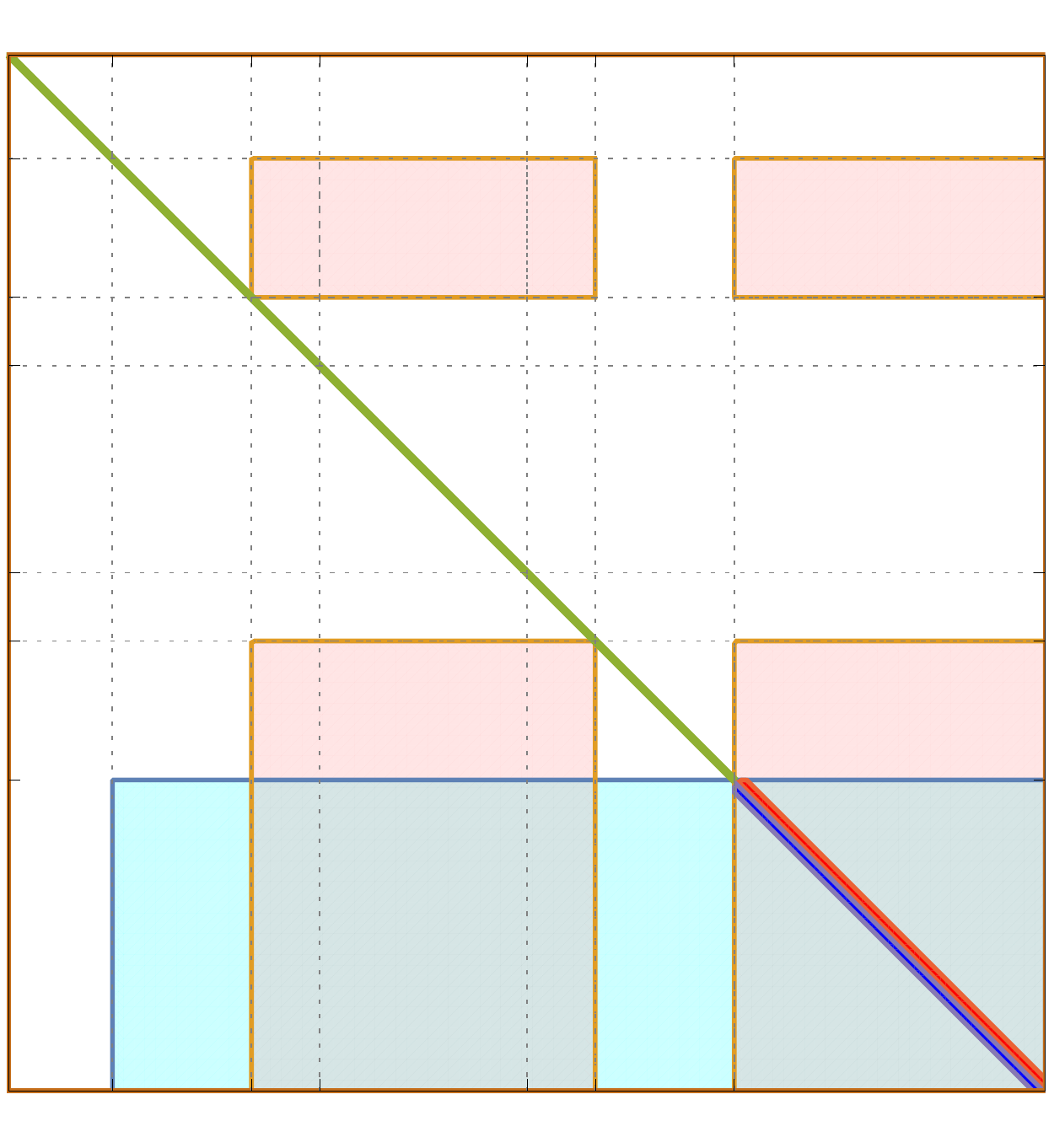}}
\put(62,16){\includegraphics[width = 0.1\textwidth]{regions_legend.pdf}}
\put(7.75,51.5){$e_1$}
\put(7.75,44.75){$e_2$}
\put(8.5,41.25){$0$}
\put(5.5,31.75){$-e_1$}
\put(7.75,28.5){$e_3$}
\put(4.25,21.75){$-2e_1$}
\put(11.75,5.25){$-2e_1$}
\put(20.5,5.25){$e_3$}
\put(23.25,5.25){$-e_1$}
\put(34.25,5.25){$0$}
\put(37.25,5.25){$e_2$}
\put(44,5.25){$e_1$}
\put(32,2){$\wp \left( a_1 \right)$}
\put(2,28){\rotatebox{90}{$\wp \left( a_2 \right)$}}
\put(68,43.5){$\kappa_1 < 0$, $\kappa_2 > 0$}
\put(68,36.75){$y_1^\pm$ in gaps, $y_2^\pm$ in bands}
\put(68,31.25){$\wp \left( a_1 \right)+\wp \left( a_2 \right) = -e_1$}
\put(68,24.75){$V = 2 \wp \left( x \right)$ solutions}
\put(68,18.75){$V = 2 \wp \left( x +\omega_2 \right)$ solutions}
\end{picture}
\vspace{-20pt}
\caption{The allowed $\wp\left( a_{1,2} \right)$ for classical string solutions in dS$_3$ when $x_1 = e_1$}
\label{fig:const_ds_sinh_trans}
\vspace{10pt}
\begin{picture}(100,60)
\put(10,5){\includegraphics[width = 0.5\textwidth]{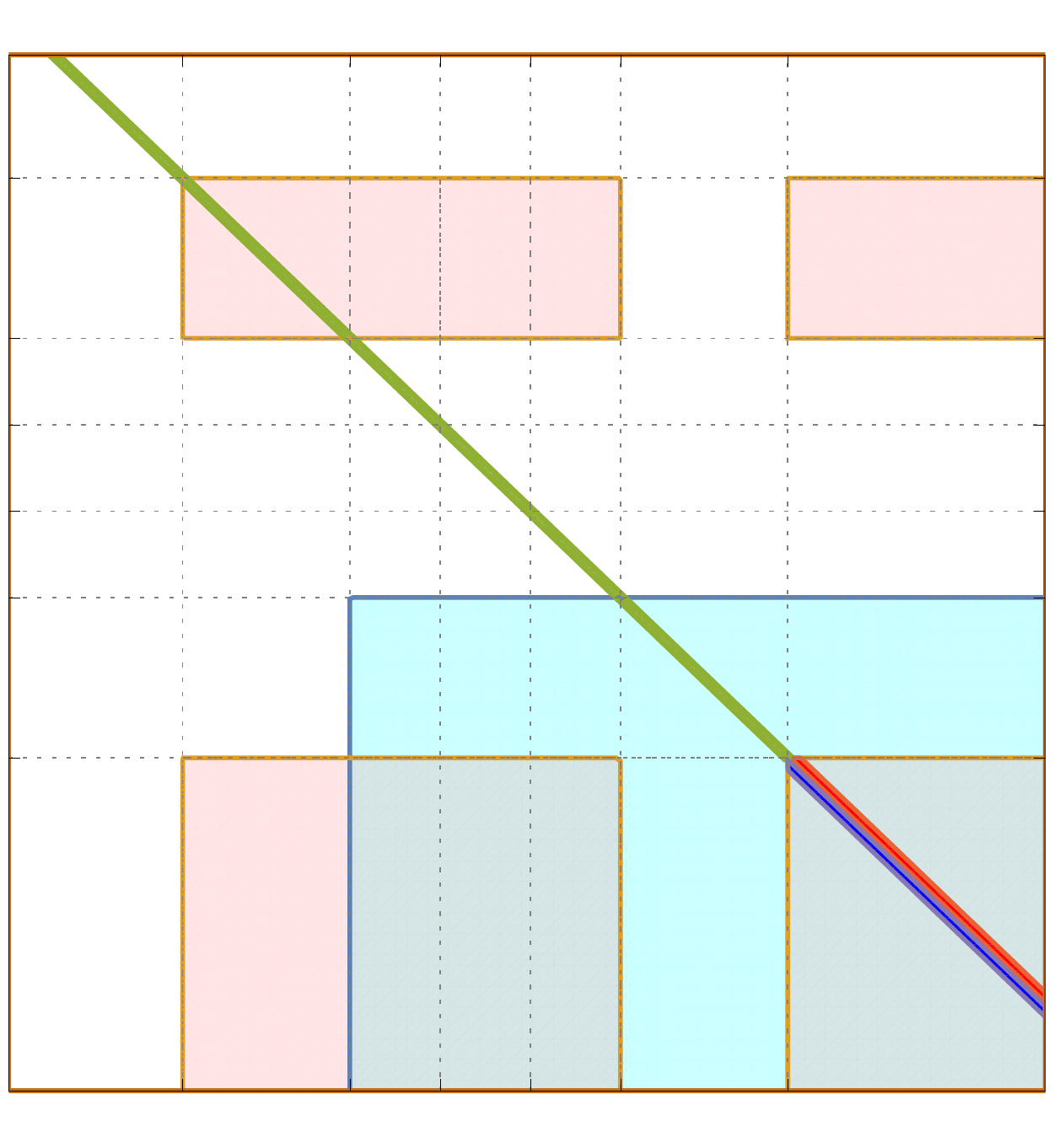}}
\put(62,16){\includegraphics[width = 0.1\textwidth]{regions_legend.pdf}}
\put(7.75,50.5){$e_1$}
\put(7.75,43){$e_2$}
\put(8.5,38.25){$0$}
\put(5.5,35){$-e_2$}
\put(7.75,23){$e_3$}
\put(4.25,30.75){$-2e_2$}
\put(22,5.25){$-2e_2$}
\put(17.75,5.25){$e_3$}
\put(28.25,5.25){$-e_2$}
\put(34.5,5.25){$0$}
\put(38.5,5.25){$e_2$}
\put(46.5,5.25){$e_1$}
\put(32,2){$\wp \left( a_1 \right)$}
\put(2,28){\rotatebox{90}{$\wp \left( a_2 \right)$}}
\put(68,43.5){$\kappa_1 < 0$, $\kappa_2 > 0$}
\put(68,36.75){$y_1^\pm$ in gaps, $y_2^\pm$ in bands}
\put(68,31.25){$\wp \left( a_1 \right)+\wp \left( a_2 \right) = -e_2$}
\put(68,24.75){$V = 2 \wp \left( x \right)$ solutions}
\put(68,18.75){$V = 2 \wp \left( x +\omega_2 \right)$ solutions}
\end{picture}
\vspace{-20pt}
\caption{The allowed $\wp\left( a_{1,2} \right)$ for classical string solutions in dS$_3$ when $x_1 = e_2$}
\label{fig:const_ds_cosh}
\end{figure}
\clearpage
}

\afterpage{
\begin{figure}[p]
\centering
\begin{picture}(100,60)
\put(10,5){\includegraphics[width = 0.5\textwidth]{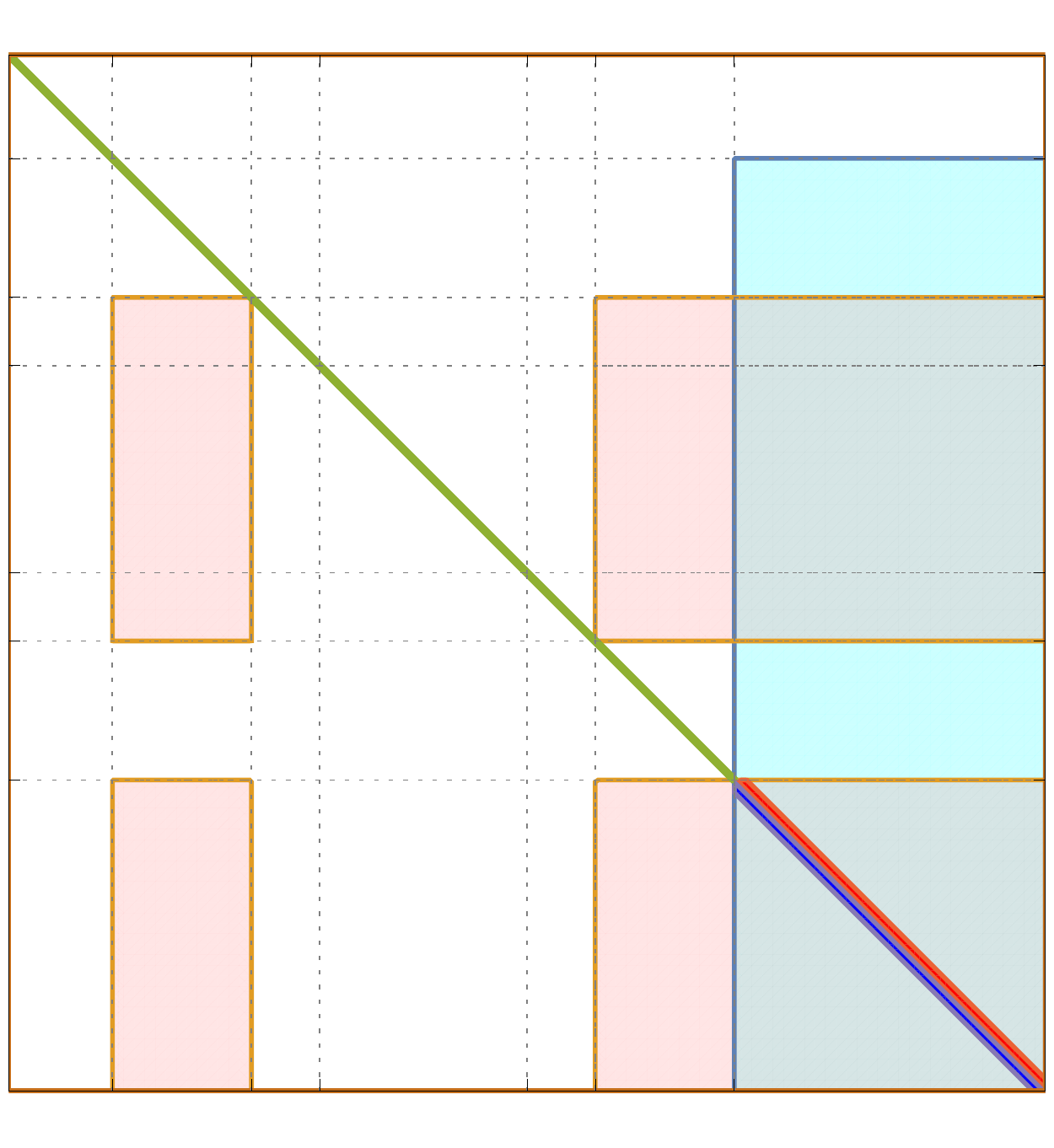}}
\put(62,16){\includegraphics[width = 0.1\textwidth]{regions_legend.pdf}}
\put(4.25,51.5){$-2e_3$}
\put(7.75,45){$e_1$}
\put(5.5,41.75){$-e_3$}
\put(8.5,31.25){$0$}
\put(7.75,28.5){$e_2$}
\put(7.75,21.75){$e_3$}
\put(14.5,5.25){$e_3$}
\put(20.75,5.25){$e_2$}
\put(24.5,5.25){$0$}
\put(32,5.25){$-e_3$}
\put(37.25,5.25){$e_1$}
\put(41.75,5.25){$-2e_3$}
\put(32,2){$\wp \left( a_1 \right)$}
\put(2,28){\rotatebox{90}{$\wp \left( a_2 \right)$}}
\put(68,43.5){$\kappa_1 < 0$, $\kappa_2 > 0$}
\put(68,36.75){$y_1^\pm$ in gaps, $y_2^\pm$ in bands}
\put(68,31.25){$\wp \left( a_1 \right)+\wp \left( a_2 \right) = -e_3$}
\put(68,24.75){$V = 2 \wp \left( x \right)$ solutions}
\put(68,18.75){$V = 2 \wp \left( x +\omega_2 \right)$ solutions}
\end{picture}
\vspace{-20pt}
\caption{The allowed $\wp\left( a_{1,2} \right)$ for classical string solutions in dS$_3$ when $x_1 = e_3$}
\label{fig:const_ds_sinh_refl}
\vspace{10pt}
\begin{picture}(100,60)
\put(10,5){\includegraphics[width = 0.5\textwidth]{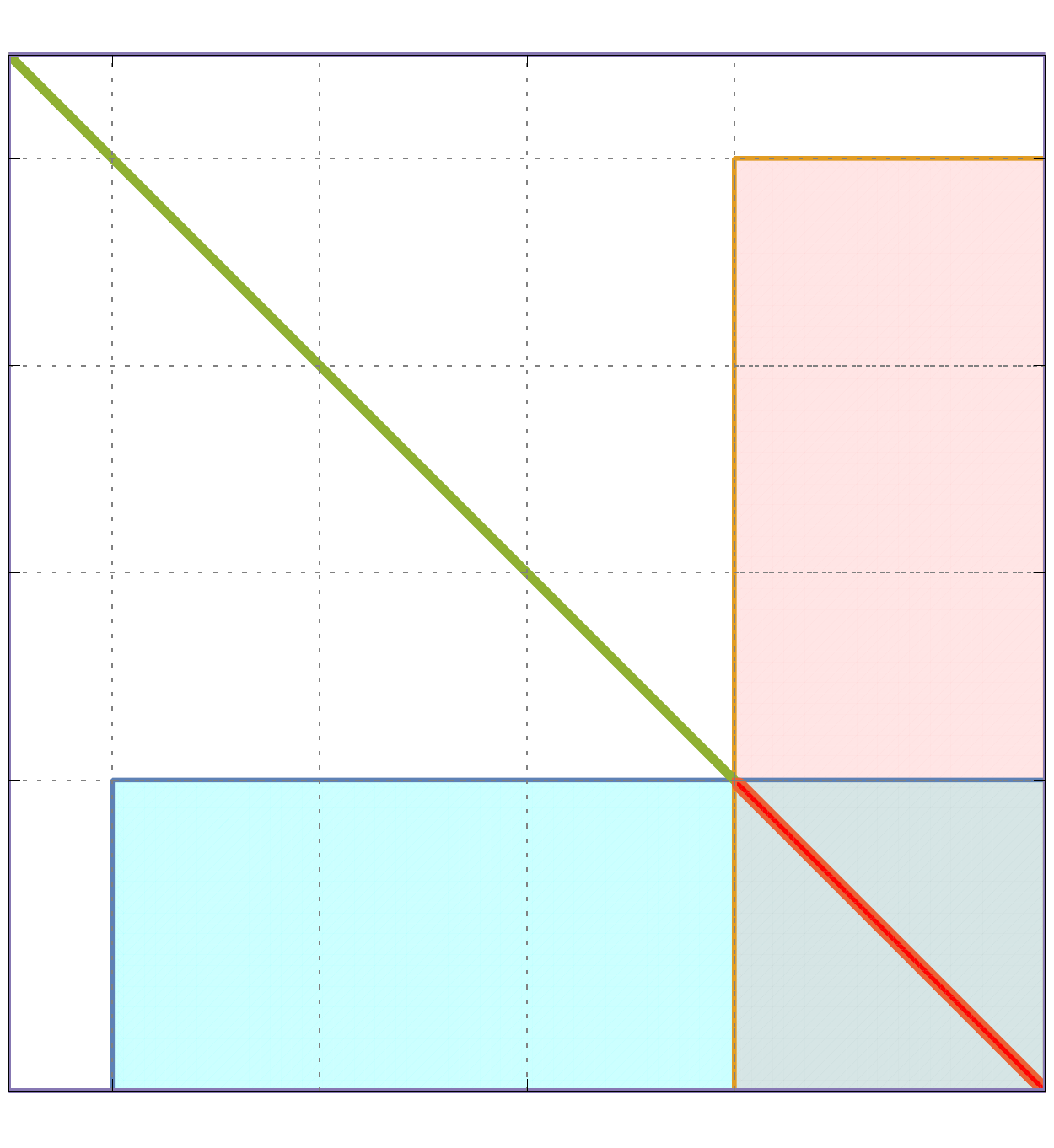}}
\put(62,16){\includegraphics[width = 0.1\textwidth]{regions_legend.pdf}}
\put(4.25,21.75){$-2e_2$}
\put(7.75,51.75){$e_2$}
\put(5.5,31.75){$-e_2$}
\put(8.5,41.25){$0$}
\put(34.25,5.25){$0$}
\put(22.25,5.25){$-e_2$}
\put(44,5.25){$e_2$}
\put(12,5.25){$-2e_2$}
\put(32,2){$\wp \left( a_1 \right)$}
\put(2,28){\rotatebox{90}{$\wp \left( a_2 \right)$}}
\put(68,43.5){$\kappa_1 < 0$, $\kappa_2 > 0$}
\put(68,36.75){$y_1^\pm$ in gaps, $y_2^\pm$ in bands}
\put(68,31.25){$\wp \left( a_1 \right)+\wp \left( a_2 \right) = -e_3$}
\put(68,24.75){$V = 2 \wp \left( x \right)$ solutions}
\put(68,18.75){$V = 2 \wp \left( x +\omega_2 \right)$ solutions}
\end{picture}
\vspace{-20pt}
\caption{The allowed $\wp\left( a_{1,2} \right)$ for classical string solutions in dS$_3$ when $x_1$ is the only real root}
\label{fig:const_ds_sinh_trans_complex}
\end{figure}
\clearpage
}

The unbounded configurations correspond to translationally invariant solutions of the sinh- or cosh-Gordon equation. The bounded solutions fall in three subclasses: when there are three real roots and $x_1$ is the largest root, they correspond to static oscillatory solutions of the sinh-Gordon equation; when $x_1$ is the intermediate root, they correspond to static solutions of the cosh-Gordon equation; and, finally, when $x_1$ is the smallest of the three real roots, they correspond to translationally invariant reflective solutions of the sinh-Gordon equation.

The string solution can be conveniently expressed using the parametrization
\begin{equation}
Y = \Lambda \left( {\begin{array}{*{20}{c}}
{\sqrt{1 - {r^2}} \sinh t}\\
{\sqrt{1 - {r^2}} \cosh t}\\
{r \cos \varphi }\\
{r \sin \varphi }
\end{array}} \right) ,
\end{equation}
in terms of static coordinates that transform the dS$_3$ metric to the form
\begin{equation}
d{s^2} =  - \left( {1 - {r^2}} \right)d{t^2} + \frac{1}{{1 - {r^2}}}d{r^2} + {r^2}d{\varphi ^2} .
\end{equation}

To simplify the presentation, we write the expressions only for the bounded solutions in these coordinates. Setting
\begin{align}
\theta_1 \left( {\xi} \right) &= - \zeta \left( {{a_1}} \right){\xi} + \ln \frac{{\sigma \left( {{\xi} + {a_1} + {\omega _2}} \right)\sigma \left( {{\omega _2}} \right)}}{{\sigma \left( {{\xi}} \right)\sigma \left( {{a_1} + {\omega _2}} \right)}} - \frac{1}{2} \ln \left( { \wp \left( {a_1} \right) - \wp \left( {\xi_0 + \omega_2} \right) } \right),\\
\theta_2 \left( {\xi} \right) &= - \arg \frac{{\sigma \left( {{\xi} + {\omega _2} + {a_2}} \right) \sigma \left( {\omega _2} \right)}}{{\sigma \left( {{\xi} + {\omega _2}} \right)\sigma \left( {{a_2}} + {\omega _2} \right)}}{e^{ - \zeta \left( {{a_2}} \right){\xi} }}
\end{align}
we obtain
\begin{equation}
\begin{split}
r &= \sqrt {\frac{{\wp \left( {{a_2}} \right) - \wp \left( {\xi _1 + {\omega _2}} \right)}}{{\wp \left( {{a_1}} \right) - \wp \left( {{a_2}} \right)}}} , \\
t &= {\ell _1}{\xi _0} + \theta_1 \left( {\xi_1} \right) ,\\
\varphi &= {\ell _2}{\xi _0} + \theta_2 \left( {\xi_1} \right) ,
\end{split}
\end{equation}
for the static solutions, and
\begin{equation}
\begin{split}
r &= \sqrt {\frac{{\wp \left( {{a_2}} \right) - \wp \left( {\xi _0 + {\omega _2}} \right)}}{{\wp \left( {{a_1}} \right) - \wp \left( {{a_2}} \right)}}} , \\
t &= {\ell _1}{\xi _1} + \theta_1 \left( {\xi_0} \right) ,\\
\varphi &= {\ell _2}{\xi _1} + \theta_2 \left( {\xi_0} \right) ,
\end{split}
\end{equation}
for the translationally invariant solutions.

As in AdS$_3$, the static solutions are rigidly rotating strings, while the translationally invariant ones are circular rotating strings with varying radius and angular velocity. The presentation of their details is beyond the scope of this paper.

\section{Discussion}
\label{sec:discussion}

We developed a method to construct classical string solutions in AdS$_3$ and dS$_3$ from a specific family of solutions of the Pohlmeyer reduced theory, namely solutions of the sinh-Gordon and cosh-Gordon equations that depend solely on one of the two world-sheet coordinates, and, thus, they are either static or translationally invariant. In all cases, these solutions admit a uniform description in terms of Weierstrass functions, facilitating their study. They are characterized by an interesting interplay between static and translationally invariant solutions and string propagation in AdS$_3$ and dS$_3$ spaces.

Our construction is based on separation of variables leading to four pairs of effective \Schrodinger problems with the same eigenvalues per component. Each pair consists of a flat potential and a periodic $n = 1$ \Lame potential. Consistent solutions fall within an ansatz that requires not one but two distinct eigenvalues. The constraints select Bloch waves with positive eigenvalues and non-normalizable states with negative eigenvalues in the gaps of the \Lame potential.

The class of elliptic string solutions that emerge in our study includes the spiky strings \cite{Kruczenski:2004wg} as well as several new solutions. They include rotating circular strings with periodically varying radius and angular velocity. Our results provide a unified framework for elliptic solutions which complements nicely the existing literature on the subject and goes beyond it. It is also interesting to understand the relation with other known rotating string solutions i.e., \cite{Arutyunov:2003uj,Arutyunov:2003za}.

The technical details rely on the inverse problem of Pohlmeyer reduction. 
For a given solution of the Pohlmeyer reduced equations, there is a continuously infinite set of distinct classical string solutions. It would be interesting to study the extension to other target space geometries, such as the sphere. Spiky string solutions are known to exist on the sphere \cite{Mosaffa:2007ty,Ishizeki:2007we}, thus it is very probable that there is an analogous treatment for them. In higher dimensional symmetric spaces, Pohlmeyer reduction results in multi-component generalizations of the sinh- or cosh-Gordon equations. An interesting question is whether there is an non-trivial extension of our techniques to those more general cases. All these will be useful for applications to strings propagating in AdS$_5 \times$S$^5$ \cite{Grigoriev:2007bu,Mikhailov:2007xr,Grigoriev:2008jq} in the framework of holography.

Finally, it will be interesting to generalize the method we have presented here to space-like surfaces and the construction of non-trivial minimal surfaces. For AdS$_4$ space, in particular, such minimal surfaces are co-dimension two and they can be related to the geometric interpretation of entanglement entropy through the Ryu-Takayanagi conjecture \cite{Ryu:2006bv}, thus providing new insight to these issues.

\acknowledgments
We would like to thank M. Axenides, E. Floratos and G. Linardopoulos for useful discussions.

\appendix
\section{Useful Formulas for the Weierstrass Functions}
\label{sec:Weierstrass_functions}

The Weierstrass function $\wp$ is a doubly periodic function of one complex variable that satisfies the equation
\begin{equation}
{\left( {\frac{{d\wp }}{{dz}}} \right)^2} = 4{\wp ^3} - {g_2}\wp  - {g_3} .
\label{eq:Weierstrass_p_equation}
\end{equation}
In this work we adopt the notation $\wp \left( {z;{g_2},{g_3}} \right)$, but in the literature one often  denotes the Weierstrass function as $\wp \left( {z;{\omega_1},{\omega_2}} \right)$, using the two half-periods $\omega_1$ and $\omega_2$.

The Weierstrass $\zeta$ function is a doubly quasi-periodic function, which obeys the defining relation
\begin{equation}
\frac{{d\zeta }}{{dz}} = - \wp .
\label{eq:Weierstrass_zeta}
\end{equation}
Finally, the Weierstrass $\sigma$ function is defined so that
\begin{equation}
\frac{1}{\sigma }\frac{{d\sigma }}{{dz}}= \zeta .
\label{eq:Weierstrass_sigma}
\end{equation}

The Weierstrass functions $\wp$, $\zeta$ and $\sigma$ have the following parity properties,
\begin{align}
\wp \left( { - z} \right) &= \wp \left( z \right) ,\\
\zeta \left( { - z} \right) &=  - \zeta \left( z \right) ,\\
\sigma \left( { - z} \right) &=  - \sigma \left( z \right) .
\end{align}

It can be shown that under a shift of the complex variable $z$ in the lattice defined by the periods $2\omega_1$ and $2\omega_2$, the Weierstrass functions transform as
\begin{align}
\wp \left( {z + 2m{\omega _1} + 2n{\omega _2}} \right) &= \wp \left( z \right) ,\label{eq:Weierstrass_period_wp}\\
\zeta \left( {z + 2m{\omega _1} + 2n{\omega _2}} \right) &= \zeta \left( z \right) + 2m\zeta \left( {{\omega _1}} \right) + 2n\zeta \left( {{\omega _2}} \right) ,\label{eq:Weierstrass_period_zeta}\\
\sigma \left( {z + 2m{\omega _1} + 2n{\omega _2}} \right) &= {\left( { - 1} \right)^{m + n + mn}}{e^{\left( {2m\zeta \left( {{\omega _1}} \right) + 2n\zeta \left( {{\omega _2}} \right)} \right)\left( {z + m{\omega _1} + n{\omega _2}} \right)}} \sigma \left( {z} \right) .\label{eq:Weierstrass_period_sigma}
\end{align}
We also note that $\zeta \left( {{\omega _1}} \right)$ and $\zeta \left( {{\omega _2}} \right)$ obey the relation
\begin{equation}
\omega _2 \zeta \left( {{\omega _1}} \right) - \omega _1 \zeta \left( {{\omega _2}} \right) = i\frac{\pi}{2}.
\label{eq:Weierstrass_zeta_special}
\end{equation}

The Weierstrass functions also obey the following homogeneity relations,
\begin{align}
\wp \left( {z;{g_2},{g_3}} \right) &= {\mu ^2}\wp \left( {\mu z;\frac{{{g_2}}}{{{\mu ^4}}},\frac{{{g_3}}}{{{\mu ^6}}}} \right) ,\label{eq:Weierstras_homogeneity_wp}\\
\zeta \left( {z;{g_2},{g_3}} \right) &= \mu \zeta \left( {\mu z;\frac{{{g_2}}}{{{\mu ^4}}},\frac{{{g_3}}}{{{\mu ^6}}}} \right) ,\\
\sigma \left( {z;{g_2},{g_3}} \right) &= \frac{1}{\mu }\sigma \left( {\mu z;\frac{{{g_2}}}{{{\mu ^4}}},\frac{{{g_3}}}{{{\mu ^6}}}} \right).
\end{align}
For the specific choice $\mu = i$, the above relations yield
\begin{align}
\wp \left( {z;{g_2},{g_3}} \right) &=  - \wp \left( {iz;{g_2}, - {g_3}} \right) ,\\
\zeta \left( {z;{g_2},{g_3}} \right) &= i\zeta \left( {iz;{g_2}, - {g_3}} \right) ,\\
\sigma \left( {z;{g_2},{g_3}} \right) &=  - i\sigma \left( {iz;{g_2}, - {g_3}} \right) ,
\end{align}
which in turn imply that $\wp$ is real, whereas $\zeta$ and $\sigma$ are imaginary on the imaginary axis of the $z$ plane.

The Weierstrass functions obey the following addition formulas,
\begin{align}
\wp \left( {z + w} \right) &=  - \wp \left( z \right) - \wp \left( w \right) + \frac{1}{4}{\left( {\frac{{\wp '\left( z \right) - \wp '\left( w \right)}}{{\wp \left( z \right) - \wp \left( w \right)}}} \right)^2} , \label{eq:Weierstrass_addition_wp}\\
\zeta \left( {z + w} \right) &= \zeta \left( z \right) + \zeta \left( w \right) + \frac{1}{2}\frac{{\wp '\left( z \right) - \wp '\left( w \right)}}{{\wp \left( z \right) - \wp \left( w \right)}} , \label{eq:Weierstrass_addition_zeta}\\
\wp \left( z \right) - \wp \left( w \right) &= - \frac{{\sigma \left( {z - w} \right)\sigma \left( {z + w} \right)}}{{{\sigma ^2}\left( z \right){\sigma ^2}\left( w \right)}}. \label{eq:Weierstrass_addition_sigma}
\end{align}
Selecting $w$ to be any of the half-periods, the last formula gives rise to the useful relation,
\begin{equation}
\wp \left( z \right) - {e_{1,3,2}} =  - \frac{{\sigma \left( {z + {\omega _{1,2,3} }} \right)\sigma \left( {z - {\omega _{1,2,3} }} \right)}}{{{\sigma ^2}\left( z \right){\sigma ^2}\left( {{\omega _{1,2,3} }} \right)}}.
\label{eq:Weierstrass_addition_sigma_special}
\end{equation}

Finally, the Weierstrass function obeys the following integral formula
\begin{equation}
\wp '\left( a \right)\int {\frac{{dz}}{{\wp \left( z \right) - \wp \left( a \right)}} = 2\zeta \left( a \right)z + \ln \frac{{\sigma \left( {z - a} \right)}}{{\sigma \left( {z + a} \right)}}} .
\label{eq:Weierstrass_integral}
\end{equation}

The periods of the function $\wp$ are related to the roots of the cubic polynomial appearing on the right hand side of the defining differential equation \eqref{eq:Weierstrass_p_equation}. These roots are denoted by $e_i$ satisfying the relation $e_1 + e_2 + e_3 = 0$. When two roots of the cubic polynomial coincide, one of the two periods diverges and the Weierstrass functions acquire simpler expressions in terms of simply periodic functions on the complex plane. In particular, when the two larger roots coincide, ${e_1} = {e_2} = {e_0}$, in which case ${e_3} =  - 2{e_0}$, the moduli $g_2$ and $g_3$ equal to $12e_0^2$ and $- 8e_0^3$, respectively. In this case the real period ${\omega _1}$ diverges and the imaginary period assumes the value ${\omega _2} = i\pi /\sqrt {12{e_0}}$. Then, the Weierstrass functions reduce to
\begin{align}
\wp \left( {z;12e_0^2, - 8e_0^3} \right) &= {e_0}\left( {1 + \frac{3}{{{{\sinh }^2}\left( {\sqrt {3{e_0}} z} \right)}}} \right) ,\label{eq:Weierstrass_wp_e1e2}\\
\zeta \left( {z;12e_0^2, - 8e_0^3} \right) &= - {e_0}z + \sqrt {3{e_0}} \coth \left( {\sqrt {3{e_0}} z} \right) ,\\
\sigma \left( {z;12e_0^2, - 8e_0^3} \right) &= \frac{{\sinh \left( {\sqrt {3{e_0}} z} \right)}}{{\sqrt {3{e_0}} }}{e^{ - \frac{1}{2}{e_0}{z^2}}} .
\end{align}

When the two smaller roots coincide, ${e_2} = {e_3} =  - {e_0}$, in which case ${e_1} = 2{e_0}$, the moduli $g_2$ and $g_3$ equal to $12e_0^2$ and $8e_0^3$, respectively. Then, in this case the imaginary period ${\omega _2}$ diverges and the real period takes the value ${\omega _1} = \pi /\sqrt {12{e_0}}$. The Weierstrass functions simplify to
\begin{align}
\wp \left( {z;12e_0^2,8e_0^3} \right) &= {e_0}\left( { - 1 + \frac{3}{{{{\sin }^2}\left( {\sqrt {3{e_0}} z} \right)}}} \right) ,\label{eq:Weierstrass_wp_e2e3}\\
\zeta \left( {z;12e_0^2,8e_0^3} \right) &= {e_0}z + \sqrt {3{e_0}} \cot \left( {\sqrt {3{e_0}} z} \right) ,\\
\sigma \left( {z;12e_0^2,8e_0^3} \right) &= \frac{{\sin \left( {\sqrt {3{e_0}} z} \right)}}{{\sqrt {3{e_0}} }}{e^{\frac{1}{2}{e_0}{z^2}}} .
\end{align}

\pagebreak


\begin{thebibliography}{1}


\bibitem{Pohlmeyer:1975nb} 
  K.~Pohlmeyer,
  ``Integrable Hamiltonian Systems and Interactions Through Quadratic Constraints'',
  Commun.\ Math.\ Phys.\  {\bf 46}, 207 (1976).
  
\bibitem{Zakharov:1973pp} 
  V.~E.~Zakharov and A.~V.~Mikhailov,
  ``Relativistically Invariant Two-Dimensional Models in Field Theory Integrable by the Inverse Problem Technique. (In Russian)'',
  Sov.\ Phys.\ JETP {\bf 47}, 1017 (1978)
  [Zh.\ Eksp.\ Teor.\ Fiz.\  {\bf 74}, 1953 (1978)].

\bibitem{Eichenherr:1979yw} 
  H.~Eichenherr and K.~Pohlmeyer,
  ``Lax Pairs for Certain Generalizations of the {Sine-Gordon} Equation'',
  Phys.\ Lett.\ B {\bf 89}, 76 (1979).

\bibitem{Pohlmeyer:1979ch} 
  K.~Pohlmeyer and K.~H.~Rehren,
  ``Reduction of the Two-dimensional O($n$) Nonlinear $\sigma$-model'',
  J.\ Math.\ Phys.\  {\bf 20}, 2628 (1979).
  
\bibitem{Eichenherr:1979uk} 
  H.~Eichenherr and J.~Honerkamp,
  ``Reduction of the {CP}$^N$ Nonlinear $\sigma$ Model'',
  J.\ Math.\ Phys.\  {\bf 22}, 374 (1981).
  
\bibitem{Eichenherr:1979ci} 
  H.~Eichenherr and M.~Forger,
  ``On the Dual Symmetry of the Nonlinear Sigma Models'',
  Nucl.\ Phys.\ B {\bf 155}, 381 (1979).
  
\bibitem{Eichenherr:1979hz} 
  H.~Eichenherr and M.~Forger,
  ``More About Nonlinear Sigma Models On Symmetric Spaces'',
  Nucl.\ Phys.\ B {\bf 164}, 528 (1980)
  [Nucl.\ Phys.\ B {\bf 282}, 745 (1987)].
  
\bibitem{Lund:1976xd} 
  F.~Lund,
  ``Note on the Geometry of the Nonlinear Sigma Model in Two-Dimensions'',
  Phys.\ Rev.\ D {\bf 15}, 1540 (1977).


\bibitem{Bakas:1993xh} 
  I.~Bakas,
  ``Conservation Laws and Geometry of Perturbed Coset Models'',
  Int.\ J.\ Mod.\ Phys.\ A {\bf 9}, 3443 (1994)
  [hep-th/9310122].
  
\bibitem{Bakas:1995bm} 
  I.~Bakas, Q.~H.~Park and H.~J.~Shin,
  ``Lagrangian Formulation of Symmetric Space Sine-Gordon Models'',
  Phys.\ Lett.\ B {\bf 372}, 45 (1996)
  [hep-th/9512030].
  
\bibitem{FernandezPousa:1996hi} 
  C.~R.~Fernandez-Pousa, M.~V.~Gallas, T.~J.~Hollowood and J.~L.~Miramontes,
  ``The Symmetric Space and Homogeneous Sine-Gordon Theories'',
  Nucl.\ Phys.\ B {\bf 484}, 609 (1997)
  [hep-th/9606032].
  
\bibitem{Miramontes:2008wt} 
  J.~L.~Miramontes,
  ``Pohlmeyer Reduction Revisited'',
  JHEP {\bf 0810}, 087 (2008)
  [arXiv:0808.3365 [hep-th]].


\bibitem{Barbashov:1980kz} 
  B.~M.~Barbashov and V.~V.~Nesterenko,
  ``Relativistic String Model in a Space-time of a Constant Curvature'',
  Commun.\ Math.\ Phys.\  {\bf 78}, 499 (1981).

\bibitem{DeVega:1992xc} 
  H.~J.~De Vega and N.~G.~Sanchez,
  ``Exact integrability of strings in D-Dimensional De Sitter Space-time'',
  Phys.\ Rev.\ D {\bf 47}, 3394 (1993).

\bibitem{Larsen:1996gn} 
  A.~L.~Larsen and N.~G.~Sanchez,
  ``Sinh-Gordon, Cosh-Gordon and Liouville Equations for Strings and Multistrings in Constant Curvature space-times'',
  Phys.\ Rev.\ D {\bf 54}, 2801 (1996)
  [hep-th/9603049].
  
\bibitem{Maldacena:1997re}
  J.~M.~Maldacena,
  ``The Large N Limit of Superconformal Field Theories and Supergravity",
  Int.\ J.\ Theor.\ Phys.\  {\bf 38}, 1113 (1999)
  [Adv.\ Theor.\ Math.\ Phys.\  {\bf 2}, 231 (1998)]
  [hep-th/9711200].
  
\bibitem{Gubser:1998bc}
  S.~S.~Gubser, I.~R.~Klebanov and A.~M.~Polyakov,
  ``Gauge Theory Correlators from Noncritical String Theory",
  Phys.\ Lett.\ B {\bf 428}, 105 (1998)
  [hep-th/9802109].
  
\bibitem{Witten:1998qj}
  E.~Witten,
  ``Anti-de Sitter Space and Holography",
  Adv.\ Theor.\ Math.\ Phys.\  {\bf 2}, 253 (1998)
  [hep-th/9802150].


\bibitem{Grigoriev:2007bu} 
  M.~Grigoriev and A.~A.~Tseytlin,
  ``Pohlmeyer Reduction of AdS$_5\times$S$^5$ Superstring Sigma Model'',
  Nucl.\ Phys.\ B {\bf 800}, 450 (2008)
  [arXiv:0711.0155 [hep-th]].
  
\bibitem{Mikhailov:2007xr} 
  A.~Mikhailov and S.~Schafer-Nameki,
  ``Sine-Gordon-like Action for the Superstring in AdS$_5\times$S$^5$'',
  JHEP {\bf 0805}, 075 (2008)
  [arXiv:0711.0195 [hep-th]].
  
\bibitem{Grigoriev:2008jq} 
  M.~Grigoriev and A.~A.~Tseytlin,
  ``On Reduced Models for Superstrings on AdS$_n\times$S$^n$'',
  Int.\ J.\ Mod.\ Phys.\ A {\bf 23}, 2107 (2008)
  [arXiv:0806.2623 [hep-th]].


\bibitem{Hofman:2006xt} 
  D.~M.~Hofman and J.~M.~Maldacena,
  ``Giant Magnons'',
  J.\ Phys.\ A {\bf 39}, 13095 (2006)
  [hep-th/0604135].

\bibitem{Chen:2006gea} 
  H.~Y.~Chen, N.~Dorey and K.~Okamura,
  ``Dyonic Giant Magnons'',
  JHEP {\bf 0609}, 024 (2006)
  [hep-th/0605155].


\bibitem{Rashkov:2008rm} 
  R.~C.~Rashkov,
  ``A Note on the Reduction of the AdS$_4\times$CP$^3$ String Sigma Model'',
  Phys.\ Rev.\ D {\bf 78}, 106012 (2008)
  [arXiv:0808.3057 [hep-th]].


\bibitem{Aharony:2008ug} 
  O.~Aharony, O.~Bergman, D.~L.~Jafferis and J.~Maldacena,
  ``$\mathcal{N} = 6$ superconformal Chern-Simons-matter Theories, M2-branes and their Gravity Duals'',
  JHEP {\bf 0810}, 091 (2008)
  [arXiv:0806.1218 [hep-th]].


\bibitem{Gubser:2002tv} 
  S.~S.~Gubser, I.~R.~Klebanov and A.~M.~Polyakov,
  ``A Semiclassical Limit of the Gauge / String Correspondence'',
  Nucl.\ Phys.\ B {\bf 636}, 99 (2002)
  [hep-th/0204051].

\bibitem{Tseytlin:2004xa} 
  A.~A.~Tseytlin,
  ``Semiclassical Strings and AdS/CFT'',
  [hep-th/0409296].

\bibitem{Plefka:2005bk} 
  J.~Plefka,
  ``Spinning Strings and Integrable Spin Chains in the AdS/CFT Correspondence'',
  Living Rev.\ Rel.\  {\bf 8}, 9 (2005)
  [hep-th/0507136].



\bibitem{Alday:2007hr} 
  L.~F.~Alday and J.~M.~Maldacena,
  ``Gluon Scattering Amplitudes at Strong Coupling'',
  JHEP {\bf 0706}, 064 (2007)
  [arXiv:0705.0303 [hep-th]].

\bibitem{Alday:2009yn} 
  L.~F.~Alday and J.~Maldacena,
  ``Null Polygonal Wilson Loops and Minimal Surfaces in Anti-de-Sitter Space'',
  JHEP {\bf 0911}, 082 (2009)
  [arXiv:0904.0663 [hep-th]].

\bibitem{Kruczenski:2002fb} 
  M.~Kruczenski,
  ``A Note on Twist Two Operators in $\mathcal{N} = 4$ SYM and Wilson Loops in Minkowski Signature'',
  JHEP {\bf 0212}, 024 (2002)
  [hep-th/0210115].

\bibitem{Bern:2005iz} 
  Z.~Bern, L.~J.~Dixon and V.~A.~Smirnov,
  ``Iteration of Planar Amplitudes in Maximally Supersymmetric Yang-Mills Theory at Three Loops and Beyond'',
  Phys.\ Rev.\ D {\bf 72}, 085001 (2005)
  [hep-th/0505205].
  
  

\bibitem{Jevicki:2007aa} 
  A.~Jevicki, K.~Jin, C.~Kalousios and A.~Volovich,
  ``Generating AdS String Solutions'',
  JHEP {\bf 0803}, 032 (2008)
  [arXiv:0712.1193 [hep-th]].
  
\bibitem{Klose:2008rx} 
  T.~Klose and T.~McLoughlin,
  ``Interacting Finite-size Magnons'',
  J.\ Phys.\ A {\bf 41}, 285401 (2008)
  [arXiv:0803.2324 [hep-th]].

\bibitem{Jevicki:2008mm} 
  A.~Jevicki and K.~Jin,
  ``Solitons and AdS String Solutions'',
  Int.\ J.\ Mod.\ Phys.\ A {\bf 23}, 2289 (2008)
  [arXiv:0804.0412 [hep-th]].
  
\bibitem{Hollowood:2009tw} 
  T.~J.~Hollowood and J.~L.~Miramontes,
  ``Magnons, their Solitonic Avatars and the Pohlmeyer Reduction'',
  JHEP {\bf 0904}, 060 (2009)
  [arXiv:0902.2405 [hep-th]].
  
\bibitem{Dorn:2009kq} 
  H.~Dorn, G.~Jorjadze and S.~Wuttke,
  ``On Spacelike and Timelike Minimal Surfaces in AdS$_n$'',
  JHEP {\bf 0905}, 048 (2009)
  [arXiv:0903.0977 [hep-th]].
    

\bibitem{Kruczenski:2004wg} 
  M.~Kruczenski,
  ``Spiky Strings and Single Trace Operators in Gauge Theories'',
  JHEP {\bf 0508}, 014 (2005)
  [hep-th/0410226].
  
\bibitem{Mosaffa:2007ty} 
  A.~E.~Mosaffa and B.~Safarzadeh,
  ``Dual Spikes: New Spiky String Solutions'',
  JHEP {\bf 0708}, 017 (2007)
  [arXiv:0705.3131 [hep-th]].

\bibitem{Kruczenski:2008bs} 
  M.~Kruczenski and A.~A.~Tseytlin,
  ``Spiky Strings, Light-like Wilson Loops and pp-wave Anomaly'',
  Phys.\ Rev.\ D {\bf 77}, 126005 (2008)
  [arXiv:0802.2039 [hep-th]].
  
\bibitem{Jevicki:2009uz} 
  A.~Jevicki and K.~Jin,
  ``Moduli Dynamics of AdS$_3$ Strings'',
  JHEP {\bf 0906}, 064 (2009)
  [arXiv:0903.3389 [hep-th]].
  
\bibitem{Bakas:2002qi} 
  I.~Bakas and C.~Sourdis,
  ``Notes on Periodic Solitons'',
  Fortsch.\ Phys.\  {\bf 50}, 815 (2002)
  [hep-th/0205007].
  
\bibitem{Arutyunov:2003uj} 
  G.~Arutyunov, S.~Frolov, J.~Russo and A.~A.~Tseytlin,
  ``Spinning Strings in AdS$_5\times$S$^5$ and Integrable Systems'',
  Nucl.\ Phys.\ B {\bf 671}, 3 (2003)
  [hep-th/0307191].

\bibitem{Arutyunov:2003za} 
  G.~Arutyunov, J.~Russo and A.~A.~Tseytlin,
  ``Spinning Strings in AdS$_5\times$S$^5$: New Integrable System Relations'',
  Phys.\ Rev.\ D {\bf 69}, 086009 (2004)
  [hep-th/0311004].
  
\bibitem{Ishizeki:2007we} 
  R.~Ishizeki and M.~Kruczenski,
  ``Single Spike Solutions for Strings on S$^2$ and S$^3$'',
  Phys.\ Rev.\ D {\bf 76}, 126006 (2007)
  [arXiv:0705.2429 [hep-th]].



\bibitem{Ryu:2006bv} 
  S.~Ryu and T.~Takayanagi,
  ``Holographic Derivation of Entanglement Entropy from AdS/CFT'',
  Phys.\ Rev.\ Lett.\  {\bf 96}, 181602 (2006)
  [hep-th/0603001].

\end{thebibliography}
\end{document}